\documentclass[preprint,aps,prd,showpacs,nofootinbib,floatfix,tightenlines]{revtex4}
\usepackage{graphicx}
\usepackage{amsmath}
\usepackage{amssymb}
\usepackage{bm}

\begin{document}

\title{\mbox{}\\[10pt]
Using Line Shapes to Discriminate \\ 
between Binding Mechanisms for the $\bm{X(3872)}$}

\author{Pierre  Artoisenet, Eric Braaten, and Daekyoung Kang}
\affiliation{Physics Department, Ohio State University, Columbus,
Ohio 43210, USA}

\date{\today}
\begin{abstract}
We construct line shapes for the $X(3872)$ that generalize 
the Flatt\'e and zero-range line shapes that have been
considered previously.
These line shapes are associated with scattering amplitudes 
that are exactly unitary for real values of the interaction
parameters and can be derived from a
renormalizable quantum field theory.
The new line shapes can be used to discriminate between the 
alternative binding mechanisms in which the $X(3872)$ is 
generated either dynamically by charm meson interactions
or by a resonance near the $D^{*0} \bar D^0$ threshold.
If the resonance is identified with the P-wave charmonium 
state $\chi_{c1}'$, the interaction parameters can be 
constrained by using charmonium phenomenology.
We analyze data on the $X(3872)$ and also data from the Belle and Babar 
Collaborations on the invariant mass distribution of the charm mesons 
from the decay $B \to K + D^{*0} \bar D^0$ up to 4000~MeV.  
Our analysis is compatible with the mechanism for the $X(3872)$
being either a fine-tuning of charm meson interactions or the 
fine-tuning of the $\chi_{c1}'$ to the $D^{*0} \bar D^0$ threshold.  
In particular, the data do not exclude a separate $\chi_{c1}'$ 
resonance between the $D^{*+} D^-$ threshold and 4000~MeV.
\end{abstract}



\maketitle


\section{Introduction}

The discovery of the $X(3872)$ resonance by the Belle Collaboration 
in 2003 \cite{Choi:2003ue}
marked the beginning of a new era in charmonium spectroscopy.
About a dozen new $c \bar c$ mesons above the open charm threshold 
have been discovered and many of them have properties that seem 
incompatible with their identification as conventional charmonium states.
This presents a serious challenge to our understanding 
of the $c \bar c$ sector of QCD.
Of all the new $c \bar c$ mesons, the $X(3872)$ is the one for which 
the most experimental information is available.  
The preponderance of this information implies
that the $X(3872)$ is a charm meson molecule 
whose constituents are a superposition of $D^{*0} \bar D^0$ and $D^0 \bar D^{*0}$.
The structure of this molecule is remarkable,
with the charm mesons almost always very well separated.
However this identification of the $X(3872)$ is not universally accepted 
within the field. The leading alternatives are 
the $^3P_1$ charmonium state $\chi_{c1}(2P)$
or a compact tetraquark $c \bar c$ meson.
Even among those who lean toward its identification as a 
charm meson molecule, the remarkable structure of the $X(3872)$ 
is not universally appreciated.

The only experimental information that is necessary to make the 
identification of the $X(3872)$ as a loosely-bound 
charm-meson molecule is the determination of its quantum numbers 
and the measurements of its mass.
The quantum numbers of the $X(3872)$ 
can be inferred to be $1^{++}$ by combining the following information:
\begin{itemize}
\item
the observation of its decay into $J/\psi \, \gamma$ \cite{Abe:2005ix,Aubert:2006aj} 
or $\psi(2S) \gamma$~\cite{Babar:2008rn}, which implies 
that it is even under charge conjugation,
\item
analyses of the momentum distributions from its decay into 
$J/\psi \, \pi^+ \pi^-$, which imply that its spin and parity are 
$1^+$ or $2^-$ \cite{Abe:2005iya,Abulencia:2006ma},
\item
either the observation of its decays into 
$D^0 \bar D^0 \pi^0$ \cite{Gokhroo:2006bt},
which disfavors $2^-$ because of angular-momentum suppression,
or the observation of its decay into 
$\psi(2S) \, \gamma$ \cite{Babar:2008rn},
which disfavors $2^-$ because of multipole suppression.
\end{itemize}
A recent analysis of decays into $J/\psi \, \pi^+ \pi^- \pi^0$
by the Babar Collaboration favors the quantum numbers $2^{-+}$, 
but does not exclude $1^{++}$ \cite{al.:2010jr}.  
In the absence of definitive evidence to the contrary, 
we will assume that the quantum numbers of the $X(3872)$ are $1^{++}$.

The mass of the $X(3872)$ can be determined
most directly by
measurements in the $J/\psi\, \pi^+ \pi^-$ decay channel.
As pointed out in Ref.~\cite{Braaten:2007dw}, measurements in the 
$D^0 \bar D^0 \pi^0$ decay channel are biased by the associated 
threshold enhancement just above the $D^{*0} \bar D^0$ threshold.
As pointed out in Ref.~\cite{Braaten:2007ft}, measurements in the 
$D^{*0} \bar D^0$ decay channel are biased by the analysis procedure 
in which $D^0 \pi^0$ with invariant mass near the mass of $D^{*0}$ 
is constrained to have invariant mass exactly equal to $M_{*0}$.
This procedure assigns an energy above the $D^{*0} \bar D^0$ threshold
to a $D^0 \bar D^0 \pi^0$ event whose energy is below the threshold.
Using the most recent measurements of the mass of the $X(3872)$
in the $J/\psi\, \pi^+ \pi^-$ 
decay channel by the Belle, CDF, Babar, and D0 Collaborations 
\cite{Abazov:2004kp,Aubert:2008gu,Belle:2008te,Aaltonen:2009vj}, 
the combined average for the position of the $X(3872)$ resonance
relative to the $D^{*0} \bar D^0$ threshold is
\begin{equation}
M_X - (M_{*0} + M_0) = -0.42 \pm 0.39~{\rm MeV} , 
\label{MX-ave}
\end{equation}
where $M_{*0}$ and $M_0$ are the masses of $D^{*0}$ and $D^0$.

The reason the quantum numbers $1^{++}$ and the measurement in 
Eq.~(\ref{MX-ave}) are sufficient to determine the nature of the $X(3872)$
is that quantum mechanics implies that an S-wave resonance 
whose energy is sufficiently close to the threshold has 
universal properties that are determined by its energy.
If the energy is below the threshold,
the state is a bound molecule consisting of pairs of 
particles that are almost always very well separated.
One of the universal predictions is a relation between the 
mean-square separation of the constituents and the binding energy $E_X$:
$\langle r^2 \rangle_X = (4 \mu E_X)^{-1}$, 
where $\mu$ is the reduced mass. 
In the case of the $X(3872)$,
the quantum numbers $1^{++}$ imply that there is an S-wave 
coupling to $D^{*0} \bar D^0$.  The tiny energy relative to the
$D^{*0} \bar D^0$ threshold implies that it is a resonant coupling.  
Thus the $X(3872)$ is an S-wave threshold resonance.
The binding energy given by Eq.~(\ref{MX-ave}) implies that the
root-mean-square separation of the charm mesons in
the $X(3872)$ is $\langle r^2 \rangle_X^{1/2} = 4.9^{+13.4}_{-1.3}$~fm. 
This huge separation of the charm mesons makes the $X(3872)$ 
a truly remarkable hadron.

One should distinguish between the nature of the $X(3872)$, 
which is a weakly-bound charm meson molecule, and its origin, 
which has to do with the binding mechanism for the molecule.
Identifying the origin of the $X(3872)$ is crucial to understanding 
its implications for the other new $c \bar c$ mesons above the
$D \bar D$ threshold.
Universality is a double-edged sword.
While it allows the nature of the $X(3872)$ to be determined 
unambiguously from limited experimental information, 
universality makes it more difficult to identify the origin of the state.  
There are two primary candidates for the binding mechanism of the 
$X(3872)$:
\begin{itemize}
\item
\underline{dynamical}.
The $X(3872)$ could be generated dynamically by the interactions between 
the charm mesons.  The interaction between $D^{*0}$
and $\bar D^0$ in the isospin-0 $1^{++}$ channel could be tuned to near 
the critical strength for the formation of a bound state.
\item
\underline{resonance}.
The $X(3872)$ could be generated by an isospin-0 resonance 
whose energy is tuned to near the 
$D^{*0} \bar D^0$ threshold.  An obvious candidate for the resonance is the 
$\chi_{c1}(2P)$, but it could also be any other type of $c \bar c$ 
meson with the appropriate quantum numbers, such as a compact tetraquark.
\end{itemize}

One set of clues to the origin of the $X(3872)$ is its decay pattern.
There are 6 decay modes that have been observed thus far:
$J/\psi \, \pi^+ \pi^-$,
$J/\psi \, \pi^+ \pi^- \pi^0$~\cite{Abe:2005ix}, 
$J/\psi \, \gamma$~\cite{Abe:2005ix,Babar:2008rn}, 
$\psi(2S) \gamma$~\cite{Babar:2008rn}
$D^0 \bar D^0 \pi^0$~\cite{Gokhroo:2006bt}, 
and $D^0 \bar D^0 \gamma$~\cite{Aubert:2007rva,Belle:2008su}.
In order to exploit this information, one would need to understand 
the pattern of branching fractions that follow from each of the binding mechanisms.
There is a well-developed phenomenology for decays of charmonium states,
but there is no corresponding phenomenology for decays of 
compact tetraquark $c \bar c$ mesons or for the inelastic scattering
of charm meson pairs.  This makes it difficult to constrain 
the origin of the $X(3872)$ using the observed branching ratios.

Another class of observables that can provide clues to the origin 
of the $X(3872)$ is the energy distribution or line shape in specific decay
channels. 
At energies much less than 8~MeV from the $D^{*0} \bar D^0$ threshold,
the line shapes in these and other channels are universal in the sense 
that they are determined only by the 
binding energy and width of the $X(3872)$ resonance \cite{Braaten:2007dw}.
The universal line shapes in the $D^0 \bar D^0 \pi^0$ and 
$D^0 \bar D^0 \gamma$ channels, which receive contributions from the 
decay of a constituent $D^{*0}$ or $\bar D^{*0}$, are different 
from those in other decay channels, such as $J/\psi \, \pi^+ \pi^-$.
Braaten and Lu presented simple analytic expressions for the universal line shapes 
that take into account the width of the $D^{*0}$ and inelastic scattering 
channels for $D^{*0} \bar D^0$ \cite{Braaten:2007dw}.
The accuracy of these line shapes in the $D^{*0} \bar D^0$ threshold
region was verified in Ref.~\cite{Hanhart:2010wh}.

There have been several theoretical analyses of the line shapes 
for $X(3872)$ produced by the decays $B \to K+X$.
The specific decay channels for which the line shapes have been 
measured by the Belle and Babar collaborations are $J/\psi \, \pi^+ \pi^-$
\cite{Acosta:2003zx,Aubert:2004ns,Abazov:2004kp,%
Aubert:2008gu,Belle:2008te,Aaltonen:2009vj},
$D^0 \bar D^0 \pi^0$~\cite{Gokhroo:2006bt}, 
and $D^{*0} \bar D^0$~\cite{Aubert:2007rva,Belle:2008su}.
Hanhart et al.\ used a Flatt\'e parameterization of the line shapes
to analyze the data on the  $J/\psi \, \pi^+ \pi^-$ and
$D^0 \bar D^0 \pi^0$ channels \cite{Hanhart:2007yq}.
They concluded that the data favored the $X(3872)$ being a virtual state 
with energy above the $D^{*0} \bar D^0$ threshold.  Their analysis 
was biased towards such a result, because they assumed that the 
line shape in $D^0 \bar D^0 \pi^0$ vanishes below the 
$D^{*0} \bar D^0$ threshold, thus ignoring any contributions from a 
resonance below the $D^{*0} \bar D^0$ threshold.
Braaten and Lu used the universal line shapes to analyze data 
on the $J/\psi \, \pi^+ \pi^-$ and $D^0 \bar D^0 \pi^0$ channels 
\cite{Braaten:2007dw}. They concluded that the data favored the 
$X(3872)$ being a bound state 
with energy below the $D^{*0} \bar D^0$ threshold.  
Zhang, Meng, and Zheng \cite{Zhang:2009bv}
followed Ref.~\cite{Hanhart:2007yq} in using the Flatt\'e line shapes
and ignoring $D^0 \bar D^0 \pi^0$ and $D^0 \bar D^0 \gamma$ events 
from a resonance below the $D^{*0} \bar D^0$ threshold.
They concluded that $X(3872)$ can be identified 
as a $^3P_1$ charmonium state  that is strongly distorted by coupled-channel effects.
Braaten and Stapleton used the universal line shapes 
to analyze data on the $J/\psi \, \pi^+ \pi^-$ and $D^{*0} \bar D^0$
channels \cite{Stapleton:2009ey}.  They pointed out that the analysis 
procedure for $D^{*0} \bar D^0$ in Refs.~\cite{Aubert:2007rva,Belle:2008su}
assigns an energy above the $D^{*0} \bar D^0$ threshold to a 
$D^0 \bar D^0 \pi^0$ or $D^0 \bar D^0 \gamma$ event from decay of a resonance 
just below the $D^{*0} \bar D^0$ threshold.  When this effect is taken into account, 
the analysis of the $D^{*0} \bar D^0$ data favors the 
$X(3872)$ being a bound state 
with energy below the $D^{*0} \bar D^0$ threshold.  
Kalashnikova and Nefediev \cite{Kalashnikova:2009gt}
followed Ref.~\cite{Hanhart:2007yq} in using the Flatt\'e line shapes
and ignoring $D^0 \bar D^0 \pi^0$ and $D^0 \bar D^0 \gamma$ events 
from a resonance below the $D^{*0} \bar D^0$ threshold.
They concluded that the Babar data prefers the
$X(3872)$ to be a virtual state with a small $^3P_1$ charmonium component
while the Belle data prefers a bound state 
with a $^3P_1$ charmonium component of about 30\%. 

Since the lines shapes very close to the $D^{*0} \bar D^0$ threshold
are universal, information about the origin of the $X(3872)$ 
can only come from the energy distributions away from the threshold.
The line shapes outside the universal region depend on the binding mechanism.
The Flatt\'e line shapes used in 
Refs.~\cite{Hanhart:2007yq,Zhang:2009bv,Kalashnikova:2009gt}
can be derived from the assumption that the charm mesons scatter 
only through their couplings to a resonance with isospin 0.
These would be the appropriate line shapes if the $X(3872)$ 
arises from a tuning of the energy of a resonance.
Alternative line shapes have been derived by Braaten and Lu 
under the assumption that the charm mesons scatter through 
zero-range interactions between the coupled channels consisting 
of neutral charm mesons and charged charm mesons \cite{Braaten:2007ft}.  
These would be the appropriate line shapes if the $X(3872)$ arises 
dynamically from interactions between the charm mesons.
The zero-range line shapes in the $J/\psi \, \pi^+ \pi^-$
and $J/\psi \, \pi^+ \pi^- \pi^0$ channels 
had been considered previously by Voloshin \cite{Voloshin:2007hh}, 
but his results were incorrect because of conceptual errors related 
to the treatment of isospin symmetry \cite{Braaten:2007ft}.

In order to discriminate between the two binding mechanisms identified above, 
it is necessary to use line shapes that allow for either possibility.
There are two criteria that we use as guiding principles 
in constructing the line shapes:
\begin{itemize}
\item
\underline{unitarity}.
The line shapes should correspond to multichannel scattering amplitudes 
that are exactly unitary for real values of the interaction parameters.
Complex deformations of the parameters can then be used to take into account 
effects of additional channels that are not treated explicitly.
\item
\underline{renormalizability}.
The scattering amplitudes should be derivable from a renormalizable 
local quantum field theory.
This guarantees that sensitivity to physics at much higher energies 
can be absorbed into the interaction parameters.
\end{itemize}
In a local quantum field theory, the production of particles 
by a short-distance process can be represented by local operators.
An advantage of a renormalizable field theory is that one can construct
renormalized operators 
whose matrix elements are insensitive to physics at much higher energies.
The line shapes from that production process are then determined 
by the interaction parameters and by the short-distance coefficients 
of those renormalized operators.

In this paper, we present a solution to the coupled-channel problem 
for pairs of neutral charm mesons $D^{*0} \bar D^0$ and $D^0 \bar D^{*0}$
and pairs of charged charm mesons $D^{*+} D^-$ and $D^+ D^{*-}$
that scatter through zero-range interactions and also through an 
isospin-0 resonance. The resulting scattering amplitudes satisfy 
the conditions of unitarity and renormalizability.
They depend on 4 interaction parameters 
and reduce to the Flatt\'e scattering amplitudes 
of Ref.~\cite{Hanhart:2007yq} and to the zero-range scattering amplitudes
of Ref.~\cite{Braaten:2007ft} in the appropriate limits.  
We also solve the renormalization problem 
for the local operators that represent the production at short distances
of pairs of neutral charm mesons, pairs of charged charm mesons,
and the resonance.  The line shapes for $X(3872)$ produced by the 
decays $B \to K + X$ depend on the interaction parameters 
and on the short-distance coefficients associated with the 
$B \to K$ transition.  We assume that the 
resonance is the $\chi_{c1}(2P)$ and we include constraints on the 
interaction parameters and on the $B \to K$ short-distance coefficients
from charmonium phenomenology.  
We use our line shapes to analyze data on the $X(3872)$ 
and data from the Belle and Babar collaborations on the production of 
$D^{*0} \bar D^0$ above the threshold up to 4000~MeV.  We try to determine
whether the data can discriminate between the two mechanisms for generating 
the $X(3872)$ that were described above.

We begin in Section~\ref{sec:widths} by establishing our notation.
In Section~\ref{sec:DDscat}, we summarize the zero-range and Flatt\'e 
scattering amplitudes that have been used in previous analyses
and we present more general scattering amplitudes 
that include both as special cases.
In Section~\ref{sec:lowEparameters}, we present experimental and 
phenomenological constraints on the
interaction parameters in the low-energy scattering amplitudes.
In Section~\ref{sec:lineshape}, we express the line shapes of the $X(3872)$ 
from $B$ meson decays in terms of short-distance coefficients associated with the 
$B \to K$ transition. In Section~\ref{sec:BtoKcoeffs}, we present experimental and 
phenomenological constraints on the short-distance coefficients.
In Section~\ref{sec:analysis}, we analyze data on $X(3872)$ 
and data on $B \to K + D^{*0} \bar D^0$  
to see whether they can discriminate between the two binding mechanisms.
We summarize our results in Section~\ref{sec:summary}.
In an Appendix, we present the quantum field theory formulation of the 
problem of charm mesons that scatter through both zero-range interactions 
and coupling to a resonance.  We determine the renormalization of the interaction 
parameters in the scattering amplitudes.
We also solve the renormalization problem for the local operators that create the 
charm meson pairs and the resonance.

\section{Notation}
\label{sec:widths}

The standard isospin multiplets for the charm mesons are
$( -D^+, D^0)$, $( \bar D^{0}, D^{-})$, $( -D^{*+}, D^{*0})$,
and $( \bar D^{*0}, D^{*-})$,
where the first and second states are the upper and lower
components of the multiplet, respectively.
The most natural charge-conjugation phase conventions
are $C D^0 = + \bar D^0$ and $C D^{*0} = - \bar D^{*0}$.
The $D^* \bar D$ channels with charge conjugation quantum number
$C=+$ are then \cite{Liu:2008fh}
\begin{subequations}
\begin{eqnarray}
(D^{*0} \bar D^0)_+ &=&  \mbox{$\frac{1}{\sqrt{2}}$}
\left( D^{*0} \bar D^0 - D^0 \bar D^{*0} \right) ,
\label{D*Dbar0}
\\
(D^{*+} D^-)_+ &=&  -\mbox{$\frac{1}{\sqrt{2}}$}
\left( D^{*+} D^- - D^+ D^{*-} \right).
\label{D*Dbar1}
\end{eqnarray}
\label{D*Dbar}
\end{subequations}
We will refer to $(D^{*0} \bar D^0)_+$ and $(D^{*+} D^-)_+$
as the neutral and charged channels, respectively. 

We use concise notation for the 
masses and widths of the charm mesons.
We denote the masses of $D^0$, $D^+$, $D^{*0}$, and $D^{*+}$
by $M_0$, $M_1$, $M_{*0}$, and $M_{*1}$, respectively.
(The numerical subscript is the absolute value of the
electric charge of the meson.)
The reduced mass for $D^{*0} \bar D^0$ is $\mu = 966.5$~MeV.
The reduced mass for $D^{*+} D^-$ is larger by about 0.3\%,
but we will ignore this difference.
We measure energies relative to the 
$D^{*0} \bar D^0$ threshold at $M_{*0} + M_0 = 3871.9$~MeV.
The energy splitting 
between the $D^{*+} D^{-}$ and $D^{*0} \bar D^0$ thresholds is 
\begin{eqnarray}
\nu_{11} \equiv (M_{*1} + M_1) - (M_{*0} + M_0) = 8.1~{\rm MeV} .
\label{nu-def}
\end{eqnarray}
The corresponding momentum scale is 
\begin{eqnarray}
\kappa_{11} \equiv \sqrt{2 \mu \nu_{11}} = 125 ~{\rm MeV} .
\label{kappa11-def}
\end{eqnarray}

The total width of the $D^{*+}$ is measured and
the total width of the $D^{*0}$ can be predicted 
from measurements of $D^*$ decays and
isospin symmetry \cite{Braaten:2007dw}.
We denote these widths by $\Gamma_{*1}$ and $\Gamma_{*0}$, respectively.
The PDG value for $\Gamma_{*1}$ \cite{Amsler:2008zzb}
and the predicted value for $\Gamma_{*0}$ are
\begin{subequations}
\begin{eqnarray}
\Gamma_{*1} &=&  96 \pm 22 \  {\rm keV} ,
\label{GamD*1}
\\
\Gamma_{*0} &=& 66 \pm 15 \  {\rm keV} .
\label{GamD*0}
\end{eqnarray}
\end{subequations}

The effects of decays of the $D^{*0}$ and $D^{*+}$ 
on charm meson scattering can be partially taken into account through 
energy-dependent widths.
In Ref.~\cite{Braaten:2007dw}, energy-dependent widths $\Gamma_{*0}(E)$
and $\Gamma_{*1}(E)$ that depend on the energy $E$ of the 
pair of charm mesons in their center-of-mass frame were defined 
by scaling the physical partial widths for the decays $D^* \to D \pi$.
The use of these energy-dependent widths above the $D^* \bar D$
thresholds was an error.
If the energy is above the $D^{*0} \bar D^0$ threshold 
at $E =0$, the appropriate width for the $D^{*0}$ is the physical width 
$\Gamma_{*0}$.  If the energy is above the $D^{*+} D^-$ threshold 
at $E= \nu_{11}$, the appropriate width for the $D^{*+}$ 
is the physical width $\Gamma_{*1}$.
Below the $D^* \bar D$ thresholds, the energy-dependent widths 
of Ref.~\cite{Braaten:2007dw} give thresholds at the correct energies 
for the 3-body $D \bar D \pi$ states.
However, as pointed out in \cite{Hanhart:2010wh},
they do not give the correct energy dependence just above these thresholds.
Moreover, the effects of the energy-dependent widths 
on the line shape of the $X(3872)$ are numerically small.
We will therefore ignore any energy dependence of the widths 
in this paper.


\section{Low-energy $\bm{D^{*} {\bar D}}$ Scattering}
\label{sec:DDscat}

In this section, we discuss the low-energy scattering of 
the charm mesons $D^*$ and $\bar D$.
We first summarize the universal scattering amplitude of  
Ref.~\cite{Braaten:2007dw},
which takes into account the large scattering length 
in the neutral channel.
We then describe the coupled-channel scattering amplitudes 
of Ref.~\cite{Braaten:2007ft},
which take into account zero-range scattering in the neutral 
and charged channels, and the Flatt\'e scattering amplitudes 
introduced in Ref.~\cite{Hanhart:2007yq},
which take into account scattering through a resonance.
Finally, we present more general scattering amplitudes that allow for  
scattering both through a resonance and through zero-range interactions.

\subsection{Universal scattering amplitude}
\label{sec:universal}

We first consider neutral charm meson pairs with scattering only in the 
channel $(D^{*0} \bar D^0)_+$ defined in Eq.~(\ref{D*Dbar0}).
The transition amplitude $\mathcal{A}(E)$ 
for the scattering of nonrelativistically normalized
charm meson pairs can be written in the form
\begin{equation}
\mathcal{A}(E) = \frac{2 \pi}{\mu} f(E) ,
\end{equation}
where $f(E)$ is the conventional nonrelativistic scattering amplitude
expressed as a function of the total energy $E$ of the charm mesons 
in the center-of-mass frame.
We measure $E$ relative to the $D^{*0} \bar D^0$ threshold. 
The universal scattering amplitude for an S-wave threshold resonance is 
\begin{equation}
f(E)  \equiv \frac{1}{- \gamma + \kappa(E)} ,
\label{f-E}
\end{equation}
where $\kappa(E)= (- 2 \mu E - i \varepsilon)^{1/2}$
and $\gamma$ is the inverse scattering length.
If $\gamma$ is a real parameter, $f(E)$ satisfies the constraints 
of unitarity for a single-channel system exactly.  
This universal scattering amplitude can be derived from a renormalizable 
nonrelativistic quantum field theory with a contact interaction 
in a single scattering channel.

The imaginary part of $f(E)$ 
can be interpreted as a spectral function for the resonance.
The spectral function associated with the scattering amplitude 
in Eq.~(\ref{f-E}) is
\begin{equation}
{\rm Im} \, f(E)  =
| f(E) |^2 
\left[ {\rm Im}  \gamma - {\rm Im} \kappa(E) \right] .
\label{ImA-optical2}
\end{equation}
If $\gamma$ is a real parameter, the spectral function reduces to
\begin{equation}
{\rm Im} \, f(E)  =
\theta(E) |f(E)|^2 \sqrt{2 \mu E}
+ \theta(\gamma) \frac{\pi \gamma}{\mu} \delta ( E + \gamma^2/(2 \mu)) .
\label{ImA-optical:1ch}
\end{equation}
There is a threshold enhancement associated with 
production of $D^{*0} \bar D^0$ and $D^0 \bar D^{*0}$
with a peak at $E = +\gamma^2/(2 \mu)$.
If $\gamma > 0$, there is also a delta function contribution
at  $E = -\gamma^2/(2 \mu)$ associated with a bound state with
binding energy $\gamma^2/(2 \mu)$.

Scattering in the $(D^{*0} \bar D^0)_+$ channel cannot be exactly 
unitary, because the $D^{*0}$ has a nonzero width and because 
the charm meson pair has inelastic scattering channels.
Following Ref.~\cite{Braaten:2007dw}, 
the dominant effects of decays of the $D^{*0}$ can be taken 
into account by replacing $\kappa(E)$ by
\begin{equation}
\kappa(E)  = \sqrt{- 2 \mu E - i \mu \Gamma_{*0}} ,
\label{kappa-E}
\end{equation}
where $\Gamma_{*0}$ is the width
of the $D^{*0}$ given in Eq.~(\ref{GamD*0}).
Following Ref.~\cite{Braaten:2007dw}, the effects of inelastic scattering 
channels for the charm-meson pair other than $D^0 \bar D^0 \pi^0$ and
$D^0 \bar D^0 \gamma$ can be taken into account by making
$\gamma$ a complex parameter with a positive imaginary part.  
In the expression for the imaginary part of  
$f(E)$ in Eq.~(\ref{ImA-optical2}), the term proportional to 
${\rm Im} \kappa(E)$ is the contribution from channels 
whose ultimate final states are $(D^0 \bar D^0 \pi^0,D^0 \bar D^0 \gamma)$, 
including $D^{*0} \bar D^0$ and $D^0 \bar D^{*0}$.
The term proportional to ${\rm Im} \gamma$ 
is the contribution from all other channels, including $J/\psi \, \pi^+ \pi^-$.
The ${\rm Im} \kappa(E)$ term has a threshold
enhancement just above the $D^{*0} \bar D^0$ threshold.
If Re$(\gamma)>0$, 
both terms have a resonant peak just below the threshold that
can be identified with the $X(3872)$.  Its
position and width are determined by the pole 
of the scattering amplitude $f(E)$ in Eq.~(\ref{f-E}).
The complex energy of the pole can be expressed as
\begin{equation}
E_{\rm pole} = - \gamma^2/(2 \mu) - i \Gamma_{*0}/2 .
\label{Epole}
\end{equation}
Quantitative constraints on the real and 
imaginary parts of $\gamma$ are presented in Section~\ref{sec:scatlength}.

\subsection{Zero-Range scattering amplitudes}
\label{sec:zero-range}

In Ref.~\cite{Braaten:2007ft},
the universal scattering amplitude of Section~\ref{sec:universal} 
was generalized to the case of coupled channels 
$(D^{*0} \bar D^0)_+$ and $(D^{*+} D^-)_+$ defined by Eqs.~(\ref{D*Dbar})
that scatter through zero-range interactions.

\subsubsection{General case}

We label the two channels $(D^{*0} \bar D^0)_+$ and $(D^{*+} D^-)_+$ 
by the indices 0 and 1, respectively.
The transition amplitudes $\mathcal{A}_{ij}(E)$ among these two channels
define scattering amplitudes $f_{ij}(E)$ that depend on the total energy $E$
relative to the $D^{*0} \bar D^0$ threshold:
\begin{equation}
\mathcal{A}_{ij}(E) = 
\frac{2 \pi}{\mu} 
f_{ij}(E) .
\label{Aij-fij}
\end{equation}
If the charm mesons scatter through a zero-range interaction,
the inverse of the $2 \times 2$ 
matrix of scattering amplitudes has the form
\begin{equation}
f(E)^{-1} = - \Lambda^{-1} + K(E) ,
\label{f-inverse}
\end{equation}
where $\Lambda$ is $2 \times 2$ symmetric matrix.
The dependence on the energy $E$ is in the diagonal matrix
\begin{equation}
K(E) = 
\left( \begin{array}{cc}
      \kappa(E) & 0 \\
      0       & \kappa_1(E)
      \end{array} \right) ,
\label{K-kappa}
\end{equation}
whose diagonal entries are 
$\kappa(E)= (- 2 \mu E - i \varepsilon)^{1/2}$ and
$\kappa_1(E)  = (- 2 \mu (E - \nu_{11}) - i \varepsilon)^{1/2}$.
If the three interaction parameters $\Lambda_{00}$, $\Lambda_{01}$, 
and $\Lambda_{11}$ are all real valued, the scattering 
amplitudes $f_{ij}(E)$ satisfy the constraints 
of unitarity for this two-channel system exactly.
They can be derived from 
a renormalizable nonrelativistic quantum field theory 
with zero-range interactions.

\subsubsection{Isospin symmetry}

The approximate isospin symmetry of QCD reduces the three 
interaction parameters $\Lambda_{ij}$ to two independent parameters
$\gamma_0$ and $\gamma_1$:
\begin{equation}
\Lambda =
(1/\gamma_0)~\mathcal{P}_0 + (1/\gamma_1)~\mathcal{P}_1,
\label{Lambda-gamma}
\end{equation}
where the matrices $\mathcal{P}_0$ and $\mathcal{P}_1$
are projectors onto the isospin 0 and 1 channels:
\begin{subequations}
\begin{eqnarray}
\mathcal{P}_0 &=&
\frac12 \left( \begin{array}{cc}
~~1 &  -1 \\
 -1 & ~~1 
      \end{array} \right) ,
\label{proj0}
\\
\mathcal{P}_1 &=&
\frac12 \left( \begin{array}{cc}
1 & 1 \\
1 & 1 
      \end{array} \right) .
\label{proj1}
\end{eqnarray}
\label{proj}
\end{subequations}
The scattering amplitudes obtained by inverting
Eq.~(\ref{f-inverse}) reduce to
\begin{subequations}
\begin{eqnarray}
f_{00}(E) &=&
\frac{- (\gamma_0 + \gamma_1) + 2 \kappa_1(E)}{D_0(E)} ,
\label{f00-E}
\\
f_{01}(E) &=&
\frac{\gamma_1 - \gamma_0}{D_0(E)} ,
\label{f01-E}
\\
f_{11}(E) &=&
\frac{- (\gamma_0 + \gamma_1) + 2 \kappa(E)}{D_0(E)} ,
\label{f11-E}
\end{eqnarray}
\label{fij-E}
\end{subequations}
where the denominator is 
\begin{equation}
D_0(E) = 2 \gamma_1 \gamma_0 
- (\gamma_1 + \gamma_0) [\kappa_1(E) +  \kappa(E)]
+ 2 \kappa_1(E) \kappa(E) .
\label{D-E}
\end{equation}
At energies $E$ far from the charm meson thresholds at 0 and $\nu_{11}$,
the difference between $\kappa_1(E)$ and $\kappa(E)$ can be neglected
and the isospin symmetry becomes exact.  By considering this limit,
we can identify $\gamma_0$ and $\gamma_1$ as the isoscalar 
and isovector inverse scattering lengths, respectively.
We will refer to the model defined by the scattering amplitudes
in Eqs.~(\ref{fij-E}) as the {\it Zero-Range model}.

\subsubsection{Optical theorem}

Scattering in the $(D^{*0} \bar D^0)_+$ and $(D^{*+} D^-)_+$ 
channels cannot be exactly unitary, because the $D^{*0}$
and $D^{*+}$ have nonzero widths and because 
the charm meson pairs have inelastic scattering channels.
Following Ref.~\cite{Braaten:2007ft}, 
the dominant effects of decays of $D^{*0}$ and $D^{*+}$ 
can be taken into account by replacing $\kappa(E)$ 
by the expression in Eq.~(\ref{kappa-E}) and by replacing 
$\kappa_1(E)$ by
\begin{equation}
\kappa_1(E)  =
\sqrt{- 2 \mu (E - \nu_{11}) - i \mu \Gamma_{*1}} ,
\label{kappa1-E}
\end{equation}
where $\Gamma_{*1}$ is the width
of the $D^{*+}$ given in Eq.~(\ref{GamD*1}).
The contribution of $\Gamma_{*1}$ to the imaginary part of 
$\kappa_1(E)$ is only important near the $D^{*+} D^-$ threshold.
Near the $D^{*0} \bar D^0$ threshold, $\kappa_1(E)$ is 
well approximated by the real quantity $\kappa_{11} = 125$~MeV.
Following Ref.~\cite{Braaten:2007ft}, the effects of 
inelastic scattering channels
other than $D \bar D \pi$ and $D \bar D \gamma$ can be taken 
into account by taking $\gamma_{0}$ and $\gamma_{1}$ 
to be complex parameters with positive imaginary parts.

The imaginary parts of the scattering amplitudes in 
Eq.~(\ref{fij-E}) can be expressed in forms that are consistent 
with the Cutkosky cutting rules:
\begin{eqnarray}
\mathrm{Im} \, f_{ij}(E) &=&
\mathrm{Im} \gamma_0~\sum_{k,l} f_{ik}(E)~\mathcal{P}_{0,kl}~f_{lj}^*(E) 
+ \mathrm{Im} \gamma_1~\sum_{k,l} f_{ik}(E)~\mathcal{P}_{1,kl}~f_{lj}^*(E)
\nonumber\\
&&
- \mathrm{Im} \kappa(E)~f_{i0}(E)~f_{0j}^*(E)
- \mathrm{Im} \kappa_1(E)~f_{i1}(E)~f_{1j}^*(E).
\label{ImAij}
\end{eqnarray}
The terms in Eq.~(\ref{ImAij}) proportional to ${\rm Im} \kappa(E)$ and 
${\rm Im} \kappa_1(E)$ are the contributions from channels 
whose ultimate final states are $(D \bar D \pi, D \bar D \gamma)$, 
including $D^* \bar D$ and $D \bar D^*$.
The terms proportional to ${\rm Im} \gamma_{0}$ 
and ${\rm Im} \gamma_{1}$ correspond to other
inelastic scattering channels with isospin 0 and 1, respectively.

\subsubsection{$D^{*0} \bar{D}^0$ threshold region}

The interaction parameters $\gamma_0$ and $\gamma_1$ can be tuned
so that there is a bound state just below the 
$D^{*0} \bar D^0$ threshold that can be identified with the $X(3872)$.
The scattering amplitudes $f_{ij}(E)$ have a pole
at a complex energy 
$E_{\rm pole}$ that satisfies Eq.~(\ref{Epole}) with
$\gamma \equiv \kappa(E_{\rm pole})$.
If the tiny difference between $\kappa_1(E_{\rm pole})$ and
$\kappa_{11}$ is neglected, the vanishing of the denominator $D_0(E)$ 
in Eq.~(\ref{D-E}) reduces to a linear equation for $\gamma$, 
whose solution is
\begin{equation}
\gamma =
\frac{2 \gamma_1 \gamma_0 - (\gamma_1 + \gamma_0) \kappa_{11}}
     {\gamma_1 + \gamma_0 - 2 \kappa_{11}}.
\label{gamma-2ch}
\end{equation}
An inverse scattering length $\gamma$ that is small compared to $\kappa_{11}$
requires a fine-tuning of $\gamma_0$ and $\gamma_1$ so that
$2 \gamma_1 \gamma_0 \approx (\gamma_1 + \gamma_0) \kappa_{11}$.
If $|\gamma_1| \gg \kappa_{11}$, $\gamma_0$ must be 
fine-tuned to near $\kappa_{11}/2 \approx +63$~MeV.
Since line shapes near the $D^{*0} \bar D^0$ threshold
are extremely sensitive to $\gamma$, it is advantageous
to take $\gamma$ to be one of the independent interaction parameters.
This can be accomplished by eliminating $\gamma_0$
in favor of $\gamma$ using 
\begin{equation}
\gamma_0 =
\frac{\gamma_1 \kappa_{11} + ( \gamma_1 - 2 \kappa_{11}) \gamma}
     {(2 \gamma_1 - \kappa_{11}) - \gamma},
\label{gamma0-2ch}
\end{equation}
which follows from Eq.~(\ref{gamma-2ch}).

The scattering amplitudes $f_{ij}(E)$ in Eqs.~(\ref{fij-E})
all have poles in the energy variable $\kappa(E)$
at $\kappa(E) = \gamma$.  The residue
of the pole of $f_{ij}(E)$ has the form $Z_i^{1/2} Z_j^{1/2}$.
The ratio of the residue factors is 
\begin{eqnarray}
\frac{Z_{1}^{1/2}}{Z_{0}^{1/2}} =
- \frac{\gamma_1 - \gamma}{\gamma_1 - \kappa_{11}} .
\label{Z1Z0}
\end{eqnarray}
If the tiny difference between 
$\kappa_1(E_{\rm pole})$ and $\kappa_{11}$ is neglected, 
the residue $Z_{0}$ is
\begin{eqnarray}
Z_{0} = \left( 1 + 
\frac{(\gamma_1 - \gamma)^2 \gamma}{(\gamma_1 - \kappa_{11})^2 \kappa_{11}}
\right)^{-1} .
\label{ZX00}
\end{eqnarray}
The behavior of the elastic scattering amplitude $f_{00}(E)$
in the entire $D^{*0} \bar D^0$ threshold region defined by
$|E| \ll \nu_{11}$ is dominated by the pole at  
$\kappa(E) = \gamma$.  It reduces in this region
to $Z_0 f(E)$, where $f(E)$ is the universal elastic scattering amplitude 
in Eq.~(\ref{f-E}). 
As $\gamma \to 0$, $Z_{0}$ approaches 1.

\subsection{Flatt\'e scattering amplitudes}
\label{sec:Flatte}

In Ref.~\cite{Hanhart:2007yq}, Hanhart, Nefediev, and Kalashnikova
proposed Flatt\'e line shapes for the $X(3872)$ resonance.
The  Flatt\'e scattering amplitudes can be derived by assuming that the
coupled channels $(D^{*0} \bar D^0)_+$ and $(D^{*+} D^-)_+$ 
scatter only through their couplings to a resonance.

\subsubsection{Isospin symmetry}

If two channels scatter only through their couplings to an isospin-0 resonance,
the entries of the $2 \times 2$ matrix of scattering amplitude 
defined by Eq.~(\ref{Aij-fij}) are
\begin{eqnarray}
f_{00}(E) = f_{11}(E) = -f_{01}(E) = f_\textrm{Flatt\'e}(E),
\label{fij-Flatte}
\end{eqnarray}
where $f_\textrm{Flatt\'e}(E)$ is the  Flatt\'e scattering amplitude:
\begin{eqnarray}
f_\textrm{Flatt\'e}(E) \equiv
\frac{-g^2/2}{E - \nu - (g^2/2)[\kappa_1(E) + \kappa(E)]} .
\label{f-Flatte}
\end{eqnarray}
The threshold functions $\kappa(E)$ and $\kappa_1(E)$
are given in Eqs.~(\ref{kappa-E}) and (\ref{kappa1-E}).
If we take $g$ to be real,
the spectral function associated with the Flatt\'e scattering 
amplitude in Eq.~(\ref{f-Flatte}) is
\begin{eqnarray}
\textrm{Im} f_\textrm{Flatt\'e}(E) =
| f_\textrm{Flatt\'e}(E) |^2
\left[ - \frac{2}{g^2} \textrm{Im} \nu 
- \textrm{Im} \kappa_1(E) - \textrm{Im} \kappa(E) \right] .
\label{Imf-Flatte}
\end{eqnarray}

The notation of Ref.~\cite{Hanhart:2007yq}
can be obtained by the substitutions 
\begin{subequations}
\begin{eqnarray}
g^2 &\longrightarrow& g,
\label{gg-g}
\\
\nu &\longrightarrow& E_f - i \Gamma(E)/2.
\label{nu-Ef}
\end{eqnarray}
\end{subequations}
In Ref.~\cite{Hanhart:2007yq}, the effects of the widths 
of the $D^{*0}$ and $D^{*+}$ were not taken into account.  
Thus the expressions for $\kappa(E)$ and $\kappa_1(E)$
were Eq.~(\ref{kappa-E}) with $\Gamma_{*0} = 0$
and Eq.~(\ref{kappa1-E}) with $\Gamma_{*1} = 0$.
The authors of Ref.~\cite{Hanhart:2007yq} 
did however allow for energy dependence in Im$\nu$, 
as indicated by Eq.~(\ref{nu-Ef}).

\subsubsection{$D^{*0} \bar{D}^0$ threshold region}

The interaction parameters $\nu$ and $g$ can be tuned
so that there is a bound state just below the 
$D^{*0} \bar D^0$ threshold that can be identified with the $X(3872)$.  
The scattering amplitude will have a
pole at a complex energy $E_{\rm pole}$ 
that satisfies Eq.~(\ref{Epole}) with $\gamma \equiv \kappa(E_{\rm pole})$.
If the tiny difference between $\kappa_1(E_{\rm pole})$ and
$\kappa_{11}$ is neglected, the vanishing of the denominator
in Eq.~(\ref{f-Flatte}) reduces to a quadratic equation 
for $\gamma$, one of whose solutions is
\begin{equation}
\gamma = - \mbox{$\frac12$} g^2 \mu
+ \sqrt{(\mbox{$\frac12$} g^2 \mu)^2 
       - \mu (2 \nu + g^2 \kappa_{11} + i \Gamma_{*0})}.
\label{gamma-Flatte}
\end{equation}
The residues of the poles of the scattering amplitudes $f_{ij}(E)$
at $\kappa(E)= \gamma$ are $Z_i^{1/2} Z_j^{1/2}$, where
\begin{equation}
Z_{0}^{1/2} = - Z_{1}^{1/2} = 
\left( 1 + \frac{\gamma}{\kappa_{11}}
+ \frac{2 \gamma}{g^2 \mu} \right)^{-1/2}.
\label{Z-Flatte}
\end{equation}
An inverse scattering length $\gamma$ that is small compared to 
$\kappa_{11}$ can be obtained by a fine-tuning of $\nu$ such that 
$|\nu + g^2 \kappa_{11}/2| \ll g^4 \mu/8, g^2 \kappa_{11}/2$.
The solution for $\gamma$ in Eq.~(\ref{gamma-Flatte}) then reduces to
$\gamma \approx - (2 \nu + g^2 \kappa_{11} + i \Gamma_{*0})/g^2$.
In this case, the elastic scattering amplitude $f_{00}(E)$ 
reduces in the entire $D^{*0} \bar D^0$ threshold region 
$|E| \ll \nu_{11}$ to $Z_0 f(E)$, where $f(E)$ is the universal scattering
amplitude $f(E)$ in Eq.~(\ref{f-E}).

An inverse scattering length $\gamma$ that is small compared to 
$\kappa_{11}$ can also be obtained by a double fine-tuning of $g$ and $\nu$ 
so that $g^2 \mu \ll \kappa_{11}$ and $\mu |\nu| \ll \kappa_{11}^2$.
In this case, the denominator in Eq.~(\ref{f-Flatte}) has two zeroes 
near the $D^{*0} \bar D^0$ threshold.
The energy dependence of $f_{00}(E)$ in the  
$D^{*0} \bar D^0$ threshold region is therefore more complicated 
than the universal scattering amplitude.

\subsubsection{Resonance far above $D^* \bar D$ threshold}

Alternatively, the parameters $\nu$ and $g$ can be chosen so that,
instead of having a pole just below the $D^{*0} \bar D^0$ threshold,
the Flatt\'e scattering amplitude in Eq.~(\ref{f-Flatte})
has a pole at an energy well above the $D^{*+} D^-$ threshold.
Such a pole could be associated with the charmonium state $\chi_{c1}(2P)$.
If the difference between $\kappa_1(E)$ and $\kappa(E)$ is neglected,
we get a quadratic equation for $\kappa$.
The solution for the complex pole is
\begin{equation}
E_{\chi} - i \Gamma_{\chi}/2 \approx  
\nu - g^4 \mu - i g^2 \sqrt{2 \mu \nu - g^4 \mu^2}.
\label{Echi:complex:R}
\end{equation}
If $\nu$ has a small imaginary part,
the energy and width of the resonance are approximately
\begin{subequations}
\begin{eqnarray}
E_{\chi} &\approx& {\rm Re} \nu - g^4 \mu,
\label{Echi:R}
\\
\Gamma_{\chi} &\approx& 
-2 {\rm Im} \nu 
+ 2 g^2 \sqrt{2 \mu ({\rm Re} \nu - g^4 \mu) + g^4 \mu^2} .
\label{Gamchi:R}
\end{eqnarray}
\label{EGam:R}
\end{subequations}
In the expression for the width, the term with the square root
is the partial width for decays into $D^* \bar D$.
The term proportional to ${\rm Im} \nu$ corresponds to 
all other decay channels. 
As the energy $E_\chi$ is increased by adjusting $\nu$,
the width $\Gamma_{\chi}$ increases as $E_\chi^{1/2}$.  
This behavior is characteristic of an ordinary resonance.
Near the resonance, the Flatt\'e scattering amplitude 
in Eq.~(\ref{f-Flatte}) can be approximated by 
\begin{equation}
f_\textrm{Flatt\'e}(E) \approx
\left( 1 - i \frac{g^2 \mu}{\sqrt{2 \mu E_\chi- i \mu \Gamma_{\chi}}} \right)^{-1}  
f_\textrm{BW}(E),
\label{fFlatte-BW}
\end{equation}
where $f_\textrm{BW}(E)$ is the Breit-Wigner scattering amplitude:
\begin{equation}
f_\textrm{BW}(E) \equiv \frac{-g^2/2}{E - E_\chi + i \Gamma_\chi/2}  .
\label{f-BW}
\end{equation}

\subsection{Zero-Range+Resonance scattering amplitudes}
\label{sec:zero-range+resonance}

We now generalize the scattering amplitudes of Sections~\ref{sec:zero-range} 
and \ref{sec:Flatte} to the case of coupled channels 
$(D^{*0} \bar D^0)_+$ and $(D^{*+} D^-)_+$ that scatter 
through a resonance as well as through zero-range interactions.
We will refer to the resonance channel as $\chi$.

\subsubsection{General case}

The transition amplitudes $\mathcal{A}_{ij}(E)$ 
in Eq.~(\ref{Aij-fij}) define a $2 \times 2$ 
matrix of scattering amplitudes $f_{ij}(E)$.
The expression in Eq.~(\ref{f-inverse}) for the inverse of that $2 \times 2$ 
matrix in the case of zero-range scattering can be generalized to one 
that also takes into account the coupling to a resonance:
\begin{equation}
f(E)^{-1} = 
- \left( \Lambda + G \frac{1}{E - \nu} G^T \right)^{-1} + K(E) ,
\label{f-inverse:res}
\end{equation}
where $\Lambda$ is a $2 \times 2$ symmetric matrix, 
$G$ is a 2-component column vector,
and $K(E)$ is the diagonal matrix in Eq.~(\ref{K-kappa}).
The propagator for this resonance is
\begin{equation}
P(E) = 
\left[ E - \nu + G^T (\Lambda - K^{-1})^{-1} G \right]^{-1} .
\label{prop:res}
\end{equation}
If the six interaction parameters $\Lambda_{00}$, $\Lambda_{01}$,
$\Lambda_{11}$, $G_0$, $G_1$, and $\nu$ are all real valued
and if $\Gamma_{*0}$ and $\Gamma_{*1}$ are set to 0 in the expressions 
for Eqs.~(\ref{kappa-E}) and (\ref{kappa1-E}),
the amplitudes $f_{ij}(E)$ satisfy the constraints of unitarity exactly.
The scattering amplitudes and the resonance propagator can be 
derived from a
renormalizable nonrelativistic quantum field theory with two 
scattering channels and a resonance that interact only 
through contact interactions.
The renormalization of this quantum field theory
is described in the Appendix.

\subsubsection{Isospin symmetry}

The approximate isospin symmetry of QCD reduces the three
interaction parameters in the matrix $\Lambda$
to two independent parameters $\gamma_0$ and $\gamma_1$
defined by Eq.~(\ref{Lambda-gamma}).
The assumption that the resonance has isospin 0 reduces the two
interaction parameters in the column vector $G$ to a single parameter $g$: 
\begin{eqnarray}
G &=&
\frac{g}{\sqrt2}
\left( \begin{array}{c}
      ~~1 \\
       -1 
      \end{array} \right) .
\label{G:isospin}
\end{eqnarray}
Thus the inverse matrix in Eq.~(\ref{f-inverse:res}) reduces to
\begin{equation}
\left( \Lambda 
      + G \frac{1}{E - \nu} G^T \right)^{-1} =
\left( \frac{1}{\gamma_0} + \frac{g^2}{E - \nu} \right)^{-1}~\mathcal{P}_0
+ \gamma_1~\mathcal{P}_1.
\label{f-inverse:isospin}
\end{equation}
The scattering amplitudes obtained by inverting the matrix in 
Eq.~(\ref{f-inverse:res}) can be obtained from those in
Eq.~(\ref{fij-E}) by replacing $\gamma_0$ by 
$[1/\gamma_0 + g^2/(E - \nu)]^{-1}$:
\begin{subequations}
\begin{eqnarray}
f_{00}(E) &=&
\frac{[-\gamma_1 - \gamma_0 + 2 \kappa_1(E)] (E - \nu + g^2 \gamma_0) 
      + g^2 \gamma_0^2}{D(E)} ,
\label{f00chi-E}
\\
f_{01}(E) &=&
\frac{(\gamma_1 - \gamma_0)(E - \nu + g^2 \gamma_0) + g^2 \gamma_0^2}{D(E)} ,
\label{f01chi-E}
\\
f_{11}(E) &=&
\frac{[- \gamma_1 - \gamma_0  + 2 \kappa(E)](E - \nu + g^2 \gamma_0) 
      + g^2 \gamma_0^2}{D(E)} ,
\label{f11chi-E}
\end{eqnarray}
\label{fijchi-E}
\end{subequations}
where the denominator is 
\begin{eqnarray}
D(E) &=& \left[ 2 \gamma_1 \gamma_0 
- (\gamma_1 + \gamma_0) [\kappa_1(E) +  \kappa(E)]
+ 2 \kappa_1(E) \kappa(E) \right] (E - \nu + g^2 \gamma_0)
\nonumber \\
&& + g^2 \gamma_0^2 [-2 \gamma_1 + \kappa_1(E) +  \kappa(E)] .
\label{DF-E}
\end{eqnarray}
The resonance propagator in Eq.~(\ref{prop:res}) is 
\begin{equation}
P(E) =
\frac{2 \gamma_1 \gamma_0 - (\gamma_1 + \gamma_0)[\kappa_1(E) + \kappa(E)]
	+ 2 \kappa_1(E) \kappa(E)}
    {D(E)} .
\label{prop:resI}
\end{equation}
At energies $E$ far from the charm meson thresholds at 0 and $\nu_{11}$,
the difference between $\kappa_1(E)$ and $\kappa(E)$ can be neglected
and the isospin symmetry becomes exact.
The scattering amplitudes in Eqs.~(\ref{fijchi-E}) 
and the resonance propagator in Eq.~(\ref{prop:resI})
depend on four independent interaction 
parameters: $\gamma_0$, $\gamma_1$, $g$, and $\nu$. 
We will refer to the model with the scattering amplitudes in 
Eqs.~(\ref{fijchi-E}) and the resonance propagator in 
Eq.~(\ref{prop:resI}) as the {\it Zero-Range+Resonance model}.


\subsubsection{Optical theorem}

The system consisting of the three channels $(D^{*0} \bar D^0)_+$, 
$(D^{*+} D^-)_+$, and $\chi$ cannot be exactly unitary, 
because the $D^{*0}$ and $D^{*+}$ have nonzero widths,
the charm meson pairs have inelastic scattering channels, 
and $\chi$ may have decay channels other than $D^* \bar D$
and $D \bar D^*$.
We can take into account the dominant effects of decays of 
$D^{*0}$ and $D^{*+}$ by replacing $\kappa(E)$ and $\kappa_1(E)$ 
by the expressions in Eqs.~(\ref{kappa-E}) and (\ref{kappa1-E}),
respectively.
We can take into account the effects of inelastic scattering channels
other than $D \bar D \pi$ and $D \bar D \gamma$ 
by taking $\gamma_{0}$ and $\gamma_{1}$
to be complex parameters with positive imaginary parts.
We can take into account the effects of decay channels for $\chi$
other than $D^* \bar D$ and $D \bar D^*$ by taking $\nu$ 
to be a complex parameter with a negative imaginary part.
We choose $g^2$ to be a real parameter.
The imaginary parts of the scattering amplitudes 
in Eq.~(\ref{fijchi-E}) can be expressed in forms 
that are consistent with the Cutkosky cutting rules:
\begin{eqnarray}
\mathrm{Im} \, f_{ij}(E) &=&
\left( \frac{|E - \nu|^2}{|E - \nu + g^2 \gamma_0|^2} \mathrm{Im} \gamma_0
- \frac{g^2 |\gamma_0|^2}{|E - \nu + g^2 \gamma_0|^2} \mathrm{Im} \nu \right)
\sum_{k,l} f_{ik}(E)~\mathcal{P}_{0,kl}~f_{lj}^*(E)
\nonumber \\
&& 
+ \mathrm{Im} \gamma_1 \sum_{k,l} f_{ik}(E)~\mathcal{P}_{1,kl}~f_{lj}^*(E)
\nonumber \\
&& 
- \mathrm{Im} \kappa(E) ~ f_{i0}(E)~f_{0j}^*(E)
- \mathrm{Im} \kappa_1(E) ~ f_{i1}(E)~f_{1j}^*(E) .
\label{Imfijchi}
\end{eqnarray}
The terms proportional to ${\rm Im} \kappa(E)$ and 
${\rm Im} \kappa_1(E)$ are the contributions from channels whose
ultimate final states are $(D \bar D \pi,D \bar D \gamma)$, 
including $D^* \bar D$ and $D \bar D^*$.
The terms proportional to ${\rm Im} \gamma_{0}$,
${\rm Im} \gamma_{1}$, 
and ${\rm Im} \nu$ correspond to 
other inelastic $D^* \bar D$ scattering channels with isospin $0$,
other inelastic $D^* \bar D$ scattering channels with isospin $1$,
and decay channels of $\chi$ with isospin $0$, respectively.

\subsubsection{Zero-range limit}

In the limits $g \to 0$ or $\nu \to \infty$, the resonance decouples
and the scattering amplitudes for the Zero-Range+Resonance model
in Eqs.~(\ref{fijchi-E}) reduce to those for the Zero-Range model
in Eqs.~(\ref{fij-E}).
The Zero-Range model is a good approximation if 
$|E - \nu| \gg g^2 |\gamma_0|$.  This energy region includes the 
$D^{*0} \bar D^0$ threshold if $|\nu| \gg g^2 |\gamma_0|$,
in which case the condition on the energy reduces to $|E| \ll |\nu|$.

The Zero-Range model actually has a larger domain of validity 
if one allows for renormalization of the parameter $\gamma_0$.
In the region $|E| \ll |\nu|,|\nu - g^2 \gamma_0|$, 
the scattering amplitudes in Eqs.~(\ref{fijchi-E})
reduce to those in Eqs.~(\ref{fij-E}) with the substitution
$\gamma_0 \to \gamma_0 \nu/(\nu-g^2 \gamma_0)$.

\subsubsection{Flatt\'e limit}

In the limits $\gamma_0,\gamma_1 \to \infty$, 
scattering proceeds only through the resonance
and the scattering amplitudes for the Zero-Range+Resonance model
in Eqs.~(\ref{fijchi-E}) reduce to the Flatt\'e scattering 
amplitudes in Eqs.~(\ref{fij-Flatte}) and (\ref{f-Flatte}).
The resonance propagator in Eq.~(\ref{prop:resI})
reduces in this limit to
$P(E) \approx (-2/g^2) f_\textrm{Flatt\'e}(E)$.
The Flatt\'e model is a good approximation in the region 
$|\gamma_1| \gg \kappa_{11}$ if the energy satisfies
$|E| \ll |\gamma_1|^2/\mu$ and 
$|E - \nu| \ll g^2 |\gamma_0|, g^2 |\gamma_1|$.
This energy region includes the 
$D^{*0} \bar D^0$ threshold if $|\nu| \ll g^2 |\gamma_0|, g^2 |\gamma_1|$.

The Flatt\'e model actually has a larger domain of validity 
if one allows for renormalization of the coupling constant $g$.
If $|\gamma_1| \gg \kappa_{11}$ and if the energy satisfies
$|E| \ll |\gamma_1|^2/\mu, |\nu - g^2 \gamma_0|$ and 
$|E - \nu| \ll |(g^2 - \nu/\gamma_0) \gamma_1|$, 
the scattering amplitudes in Eqs.~(\ref{fijchi-E})
reduce to those in Eqs.~(\ref{fij-Flatte}) with the substitution
$g^2 \to g^2 - \nu/\gamma_0$. This energy region includes the 
$D^{*0} \bar D^0$ threshold if 
$|\gamma_1|\gg |\gamma_0 \nu/(\nu - g^2 \gamma_0)|$.

\subsubsection{$D^{*0} \bar{D}^0$ threshold region}

The interaction parameters $\gamma_0$, $\gamma_1$, $g$, and $\nu$ can be tuned
so that there is a bound state just below the 
$D^{*0} \bar D^0$ threshold that can be identified with the $X(3872)$.  
The scattering amplitudes will have 
poles at a complex energy $E_{\rm pole}$ 
that satisfies Eq.~(\ref{Epole}) with $\gamma \equiv \kappa(E_{\rm pole})$.
The denominator defined in Eq.~(\ref{DF-E}) must vanish at $E_{\rm pole}$.
If the tiny difference between $\kappa_1(E_{\rm pole})$ and
$\kappa_{11}$ is neglected, the equation $D(E_{\rm pole}) = 0$
reduces to a cubic polynomial equation for $\gamma$, which can be written
\begin{eqnarray}
\left[ \gamma_1 \kappa_{11} + (\gamma_1 - 2 \kappa_{11}) \gamma \right]
\left[ 2 \mu (\nu - g^2 \gamma_0) + \gamma^2 + i \mu \Gamma_{*0} \right]
\nonumber \\
= \gamma_0 \left( 2 \gamma_1 - \kappa_{11} - \gamma \right) 
\left( 2 \mu \nu +\gamma^2 +  i \mu \Gamma_{*0} \right).
\label{eq:gamZRR}
\end{eqnarray}
The general behavior of the three poles has been analyzed 
in Ref.~\cite{Baru:2010ww} for the case $ \Gamma_{*0} = 0$.
The approximate fine tuning required to obtain a small inverse scattering length 
can be obtained by setting $\gamma = 0$ and $\Gamma_{*0} = 0$:
\begin{equation}
\gamma_1 \kappa_{11} (\nu - g^2 \gamma_0) \approx 
\gamma_0 (2 \gamma_1 - \kappa_{11}) \nu .
\label{eq:tuneres}
\end{equation}
If $|\gamma_1| \gg \kappa_{11}$, this fine-tuning condition 
reduces to $1/\gamma_0 - g^2/\nu \approx 2/\kappa_{11}$. 
An inverse scattering length $\gamma$ 
that is small compared to $\kappa_{11}$ can be obtained 
by fine-tuning $\gamma_0$ to near $\kappa_{11}/2 \approx +63$~MeV
with $|\nu|/g^2  \gg \kappa_{11}$ or
by fine-tuning $\nu$ to near $- g^2 \kappa_{11}/2$
with $|\gamma_0| \gg \kappa_{11}$.

Since the line shapes near the $D^{*0} \bar D^0$ threshold
are extremely sensitive to $\gamma$, it is advantageous
to take $\gamma$ to be one of the independent interaction parameters.
The equation $D(E_{\rm pole}) = 0$ can be solved for $\gamma_0$ 
as a function of $\gamma_1$, $g$, $\nu$, and $\gamma$.
If the tiny difference between $\kappa_1(E_{\rm pole})$ and
$\kappa_{11}$ is neglected, the solution is
\begin{equation}
\gamma_0 =
\left( \frac{(2 \gamma_1 -\kappa_{11}) - \gamma}
    {\gamma_1 \kappa_{11} + (\gamma_1 - 2 \kappa_{11}) \gamma}
+ \frac{g^2}{\nu - E_\textrm{pole}} \right)^{-1}.
\label{eq:gamma0}
\end{equation}

If the equation for $\gamma$ in Eq.~(\ref{eq:gamZRR}) 
is expanded to first order in the 
imaginary parts of all the variables, one can solve for 
the imaginary part of $\gamma$:
\begin{equation}
\textrm{Im} \gamma \approx
\frac12 \left( \frac{\gamma_1 \kappa_{11}}
     {(\gamma_1 - \kappa_{11}) \gamma_0} \right)^2 \textrm{Im} \gamma_0
+ \frac12 \left( \frac{\kappa_{11}}
     {\gamma_1 - \kappa_{11}} \right)^2 \textrm{Im} \gamma_1
- \frac12 \left( \frac{g \gamma_1 \kappa_{11}}
     {(\gamma_1 - \kappa_{11}) \nu} \right)^2 
     \left( \textrm{Im} \nu + \mbox{$\frac12$} \Gamma_{*0} \right).
\label{Imgamma-ZR+R}
\end{equation}
We have simplified the coefficients by setting $\gamma = 0$.
We have also used the fine-tuning condition in Eq.~(\ref{eq:tuneres})
to express the coefficients in a manifestly positive form.

The scattering amplitudes $f_{ij}(E)$ in Eqs.~(\ref{fijchi-E})
have poles at $\kappa(E) = \gamma$ with residues $Z_i^{1/2} Z_j^{1/2}$.  
If the tiny difference between 
$\kappa_1(E_{\rm pole})$ and $\kappa_{11}$ is neglected, 
the ratio of the residue factors is
\begin{eqnarray}
\frac{Z_{1}^{1/2}}{Z_{0}^{1/2}} &=&
- \frac{\gamma_1 - \gamma}{\gamma_1 - \kappa_{11}} .
\label{Z1/Z0:ZRR}
\end{eqnarray}
This ratio does not depend on the resonance parameters 
$\nu$ and $g$.  The residue $Z_{0}$  is
\begin{eqnarray}
Z_{0} =
\left[ 1 + 
\frac{(\gamma_1 - \gamma)^2 \gamma}{(\gamma_1 - \kappa_{11})^2 \kappa_{11}} 
+ \frac{g^2 [\gamma_1 \kappa_{11} 
       + (\gamma_1 - 2 \kappa_{11}) \gamma]^2 \gamma}
      {2 \mu (\gamma_1 - \kappa_{11})^2(\nu - E_\textrm{pole})^2}
\right]^{-1} .
\label{Z0:ZRR}
\end{eqnarray}
As $\gamma \to 0$, this residue approaches 1.
If the small value of $\gamma$ arises from 
either the fine tuning $\gamma_0 \approx \kappa_{11}/2$
or the fine tuning $\nu \approx - g^2 \kappa_{11}/2$,
the scattering amplitudes $f_{ij}(E)$
in the entire $D^{*0} \bar D^0$ threshold region defined by
$|E| \ll \nu_{11} = 8.1$~MeV are dominated by the pole at  
$\kappa(E) = \gamma$.   They reduce 
to the universal scattering amplitude $f(E)$
given in Eq.~(\ref{f-E}) multiplied by residue factors:
\begin{eqnarray}
f_{ij}(E) \approx Z_i^{1/2}~f(E)~Z_j^{1/2}. 
\label{fij-thresh}
\end{eqnarray}

An inverse scattering length $\gamma$ 
that is small compared to $\kappa_{11}$ can also be obtained
by a double fine-tuning of $\gamma_0$ and $\nu$
so that they satisfy $|\nu| \ll g^2 \kappa_{11}$,
$|\gamma_0| \ll \kappa_{11}$, and $|\nu - g^2 \gamma_0| \ll |\nu|$.
In this case, the residue $Z_0$ in Eq.~(\ref{Z0:ZRR}) 
can be significantly smaller than 1.  The energy dependence of $f_{ij}(E)$
can also differ significantly from that of the 
universal amplitude in Eq.~(\ref{f-E}).

\subsubsection{Resonance far above $D^* \bar D$ threshold}

One region of parameter space in which the amplitudes simplify is
when the resonance parameter $\nu$ is much larger than the energy scale 
$\nu_{11} = 8.1$~MeV associated with isospin splitting. 
In this case, the existence of the $X(3872)$ requires the 
fine-tuning $\gamma_0 \approx \kappa_{11}/2$.
In addition to the $X(3872)$ resonance just below the $D^{*0} \bar D^0$ 
threshold, there is a second resonance 
$\chi$ well above the $D^{*+} D^-$ threshold.
This second resonance could be identified with
the P-wave charmonium state $\chi_{c1}'$.

We first consider this system at energies in the 
$D^* \bar D$ threshold region $|E| \lesssim \nu_{11}$.  In this region,
the scattering amplitudes in Eqs.~(\ref{fijchi-E}) reduce to those in 
Eqs.~(\ref{fij-E}) for the Zero-Range model
with the substitution $\gamma_0 \to \gamma_0 \nu/(\nu - g^2 \gamma_0)$.

We next consider this system at energies in the 
$\chi$ resonance region.  In this region,
the difference between $\kappa(E)$ and $\kappa_1(E)$ can be neglected 
and the resonance propagator in Eq.~(\ref{prop:resI}) reduces to 
\begin{equation}
P(E) \approx 
\frac{1}{E - \nu + g^2/[1/\gamma_0 - 1/\kappa(E)]}.
\label{prop:resIsym}
\end{equation}
It has a pole at a complex energy $E_{\chi} - i \Gamma_{\chi}/2$ 
near $\nu - g^2 \gamma_0$.
The position of the pole can be calculated by iterating
around this approximate solution.
The solution to first order in $\gamma_0/\kappa(E_{\chi})$ is
\begin{equation}
E_{\chi} - i \Gamma_{\chi}/2 \approx  \nu - g^2 \gamma_0
- i \frac{g^2 \gamma_0^2}{\sqrt{2 \mu (\nu - g^2 \gamma_0)}} .
\label{Echi:complex}
\end{equation}
If $\gamma_0$ and $\nu$ have small imaginary parts,
the energy and width of the resonance are approximately
\begin{subequations}
\begin{eqnarray}
E_{\chi} &\approx& {\rm Re} (\nu  - g^2 \gamma_0) ,
\label{Echi:F}
\\
\Gamma_{\chi} &\approx& 
-2 {\rm Im} \nu + 2g^2{\rm Im}\gamma_0
+ \frac{2 g^2 ({\rm Re}\gamma_0)^2}{\sqrt{2 \mu {\rm Re}(\nu - g^2 \gamma_0)}}.
\label{Gamchi:F}
\end{eqnarray}
\label{EGam:F}
\end{subequations}
The last term in Eq.~(\ref{Gamchi:F}) is the 
partial width for decays into $D^* \bar D$ and $D \bar D^*$.
The terms proportional to ${\rm Im} \nu$ and ${\rm Im}\gamma_{0}$ 
correspond to other decay channels of $\chi_{c1}'$ and to inelastic 
isospin-$0$ scattering channels for charm meson pairs, respectively. 
As the energy $E_\chi$ is increased by adjusting $\nu$,
the partial width into $D^* \bar D$ and $D \bar D^*$  
decreases to 0 as $E_\chi^{-1/2}$.  
This behavior is characteristic of a Feshbach resonance \cite{feshbach}.

The scattering amplitudes also simplify near the resonance.
Using the approximation $|\gamma_0| \ll |2 \mu E_\chi|^{1/2}$,
they reduce to
\begin{equation}
f_{00}(E) \approx f_{11}(E) \approx - f_{01}(E) \approx 
- \frac{\gamma_0^2}{2 \mu E_\chi}~f_\textrm{BW}(E)  ,
\label{fij-hinu}
\end{equation}
where $f_\textrm{BW}(E)$ is the Breit-Wigner amplitude
in Eq.~(\ref{f-BW}).


\section{Estimates of the interaction parameters}
\label{sec:lowEparameters}

The scattering amplitudes for the Zero-Range+Resonance model
in Section~\ref{sec:zero-range+resonance} depend on the 
interaction parameters $\gamma$, $\gamma_1$, $g$, 
and $\nu$. In this section, we analyze 
the constraints on these parameters.

\subsection{Inverse scattering length}
\label{sec:scatlength}

The complex inverse scattering length $\gamma$ can be determined 
from measurements of the position and width of the $X(3872)$ resonance
in the $J/\psi\, \pi^+ \pi^-$ decay channel.
The position of the $X(3872)$ resonance relative to the 
$D^{*0} \bar D^0$ threshold is given in Eq.~(\ref{MX-ave}).
There are only upper limits on the width of the $X(3872)$ resonance
in the $J/\psi\, \pi^+ \pi^-$ decay channel.
The upper limit from combining the results of the Belle and Babar 
Collaborations \cite{Choi:2003ue,Aubert:2008gu} is 
\begin{equation}
\Gamma_X <  2.2~{\rm MeV}~~~~~(90\%~{\rm C.L.}) . 
\label{GamX-lim}
\end{equation}
A lower bound on the width $\Gamma_X$ is the width $\Gamma_{*0}$ 
of the constituent $D^{*0}$, which is given in Eq.~(\ref{GamD*0}).
This contribution to the width of the $X(3872)$ can be identified 
with the decay modes $(D^0 \bar D^0 \pi^0,D^0 \bar D^0 \gamma)$.

We denote the real and imaginary parts of the complex inverse 
scattering length by $\gamma_{\rm re}$ and $\gamma_{\rm im}$:
\begin{equation}
\gamma = \gamma_{\rm re} + i \gamma_{\rm im}.
\label{gamma-reim}
\end{equation}
An alternative pair of variables that can in principle be measured 
directly are the peak position $E_{\rm max}$ of the resonance 
and its full width at half-maximum $\Gamma_{\rm fwhm}$.
The variables $E_{\rm max}$ and $\Gamma_{\rm fwhm}$ are functions of 
$\gamma_{\rm re}$, $\gamma_{\rm im}$, and the $D^{*0}$ width $\Gamma_{*0}$.
They can be expanded in powers of $\Gamma_{*0}$ \cite{Stapleton:2009ey}:
\begin{subequations}
\begin{eqnarray}
E_{\rm max} &=& 
- \frac{\gamma_{\rm re}^2}{2 \mu} 
- \frac{\gamma_{\rm im}}{2\gamma_{\rm re}}\Gamma_{*0} + \ldots ,
\label{Emax}
\\
\Gamma_{\rm fwhm} &=& 
\frac{2 \gamma_{\rm re} \gamma_{\rm im}}{\mu} + \Gamma_{*0} + \ldots .
\label{Gammafwhm}
\end{eqnarray}
\end{subequations}
In the expression for $\Gamma_{\rm fwhm}$ in Eq.~(\ref{Gammafwhm}), 
the second term $\Gamma_{*0}$ can be identified with the partial width 
for decay into $(D^0 \bar D^0 \pi^0,D^0 \bar D^0 \gamma)$
while the first term can be identified with the partial width
into other decay modes.

The result in Eq.~(\ref{MX-ave}) can be interpreted as a measurement 
of $E_{\rm max}$.  Keeping only the leading term in the expansion 
for $E_{\rm max}$ in Eq.~(\ref{Emax}), 
we obtain a determination of $\gamma_{\rm re}$:
\begin{equation}
\gamma_{\rm re} = 28^{+12}_{-20}~{\rm MeV}.
\label{gamma-re}
\end{equation}
The result in Eq.~(\ref{GamX-lim}) can be interpreted 
as an upper limit on $\Gamma_{\rm fwhm}$.
Keeping only the first two terms in the expansion 
for $\Gamma_{\rm fwhm}$ in Eq.~(\ref{Gammafwhm}), 
we obtain an upper limit on the product 
of $\gamma_{\rm re}$ and $\gamma_{\rm im}$:
\begin{equation}
0 < \gamma_{\rm re} \gamma_{\rm im} < (32~{\rm MeV})^2.
\label{gamma-re*im}
\end{equation}

\subsection{Resonance parameters}
\label{sec:resonance}

The resonance parameters are the energy variable $\nu$ 
and the coupling constant $g$.  
If $\nu$ is much larger than the energy scale 
$\nu_{11} = 8.1$~MeV of isospin splitting, the  
Zero-Range+Resonance model predicts a narrow resonance $\chi$ 
whose energy is well above the $D^{*+} D^-$ threshold. 
This resonance could be identified with the P-wave charmonium
state $\chi_{c1}' \equiv \chi_{c1}(2P)$ or a $1^{++}$ $c \bar c$ 
tetraquark meson or some other $1^{++}$ meson.
If $\chi$ is identified with the $\chi_{c1}'$, we can take advantage 
of the well-developed charmonium phenomenology based on
quark potential models to constrain the resonance parameters.

Predictions from potential models 
for the mass of $\chi_{c1}'$ range from about 3920~MeV to about 
4010~MeV \cite{Godfrey:1985xj,Zeng:1994vj,Ebert:2002pp,Barnes:2003vb,%
Eichten:2004uh,Barnes:2005pb,Eichten:2005ga,Radford:2007vd}.
These predictions are all 50~MeV or more higher than the mass of the 
$X(3872)$.  
Since the effects of couplings of charmonium states to pairs 
of charm mesons are not well understood, we cannot exclude the possibility
that they shift the mass of the $\chi_{c1}'$ down into the 
$D^* \bar D$ threshold region.
We will  take the real part of $\nu$ to be an adjustable parameter.

\begin{table}[t]
\begin{tabular}{l|cc|c}
Reference  &  ~~~~~~Mass~(MeV)~~~~~~ &  Partial width~(MeV) & $g$ \\
\hline
BG  \cite{Barnes:2003vb}  & 3953 & 118 &~~~0.37~~~\\
ELQ \cite{Eichten:2004uh} & 3968 & 150 & 0.40 \\
BGS \cite{Barnes:2005pb}  & 3925 & 165 & 0.46 \\
ELQ \cite{Eichten:2005ga} & 3920 &  81 & 0.35 \\
\end{tabular}
\caption{Results from coupled-channel potential models 
for the mass of the $\chi_{c1}(1P)$, its partial width into 
$D^* \bar D$ and $D \bar D^*$, and the  
coupling constant $g$ inferred from Eqs.~(\ref{EGam:R}).}
\label{tab:gchi}
\end{table}

We can use potential models to estimate the coupling constant $g$.
Conventional potential models contain no information about charm mesons.
Coupled-channel potential models include additional interactions that 
couple a charm-quark and antiquark to pairs of charm mesons.
These models can be used to calculate the partial widths for decays of
charmonium states into pairs of charm mesons.  The partial width 
of $\chi_{c1}'$ into $D^* \bar D$ and $D \bar D^*$ has been calculated 
using the $^3P_0$ model \cite{Barnes:2003vb,Barnes:2005pb}
and the CCC model \cite{Eichten:2004uh,Eichten:2005ga}.
The results for the mass of the $\chi_{c1}'$ and its partial
width into $D^* \bar D$ and $D \bar D^*$ are given in Table~\ref{tab:gchi}.
Although these coupled-channel potential models allow scattering of charm
mesons, there is no reason to expect the scattering lengths to be 
much larger than the range of charm meson interactions
in the absence of the fine-tuning of a charmonium state to the 
$D^{*0} \bar D^0$ threshold.  Thus the appropriate limit of the
Zero-Range+Resonance model in Sec.~\ref{sec:zero-range+resonance} 
is the Flatt\'e limit $\gamma_0,\gamma_1 \to \infty$, in which the 
charm mesons scatter only through their coupling to the resonance.  
The energy $E_\chi$ and the width $\Gamma_\chi$ 
of the resonance in this limit are given in Eqs.~(\ref{EGam:R}).
By fitting these expressions for the energy and the width,
we obtain the coupling constants listed in Table~\ref{tab:gchi}.  
The average value is
\begin{equation}
g = 0.40,
\label{g-ave}
\end{equation}
and the variations are less than 15\%.
We will use the value in Eq.~(\ref{g-ave}) in the numerical  
analysis in Section~\ref{sec:analysis}.

The effects of other decay modes of the $\chi_{c1}'$ 
besides $D^* \bar D$ and $D \bar D^*$ can be taken into account 
through the imaginary part of the parameter $\nu$.
The next most important decay modes are expected to be the 
radiative transitions $\chi_{c1}' \to J/\psi \, \gamma$ and 
$\chi_{c1}' \to \psi(2S) \gamma$.  The partial widths scale like the 
cube of the photon energy, so they depend on the mass of the 
$\chi_{c1}'$.  They have been calculated using quark potential models
\cite{Barnes:2003vb,Eichten:2004uh,Barnes:2005pb,Eichten:2005ga,Radford:2007vd}.
If the mass of the $\chi_{c1}'$ is close to 3872~MeV,
the partial widths are roughly 10~keV for $J/\psi \, \gamma$ and 
roughly 60~keV for $\psi(2S) \gamma$.
These are small enough that the imaginary part of $\nu$ 
will not have a dramatic effect on the line shapes 
near the $D^{*0} \bar D^0$ threshold.
We will therefore set Im$\nu = 0$.

\subsection{Charm meson scattering parameters}
\label{sec:charmscat}

The charm meson scattering parameters are $\gamma_0$ and $\gamma_1$.
We first consider the natural scale for these parameters. 
Since the low-energy scattering of charm mesons 
is dominated by pion exchange, the obvious estimate for the range 
of the interaction is $1/m_\pi$.  The corresponding estimate 
of the natural scale for $\gamma_0$ and $\gamma_1$ is $m_\pi \approx 140$~MeV.
Suzuki has suggested that the range of the interaction could 
be much larger than $1/m_\pi$ \cite{Suzuki:2005ha}.
The denominator of the propagator for the exchange of a 
$\pi^0$ of momentum $\bm{q}$ between $D^{*0}$ and $\bar D^0$ is
$q^2 + m_{\pi^0}^2 - (M_{*0}-M_0)^2$.  
There is a near cancellation between the terms $m_{\pi^0}^2$
and $(M_{*0}-M_0)^2$, because $M_{*0}-M_0$
is larger than $m_{\pi^0}$ by only 7.1~MeV. 
If we define an effective mass 
$m$ by expressing the denominator as $q^2 + m^2$, 
this effective mass is pure imaginary: $m^2 = -(44~\textrm{MeV})^2$.
One might be tempted to take $|m|$ as the natural scale for 
$\gamma_0$ and $\gamma_1$, but this would be incorrect.  
This can be seen by considering the limit $m_{\pi^0} \to M_{*0} - M_0$,
which implies $m \to 0$.  In this limit, the potential from the exchange of 
$\pi^0$ reduces to a $1/r^3$ potential that couples the 
S-wave and D-wave components of the wavefunction.
Although this is a long-range potential, it is not a scale-invariant 
potential.  The coefficient $C_3$ in the potential $C_3/r^3$ provides a scale.
In Ref.~\cite{Thomas:2008ja}, this coefficient was expressed as 
$C_3 = g^2/(4 \pi f_\pi^2)$, where $g/f_\pi$ is the coupling constant for the  
$D^*-D \pi$ interaction.  The corresponding momentum scale is 
$(2 \mu C_3)^{-1} = 360$~MeV.  This is the natural scale for 
$\gamma_0$ and $\gamma_1$ when $m=0$. 

Phenomenological estimates of $\gamma_0$ and $\gamma_1$ 
can be obtained from any model
that can be used to calculate the scattering amplitudes for charm mesons.
One such class of models is meson potential models,
in which the degrees of freedom are mesons and their interactions 
are defined by potentials.  The simplest such model is one in which the
charm mesons interact only through the exchange of pions
\cite{Tornqvist:1993ng,Tornqvist:2004qy,Thomas:2008ja}.
More elaborate meson potential models include the effects of the exchange 
of other mesons \cite{Liu:2008fh,Liu:2008tn}
or the exchange of quarks \cite{Swanson:2004pp}.
These models typically require an ultraviolet cutoff
to regularize singularities in the potentials at short distances. 
The scattering amplitudes can be calculated by solving the Schr\"odinger
equation.
Another class of models that can be used to calculate the scattering
amplitudes are meson scattering models,
which are defined by scattering parameters.  
In the simplest such models, such as the universal theory described in 
Section~\ref{sec:universal} and the Zero-Range model described in 
Section~\ref{sec:zero-range},
the only degrees of freedom are charm mesons.
Other hadrons can also be included 
as degrees of freedom in meson scattering models.
The Flatt\'e model in Section~\ref{sec:Flatte} and the  
Zero-Range+Resonance model in Section~\ref{sec:zero-range+resonance} include a
resonance that can be identified with the charmonium state $\chi_{c1}'$.
More elaborate meson scattering models that include many other mesons
as degrees of freedom have also been considered \cite{Gamermann:2009fv}.
In meson scattering models, the scattering amplitudes 
are calculated by solving integral equations, 
such as the Lippmann-Schwinger equation.
These models typically require an ultraviolet cutoff
to regularize singularities at large momenta. 
If the model is renormalizable, like those described in 
Sections~\ref{sec:universal}, \ref{sec:zero-range}, \ref{sec:Flatte}, 
and \ref{sec:zero-range+resonance}, all dependence on the cutoff can be absorbed 
into the scattering parameters.
Although many meson potential models and meson scattering 
models have been applied to the $X(3872)$, charm meson scattering 
lengths have not been calculated in these models.
Thus the published results for these models do not provide 
any direct estimates of the scattering 
parameters $\gamma_0$ and $\gamma_1$.

A particularly convenient observable for constraining $\gamma_0$ 
and $\gamma_1$ is
the ratio of the residues of the poles in the elastic scattering
amplitudes $f_{00}(E)$ for $D^{*0} \bar D^0$ 
and $f_{11}(E)$ for $D^{*+} D^-$ at the $X(3872)$ resonance.
We denote the residue of the pole in $f_{ij}(E)$ at $\kappa(E) = \gamma$ 
by $Z_{i}^{1/2}Z_{j}^{1/2}$.  
The quantity $|Z_{1}/Z_{0}|^{1/2}$ can be interpreted as the 
ratio of the wavefunctions at the origin for the
$(D^{*+} D^-)_+$ and $(D^{*0} \bar D^0)_+$ components of the 
$X(3872)$ \cite{Gamermann:2009uq}.
For the Zero-Range+Resonance model,
the ratio $Z_{1}^{1/2}/Z_{0}^{1/2}$ is given in Eq.~(\ref{Z1/Z0:ZRR}).
It is determined primarily by $\gamma_1$, because $|\gamma| \ll \kappa_{11}$.
It does not depend on the resonance parameters $g$ and $\nu$.  
This ratio has been calculated in a meson
scattering model whose degrees of freedom are $SU(4)$ multiplets 
of pseudoscalar and vector mesons \cite{Gamermann:2009fv}.
The parameters of the model were fine-tuned so that the 
binding energy of the $X(3872)$ is 0.4~MeV.
The absolute value of the ratio of the residues was calculated to be 
$|Z_{11}/Z_{00}|^{1/2} = 0.9923$.
If we insert the expression for the ratio of the residues
in Eq.~(\ref{Z1/Z0:ZRR}),  we obtain two solutions for $\gamma_1$:
$+76$~MeV and $-12,500$~MeV.  We expect $\gamma_1$ to be large 
compared to $\kappa_{11} = 125$~MeV and positive,
because the pion-exchange interaction is repulsive in the 
$C=+$ isospin-1 $D^* \bar D$ channel.  Neither of the two solutions for
$\gamma_1$ are consistent with this expectation.  It is possible 
that the model of \cite{Gamermann:2009fv} does not describe scattering 
in the isospin-1 channel with sufficient accuracy.

\section{Line shapes of $\bm{X(3872)}$ in $\bm{B}$ Meson Decay}
\label{sec:lineshape}

In this section, we consider the line shapes of the $X(3872)$ 
resonance in the decay $B \to K + X$.
We first present a general formulation of the problem of production
by a short-distance process in terms 
of an effective field theory that describes the 
$D^* \bar D$ threshold region.
We summarize the universal line shapes of Ref.~\cite{Braaten:2007dw},
which take into account only the large scattering length in the 
neutral channel.
We summarize the line shapes of Ref.~\cite{Braaten:2007ft},
which take into account zero-range scattering in the neutral 
and charged channels.
We then describe the Flatt\'e line shapes 
introduced in Ref.~\cite{Hanhart:2007yq},
which take into account scattering through a resonance.
Finally, we present more general line shapes that allow 
for both scattering through a resonance
and through zero-range interactions.

\subsection{Effective field theory formulation}
\label{sec:EFT}

Our starting point for the derivation
of the line shapes produced by the decay $B^+ \to K^+ + X$
is the optical theorem for the width of the $B^+$:
\begin{equation}
\Gamma [B^+] = 
\frac 1 {M_B} \, {\rm Im} \, {\mathcal A}[ B^+ \to B^+] ,
\label{GamB:optical}
\end{equation}
where $i {\mathcal A} [ B^+ \to B^+]$ is the one-meson-irreducible 
forward amplitude for $B^+$ at leading order in the
electroweak interactions and to all orders in QCD interactions. 
The imaginary part of this amplitude 
has a contribution from the intermediate
state $K^+ + X(3872)$.
In addition to the $X(3872)$ itself, there are other
sets of particles with
the quantum numbers $J^{PC} = 1^{++}$ 
that have enhanced production rates near the 
$D^{*0} \bar D^0$ threshold.  We will denote these states 
collectively by the symbol $(1^{++})$.
We will use the phrase $X(3872)$ {\it resonance} to refer 
specifically to the peak in the energy distribution 
just below the $D^{*0} \bar D^0$ threshold. 

Decays of the $B^+$ proceed through weak interactions 
that are mediated by the $W$ boson.  Because the mass of the $W$ 
is so much larger than that of the $B$ meson, 
the decays can be described completely within QCD 
using effective field theory methods.  The effects of $W$ exchange 
can be reproduced by an effective weak Hamiltonian.
The leading terms that contribute to $B^+ \to K^+ + X$ are 
current-current operators:
\begin{equation}
{\cal H}_{\rm weak} = 
C_{cc}~\bar b \gamma^\mu (1 - \gamma_5) c~\bar c \gamma_\mu (1 - \gamma_5) s
+ C_{nc}~\bar b \gamma^\mu (1 - \gamma_5) s~\bar c \gamma_\mu (1 - \gamma_5) c,
\label{Hweak}
\end{equation}
where $C_{cc}$ and $C_{nc}$ are short-distance coefficients.
The forward amplitude in Eq.~(\ref{GamB:optical})
can be expressed as the expectation value in the $B^+$ meson of the 
Fourier transform of a bilocal QCD operator:
\begin{equation}
{\mathcal A}[ B^+ \to B^+] = 
i \int d^4x~
\langle B^+ | {\rm T} {\cal H}_{\rm weak}^\dagger(x) 
                         {\cal H}_{\rm weak}(0) | B^+ \rangle.
\label{ABB}
\end{equation}

The $K^+$ and $X(3872)$ produced by the decay of $B^+$
have recoil momenta of 1140 MeV, which is large compared to 
the momentum scale associated with the resonance.
If the $X(3872)$ can be described using an effective field theory 
(EFT) for charm mesons, the decay $B^+ \to K^+ + X(3872)$
can also be described completely within that EFT.
In the rest frame of the $X(3872)$, the $B^+ \to K^+$ transition
acts like a point source for pairs of charm mesons and possibly
other degrees of freedom described by the EFT.
As far as the resonance is concerned, the effects of the 
$B^+ \to K^+$ transition can be reproduced by local EFT operators 
${\cal O}_i^m$ acting on the EFT vacuum.
The superscript $m$ is a Cartesian vector index and 
the subscript $i$ labels the various operators.
The contribution to the forward amplitude
in Eq.~(\ref{ABB}) from transitions to $K^+$ that create
$1^{++}$ states with invariant mass in the $D^* \bar D$ threshold region
can be expressed in terms of expectation values in the 
EFT vacuum of the Fourier transforms of bilocal EFT operators:
\begin{eqnarray}
{\mathcal A}_{(1^{++})} &=& 
\int \frac {d^4 P_K}{(2 \pi)^4} 
\frac{i}{P_K^2 - m_K^2 + i \varepsilon}
\sum_{i,j} (C^{j,n})^* C^{i,m}
\nonumber
\\
&& \times
i \int \!\! dt~e^{i E t} \int \!\! d^3r~
\langle 0 | {\rm T} {\cal O}_j^{n \dagger}(\bm{r},t) 
                        {\cal O}_i^{m}(\bm{0},0) | 0 \rangle.
\label{ABB-EFT}
\end{eqnarray}
The energy $E$ is defined by $(P_B-P_K)^2 = (M_{*0} + M_{0} + E)^2$,
where $P_B$ and $P_K$ are the 4-momenta of the $B^+$ and $K^+$.
The integral over the 4-momentum $P_K$ of the $K^+$ 
is implicitly restricted to the 
region near the $D^* \bar D$ threshold where $E$ is small.  
The short-distance coefficients $C^{i,m}$ specify the linear
combinations of EFT operators $\sum_i C^{i,m}{\cal O}_i^{m}$
(with an implied sum on the vector index $m$)
that reproduce the effects of the $B^+ \to K^+$ transition.

The short-distance coefficients $C^{i,m}$ can 
in principle be determined by matching 
matrix elements of ${\cal H}_{\rm weak}$ in QCD with
matrix elements of $\sum_{i} C^{i,m} {\cal O}_i^m$ in the EFT.
If we consider a $D^{*0} \bar D^0$ state in the 
$D^{*} \bar D$ threshold region, the matching condition is
\begin{eqnarray}
\langle K^+ D^{*0} \bar D^0 | {\cal H}_{\rm weak}(0) | B^+ \rangle =
\sum_{i} C^{i,m} 
\langle D^{*0} \bar D^0 | {\cal O}_i^{m}(\bm{0},0) | 0 \rangle.
\label{HOmatch}
\end{eqnarray}
Although we cannot calculate the QCD matrix element on the left side, 
we can use the Lorentz invariance of QCD to determine the tensor
structure of the coefficients $C^{i,m}$.  We choose the charm meson
pair $D^{*0} \bar D^0$ to have zero relative momentum.
The QCD matrix element on the left side of Eq.~(\ref{HOmatch}) 
must be linear in the 
polarization 4-vector $\epsilon^\mu$ of the $D^{*0}$ and it can depend only on 
the external 4-momenta $P_B$, $P_K$, and $P_X \equiv P_B - P_K$
of the $B^+$, $K^+$, and $D^{*0} \bar D^0$ pair.
Since the operator ${\cal H}_{\rm weak}$ is a Lorentz scalar and 
$P_X \cdot \epsilon = 0$, the QCD matrix element 
must be $P_B \cdot \epsilon$ multiplied by a constant.
The EFT matrix element on the right side of Eq.~(\ref{HOmatch}) 
must be linear in the polarization 3-vector $\bm{\epsilon}$ 
of the $D^{*0}$ in the $D^{*0} \bar D^0$ center-of-mass frame 
and there are no 3-momenta that it can depend on.
Since the operator ${\cal O}_i^m$ is a Cartesian vector 
with index $m$, the EFT matrix element on the left side 
of Eq.~(\ref{HOmatch}) must be $\epsilon^m$ multiplied 
by a constant.  By matching the two sides of Eq.~(\ref{HOmatch}),
we conclude that the short-distance coefficients must have the form
\begin{equation}
C^{i,m} = \frac{1}{\sqrt{M_B}} C_{B^+}^{K^+,i} P_B^\mu L_{\mu}^m,
\label{Cnj-match}
\end{equation}
where the coefficients $C_{B^+}^{K^+,i}$ are complex numbers
and $L_{\mu}^m$ is the boost 
matrix from a general Lorentz frame to the center-of-mass frame of the 
$D^{*0} \bar D^0$ pair. 
The prefactor $M_B^{-1/2}$ has been inserted for later convenience.
The boost matrix satisfies $P_X^\mu L_{\mu}^m = 0$ 
and the tensor identities
\begin{subequations}
\begin{eqnarray}
\delta^{mn} L_{\mu}^m L_{\nu}^n
&=& -g_{\mu \nu} + P_{X \mu} P_{X \nu}/P_X^2 ,
\label{deltaLL}
\\
g^{\mu \nu} L_{\mu}^m L_{\nu}^n 
&=& - \delta^{mn}.
\label{gLL}
\end{eqnarray}
\end{subequations}

The imaginary part of ${\mathcal A}_{(1^{++})}$ can be obtained 
from Eq.~(\ref{ABB-EFT}) by using cutting rules.
The relevant cuts run through the $K^+$ propagator 
and through the EFT Green's function:
\begin{eqnarray}
{\rm Im} {\mathcal A}_{(1^{++})} &=& 
\frac{1}{M_B} \sum_{i,j} C_{B^+}^{K^+,i} (C_{B^+}^{K^+,j})^*
\int \frac {d^4 P_K}{(2 \pi)^4} ~
2 \pi \delta(P_K^2 - m_K^2) \, P_B^\mu P_B^\nu L_{\mu}^m L_{\nu}^n 
\nonumber
\\
&& \times
{\rm Im} \, i \int \!\! dt~e^{i E t} \int \!\! d^3r~
\left \langle 0 | {\rm T} {\cal O}_j^{n \dagger}(\bm{r},t) 
                        {\cal O}_i^{m}(\bm{0},0) | 0 \right \rangle.
\label{ImABB-EFT}
\end{eqnarray}
Rotational invariance implies that the EFT matrix element is 
proportional to $\delta^{mn}$.  Using Eq.~(\ref{deltaLL}),
the integrand of the integral over $P_K$ in Eq.~(\ref{ImABB-EFT})
becomes Lorentz invariant.  It can be reduced to an integral 
over $(P_B-P_K)^2$ or, equivalently, $E$:
\begin{eqnarray}
\int \frac {d^4 P_K}{(2 \pi)^4} ~ 2 \pi \delta(P_K^2 - m_K^2) ~ 
P_B^\mu P_B^\nu (L_{\mu}^m L_{\nu}^n) \delta^{mn} = 
\int dE~\frac {\lambda^{3/2} (M_B, m_K, M_{*0}+M_0+E)}
{32 \pi^2 M^2_B (M_{*0} +M_0+E)} .
\label{intPK}
\end{eqnarray}
In the $D^* \bar D$ threshold region, the energy $E$ 
is negligible compared to $M_{*0} + M_0$.
The amplitude in Eq.~(\ref{ImABB-EFT}) therefore reduces to
\begin{eqnarray}
{\rm Im} {\mathcal A}_{(1^{++})} &=& 
\frac {\lambda^{3/2} (M_B, m_K, M_{*0}+M_0)}
{96 \pi^2 M_B^3 (M_{*0} +M_0)}
\sum_{i,j} C_{B^+}^{K^+,i} (C_{B^+}^{K^+,j})^*
\int dE~{\rm Im} F_{ij}(E),
\label{ImABB-EFT:2}
\end{eqnarray}
where $F_{ij}(E)$ is the EFT matrix element:
\begin{equation}
F_{ij}(E) = 
i \int \!\! dt~e^{i E t} \int \!\! d^3r~
\langle 0 | {\rm T} {\cal O}_j^{m \dagger}(\bm{r},t) 
                        {\cal O}_i^{m}(\bm{0},0) | 0 \rangle.
\label{Fij}
\end{equation}
Note that there is an implied sum over the repeated 
Cartesian vector index $m$.
Inserting Eq.~(\ref{ImABB-EFT:2}) into Eq.~(\ref{GamB:optical}), 
we get a factorization formula 
for the inclusive energy distribution summed over $1^{++}$ states:
\begin{equation}
\frac{d\Gamma}{dE}[ B^+ \to K^+ + (1^{++})] = \Gamma_{B}^{K} \, 
\sum_{i,j} C_{B^+}^{K^+,i} (C_{B^+}^{K^+,j})^* \, {\rm Im} \, F_{ij}(E) .
\label{lsh:gen}
\end{equation}
The prefactor has dimensions of energy:
\begin{equation}
\Gamma^{K}_{B} = 
\frac {\lambda^{3/2} (M_B, m_K, M_{*0}+M_0)}
{96 \pi^2 M_B^4 (M_{*0} +M_0)} .
\label{Gammaij}
\end{equation}
If  the short-distance coefficients 
$C_{B^+}^{K^+,i}$ are chosen to be dimensionless,
the functions $F_{ij}(E)$ have the same dimensions (momentum)$^{-1}$
as a scattering amplitude.

\subsection{Universal line shapes}
\label{sec:LSuniversal}

In Ref.~\cite{Braaten:2007dw}, Braaten and Lu derived expressions for 
the universal line shapes for $1^{++}$ states near the
$D^{*0} \bar D^0$ threshold produced by the
decays $B \to K + (1^{++})$ that take into account the large scattering length
in the neutral channel
$(D^{*0} \bar D^0)_+$. The $D^{*0} \bar D^0$ threshold region can be 
described by an effective field theory for the neutral charm mesons
$D^0$, $D^{*0}$, $\bar D^0$, and $\bar D^{*0}$ in which the only 
interaction is zero-range scattering in the $(D^{*0} \bar D^0)_+$ channel.
The scattering amplitude $f(E)$ is given in Eq.~(\ref{f-E}),
with $\kappa(E)$ given by Eq.~(\ref{kappa-E}).
It depends on a single interaction parameter $\gamma$.

The effects of the $B^+ \to K^+$ transition can be reproduced 
in this effective field theory by a local operator ${\cal O}_0^{m}$
with a Cartesian vector index $m$
that creates a pair of charm mesons in the $(D^{*0} \bar D^0)_+$ channel.
The Green's function defined by Eq.~(\ref{Fij}) for this operator
is proportional to the scattering amplitude $f(E)$.
The normalization of the operator ${\cal O}_0^n$
can be chosen so that $F_{00}(E)$ is exactly $f(E)$.
The resulting factorization formula for the inclusive energy
distribution of $1^{++}$ states produced by the transition 
$B^+ \to K^+$ is
\begin{equation}
\frac {d \Gamma}{dE} [ B^+ \to K^+ + (1^{++}) ] = 
\Gamma^{K}_{B} \, \big|C_{B^+}^{K^+,0} \big|^2 \, {\rm Im} \, f(E) ,
\label{lsh1:nc}
\end{equation}
where $C_{B^+}^{K^+,0}$ is a complex short-distance coefficient.
The imaginary part of the scattering amplitude in Eq.~(\ref{lsh1:nc}) 
is given by the optical theorem in Eq.~(\ref{ImA-optical2}).
The term proportional to Im$\kappa(E)$ is the contribution 
from channels whose ultimate final states are
$(D^0 \bar D^0 \pi^0,D^0 \bar D^0 \gamma)$.
The term proportional to ${\rm Im} \gamma$ 
is the contribution from all other channels.

The inclusive energy distribution for $1^{++}$ states
produced by the transition $B^0 \to K^0$ 
is given by an expression identical to
Eq.~(\ref{lsh1:nc}) except that $C_{B^+}^{K^+,0}$ is replaced by 
an independent short-distance coefficient $C_{B^0}^{K^0,0}$.
Since the rate in Eq.~(\ref{lsh1:nc}) does not depend 
on the phase of $C_{B^+}^{K^+,0}$, the short-distance 
coefficients for the $B^+ \to K^+$ and $B^0 \to K^0$ transitions
depend on only two independent real constants.

\subsection{Zero-range line shapes}
\label{sec:LSzero-range}

In Ref.~\cite{Braaten:2007ft},
the universal line shapes in Section~\ref{sec:LSuniversal} 
were generalized to the Zero-Range model
of Section~\ref{sec:zero-range}. 
This model can be derived from a renormalizable quantum field theory 
for neutral and charged charm mesons with zero-range interactions 
among the channels $(D^{*0} \bar D^0)_+$ and $(D^{*+} D^-)_+$.
The effects of the $B^+ \to K^+$ transition can be reproduced 
by a linear combination of two local operators 
that create pairs of charm mesons in the 
$(D^{*0} \bar D^0)_+$ and $(D^{*+} D^-)_+$ channels.
The two operators mix under renormalization.
For an appropriate choice of the renormalized operators 
${\cal O}_0^n$ and ${\cal O}_1^n$,
the Green's functions $F_{ij}(E)$ defined by Eq.~(\ref{Fij}) 
are simply the scattering amplitudes $f_{ij}(E)$.
The resulting factorization formula for the inclusive energy
distribution of $1^{++}$ states produced by the transition 
$B^+ \to K^+$ is
\begin{equation}
\frac{d\Gamma}{dE}[ B^+ \to K^+ + (1^{++})] =
\Gamma^{K}_{B} \, \sum_{i,j}  C_{B^+}^{K^+,i} (C_{B^+}^{K^+,j})^*\, 
{\rm Im} \, f_{ij}(E) ,
\label{lsh2:cc}
\end{equation}
where the sums are over $i,j \in \{ 0,1 \}$.
The short-distance coefficients $C_{B^+}^{K^+,i}$ can be complex.
Since the rate in Eq.~(\ref{lsh2:cc}) does not depend on a
common phase in $C_{B^+}^{K^+,0}$ and $C_{B^+}^{K^+,1}$,
the short-distance coefficients are determined by 
three independent real parameters.
The scattering amplitudes $f_{ij}(E)$ are given in Eqs.~(\ref{fij-E})
and expressions for their imaginary parts are given in Eq.~(\ref{ImAij}).
Note that Eq.~(\ref{ImAij}) gives two different decompositions 
of the imaginary part of the function $f_{10}(E) = f_{01}(E)$ 
into terms linear in the imaginary parts of $\kappa(E)$, $\kappa_1(E)$,
$\gamma_0$, and $\gamma_1$.  
To obtain an expression for $d\Gamma/dE$
that is consistent with the Cutkosky cutting rules,
these two different expressions must be inserted for 
${\rm Im} f_{01}(E)$ and ${\rm Im} f_{10}(E)$ in Eq.~(\ref{lsh2:cc}).

The inclusive energy distribution for $1^{++}$ states
produced by the transition 
$B^0 \to K^0$ is given by an expression identical to
Eq.~(\ref{lsh2:cc}) except that the short-distance coefficients 
$C_{B^+}^{K^+,i}$ are replaced by $C_{B^0}^{K^0,i}$.
Since the effective weak hamiltonian ${\cal H}_{\rm weak}$ 
in Eq.~(\ref{Hweak}) is an isospin scalar,
the operator $\sum_{i} C^{i,m} {\cal O}_i^m$  
must also be an isospin scalar.
The approximate isospin symmetry of QCD can then be used 
to relate $C_{B^0}^{K^0,i}$ to $C_{B^+}^{K^+,i}$ \cite{Braaten:2007ft}:
\begin{subequations}
\begin{eqnarray}
C_{B^0}^{K^0,0} &=& C_{B^+}^{K^+,1},
\label{C:BK0}
\\
C_{B^0}^{K^0,1} &=& C_{B^+}^{K^+,0}.
\label{C:BK1}
\end{eqnarray}
\label{CBK:isospin}
\end{subequations}
Thus the two short-distance coefficients for
the $B^0 \to K^0$ transition are determined by the same 
three independent real constants as the short-distance coefficients 
for the $B^+ \to K^+$ transition.
In Ref.~\cite{Braaten:2007ft}, the expressions for $C_{B^0}^{K^0,i}$
had the opposite signs from those in Eqs.~(\ref{CBK:isospin}).
This error did not affect any of the physical results 
in Ref.~\cite{Braaten:2007ft}.

\subsection{Flatt\'e line shapes}
\label{sec:LSFlatte}

In Ref.~\cite{Hanhart:2007yq}, Hanhart, Nefediev, and Kalashnikova
proposed line shapes for the $X(3872)$ resonance 
in the $D^{*0} \bar D^0$ threshold region that correspond 
to the Flatt\'e parameterization of the $D^{*0} \bar D^0$
elastic scattering amplitude in Eq.~(\ref{f-Flatte}).
Their line shapes in the $D^{*0} \bar D^0$ and 
$J/\psi \, \pi^+ \pi^-$ decay channels are
\begin{subequations}
\begin{eqnarray}
\frac{d \Gamma}{dE} [ B^+ \to K^+ (D^{*0} \bar D^0,D^0 \bar D^{*0})] &=& 
\frac{2 {\cal B} \Gamma[B^+]}{\pi g^2}
|f_\textrm{Flatt\'e}(E)|^2 \sqrt{2 \mu E}~\theta(E),
\label{HKN:lsJpsi}
\\
\frac{d \Gamma}{dE} [ B^+ \to K^+ J/\psi \, \pi^+ \pi^-] &=& 
\frac{2 {\cal B} \Gamma[B^+]}{\pi g^4}
|f_\textrm{Flatt\'e}(E)|^2~\Gamma_{J/\psi  \pi \pi}(E),
\label{HKN:lsDD}
\end{eqnarray}
\label{HKN:ls}
\end{subequations}
where $\Gamma[B^+]$ is the total width of the $B^+$ and 
${\cal B}$ is a short-distance coefficient associated with the 
$B^+ \to K^+$ transition.  The energy dependence of
the function $\Gamma_{J/\psi  \pi \pi}(E)$ is the decay rate 
of a resonance of energy $E$ into $J/\psi \, \pi^+ \pi^-$
through the decay into $J/\psi \, \rho$ followed by 
$\rho \to \pi^+ \pi^-$.

The Flatt\'e line shapes can be derived in the effective field 
theory framework of Section~\ref{sec:EFT} by assuming that the 
short-distance process is the creation of the resonance 
through an operator ${\cal O}_2^m$.  
The Green's function $F_{22}(E)$ is proportional to the propagator 
of the resonance.  The constant factor can be chosen so that  
$F_{22}(E)$ is equal to the Flatt\'e scattering amplitude 
$f_\textrm{Flatt\'e}(E)$ in Eq.~(\ref{f-Flatte}).
The inclusive energy distribution 
in Eq.~(\ref{lsh:gen}) is then
\begin{equation}
\frac{d\Gamma}{dE}[ B^+ \to K^+ + (1^{++})] =
\Gamma^{K}_{B} \, \big| C_{B^+}^{K^+,2}\big|^2 \,
{\rm Im} \, f_\textrm{Flatt\'e}(E) ,
\label{lsh2:Flatte}
\end{equation}
where $C_{B^+}^{K^+,2}$ is the short-distance coefficient 
for the operator that creates the resonance.
The imaginary part of the Flatt\'e scattering amplitude is given 
in Eq.~(\ref{Imf-Flatte}).  

The line shapes in 
Eqs.~(\ref{HKN:lsJpsi}) and (\ref{HKN:lsDD}) can be obtained by 
replacing $\Gamma^{K}_{B} | C_{B^+}^{K^+,2}|^2$ by 
$2 {\cal B} \Gamma[B^+]/(\pi g^2)$ and by making the substitutions
\begin{subequations}
\begin{eqnarray}
- \textrm{Im} \kappa(E) &\longrightarrow& \sqrt{2 \mu E}~\theta(E),
\label{Imk:HKN}
\\
- \textrm{Im} \nu &\longrightarrow& \Gamma_{J/\psi \pi \pi}(E)/2.
\label{Imnu:HKN}
\end{eqnarray}
\label{Imknu:HKN}
\end{subequations}
The line shape for $D^{*0} \bar D^0$ in Eq.~(\ref{HKN:lsJpsi})
vanishes for $E<0$.  Thus the substitution in
Eq.~(\ref{Imk:HKN}) discards all contributions to  
$(D^0 \bar D^0 \pi^0,D^0 \bar D^0 \gamma)$ final states 
from the $X(3872)$ resonance below the $D^{*0} \bar D^0$ threshold.

\subsection{Zero-range+resonance line shapes}
\label{sec:LSzero-range+resonance}

We now generalize the line shapes of Sections~\ref{sec:LSzero-range}
and \ref{sec:LSFlatte} to the Zero-Range+Resonance model
of Section~\ref{sec:zero-range+resonance}, in which the coupled
channels $(D^{*0} \bar D^0)_+$ and $(D^{*+} D^-)_+$
scatter through a resonance as well as through zero-range interactions.  

\subsubsection{General case}

In the Zero-Range+Resonance model,
the scattering amplitudes $f_{ij}(E)$ are given by Eq.~(\ref{f-inverse:res})
and the resonance propagator $P(E)$ is given in Eq.~(\ref{prop:res}).
They depend on the three independent entries of the symmetric matrix 
$\Lambda$, the two components of the column vector $G$, and $\nu$.
As described in the Appendix, this model can be 
derived from a renormalizable field theory for the neutral and charged
charm mesons and the resonance $\chi$.

The effects of the $B^+ \to K^+$ transition can be reproduced 
in this model by a linear combination of three local operators: 
two of them create pairs of charm mesons in the 
$(D^{*0} \bar D^0)_+$ and $(D^{*+} D^-)_+$ channels
and the third creates the resonance.
The solution to the renormalization problem for these operators is 
presented in the Appendix.
Two of the renormalized operators ${\cal O}_0^n$ and ${\cal O}_1^n$
can be chosen so that the Green's functions $F_{ij}(E)$  
defined in Eq.~(\ref{Fij}) for $i,j = 0,1$
are just the scattering amplitudes:
\begin{equation}
F_{ij}(E) = f_{ij}(E), \hspace{1cm} i,j \in \{ 0,1\}.
\label{Fij-F}
\end{equation}
The remaining operator ${\cal O}_2^n$ can be chosen
so that $F_{22}(E)$ is the resonance propagator $P(E)$
given in Eq.~(\ref{prop:res}) multiplied by a constant: 
\begin{eqnarray}
F_{22}(E) &=& - (G^T G/2) P(E).
\label{FRR}
\end{eqnarray}
The multiplicative factor of $G^T G$ has been inserted only to ensure
that $F_{22}(E)$ has the same dimensions as a scattering amplitude.
The two remaining Green's functions $F_{02}(E)$ and $F_{12}(E)$ 
are then given by
\begin{eqnarray}
F_{i2}(E) &=& (G^T G/2)^{1/2}
 \left( f(E)~
        [(E- \nu)\Lambda + G G^T]^{-1}~
        G \right)_i .
\label{FiR}
\end{eqnarray}

The inclusive energy distribution for $1^{++}$ states produced by the
transition $B^+ \to K^+$ 
is given by the factorization formula in Eq.~(\ref{lsh:gen}),
where the sums are over $i,j \in \{ 0,1,2 \}$.
The short-distance coefficients $C_{B^+}^{K^+,i}$ can be complex.
Since the rate in Eq.~(\ref{lsh:gen}) is not affected by a common 
phase in $C_{B^+}^{K^+,i}$, these three coefficients are determined
by 5 independent real constants.

\subsubsection{Isospin symmetry}

Using the approximate isospin symmetry of QCD,
the matrix $\Lambda$ can be expressed in terms of two independent 
parameters $\gamma_0$ and $\gamma_1$ using Eq.~(\ref{Lambda-gamma})
and the column vector $G$ can be expressed in terms of a single
coupling constant $g$ using Eq.~(\ref{G:isospin}).
The scattering amplitudes $f_{ij}(E)$ are given in Eqs.~(\ref{fijchi-E}).
The remaining Green functions $F_{ij}(E)$  
in Eqs.~(\ref{FiR}) and (\ref{FRR}) reduce to
\begin{subequations}
\begin{eqnarray}
F_{02}(E) &=& 
\frac{g^2 \gamma_0 [-\gamma_1 + \kappa_1(E)]}{D(E)} ,
\label{F0R}
\\
F_{12}(E) &=& 
- \frac{g^2 \gamma_0 [-\gamma_1 + \kappa(E)]}{D(E)} ,
\\
\label{F1R}
F_{22}(E) &=& 
\frac{-g^2 D_0(E)}{2 D(E)} ,
\label{F22R}
\end{eqnarray}
\end{subequations}
where the denominator  $D(E)$ is given in Eq.~(\ref{DF-E})
and $D_0(E)$ is given in Eq.~(\ref{D-E}).

The inclusive energy distribution for $1^{++}$ states produced by the
transition $B^0 \to K^0$ 
is given by a factorization formula identical to
Eq.~(\ref{lsh:gen}) except that the short-distance coefficients 
$C_{B^+}^{K^+,i}$ are replaced by $C_{B^0}^{K^0,i}$.
Isospin symmetry implies the relations between $C_{B^0}^{K^0,i}$ 
and $C_{B^+}^{K^+,i}$ for $i=0,1$ in Eqs.~(\ref{CBK:isospin}).
The assumption that the resonance has isospin 0 
implies a similar relation between the short-distance coefficients for $i=2$:
\begin{equation}
C_{B^+}^{K^+,2}= C_{B^0}^{K^0,2}.
\label{C3:isospin}
\end{equation}
Thus the three complex short-distance coefficients for the
$B^0 \to K^0$ transition are determined by the same 
five independent real constants as the short-distance coefficients 
for the $B^+ \to K^+$ transition.

\subsubsection{Optical theorem}

The inclusive energy distribution in Eq.~(\ref{lsh:gen})
depends on the imaginary parts of the amplitudes $F_{ij}(E)$.
Expressions for the imaginary parts of $F_{ij}(E)$ for 
$i,j \in \{0,1\}$ that are consistent with the Cutkosky cutting rules
are given in Eqs.~(\ref{Imfijchi}).  We also need expressions for the 
imaginary parts of $F_{02}(E)$, $F_{12}(E)$, and $F_{22}(E)$
that are consistent with the Cutkosky cutting rules.
The imaginary parts of $F_{i2}(E) = F_{2i}(E)$ can be expressed as
\begin{subequations}
\begin{eqnarray}
{\rm Im} F_{i2}(E) &=& 
\frac{g^2}{2} {\rm Im}\left[ f_{i0}(E) - f_{i1}(E) \right]
\left( \frac{  \gamma_0}{E - \nu + g^2 \gamma_0} \right)^* 
\nonumber \\
&& + \frac{g^2}{2} \left[ f_{i0}(E) - f_{i1}(E) \right]
\left( \frac{E-\nu}{|E - \nu + g^2 \gamma_0|^2} {\rm Im} \gamma_0
+ \frac{\gamma_0}{|E - \nu + g^2 \gamma_0|^2} {\rm Im} \nu
\right) ,
\nonumber \\
\label{Im_FiR}
\\
{\rm Im} F_{2i}(E) &=& \frac{g^2}{2} 
\left( \frac{  \gamma_0}{E - \nu + g^2 \gamma_0} \right) 
{\rm Im}\left[ f_{0i}(E) - f_{1i}(E) \right] 
\nonumber \\
&& + \frac{g^2}{2} 
\left( \frac{E-\nu}{|E - \nu + g^2 \gamma_0|^2} {\rm Im} \gamma_0
+ \frac{\gamma_0}{|E - \nu + g^2 \gamma_0|^2}  {\rm Im} \nu \right) 
\left[ f_{0i}^*(E) - f_{1i}^*(E) \right] .
\nonumber \\
\label{Im_FRi}
\end{eqnarray}
\label{Im_FiRRi}
\end{subequations}
The imaginary parts of $f_{i0}(E)$, $f_{i1}(E)$, $f_{0i}(E)$, 
and $f_{1i}(E)$ can be decomposed into terms linear in the
imaginary parts of $\kappa(E)$, $\kappa_1(E)$, $\gamma_0$, 
$\gamma_1$, and $\nu$ by using Eqs.~(\ref{Imfijchi}).
Note that Eqs.~(\ref{Im_FiR}) and (\ref{Im_FRi})
give two different decompositions of the imaginary parts of
the function $F_{i2}(E) = F_{2i}(E)$.
To obtain an expression for $d\Gamma/dE$
that is consistent with the Cutkosky cutting rules,
these two different expressions must be inserted for 
${\rm Im} F_{i2}(E)$ and ${\rm Im} F_{2i}(E)$ in Eq.~(\ref{lsh:gen}).
The imaginary part of $F_{22}(E)$ can be expressed as
\begin{eqnarray}
{\rm Im} \,F_{22}(E) &=& 2 |F_{22}|^2
\Bigg[  
- \frac{1}{g^2}{\rm Im}\nu
\nonumber \\
& &
+ \left|\frac{-\gamma_1[\kappa(E)+\kappa_1(E)]+2\kappa_1(E)\kappa(E)}{D_0(E)} \right|^2 
		{\rm Im}\gamma_0 
+ \left|\frac{\gamma_0[\kappa(E)-\kappa_1(E)]}{D_0(E)}\right|^2
		{\rm Im}\gamma_1
\nonumber \\
& &
- 2 \left|\frac{\gamma_0[-\gamma_1+\kappa_1(E)]}{D_0(E)}\right|^2
		{\rm Im}\kappa(E)
- 2 \left|\frac{\gamma_0[-\gamma_1+\kappa(E)]}{D_0(E)}\right|^2
	{\rm Im}\kappa_1(E)
\Bigg] ,
\label{Im_FRR}
\end{eqnarray}
where $D_0(E)$ is the denominator given in Eq.~(\ref{D-E}).

\subsubsection{$D^{*0} \bar{D}^0$ threshold region}

The interaction parameters $\gamma_0$, 
$\gamma_1$, $\nu$, and $g$ can be tuned so that there is a 
bound state just below the $D^{*0} \bar D^0$ threshold
that can be identified with the $X(3872)$.
The amplitudes $F_{ij}(E)$ have poles at $\kappa(E) = \gamma$
with residues $Z_i^{1/2} Z_j^{1/2}$.
The residue factors $Z_0^{1/2}$ and $Z_1^{1/2}$ are given by 
Eqs.~(\ref{Z1/Z0:ZRR}) and (\ref{Z0:ZRR}).
The ratio of $Z_2^{1/2}$ and $Z_0^{1/2}$ is
\begin{eqnarray}
\frac{Z_2^{1/2}}{Z_0^{1/2}} = 
- \frac{g^2 [\gamma_1 \kappa_{11} + (\gamma_1 - 2 \kappa_{11}) \gamma]}
       {2 (\gamma_1 - \kappa_{11}) (\nu - E_\textrm{pole})} . 
\label{Z2/Z0}
\end{eqnarray}
If the small value of $\gamma$ 
is obtained by a single fine-tuning of $\gamma_0$ or $\nu$,
the behavior of these amplitudes 
in the entire $D^{*0} \bar D^0$ threshold region 
$|E| \ll \nu_{11} \approx 8.1$~MeV is dominated by the pole
at  $\kappa(E) = \gamma$.  
They all reduce to the universal scattering amplitude
$f(E)$ in Eq.~(\ref{f-E}) multiplied by residue factors:
\begin{eqnarray}
F_{ij}(E) \approx Z_i^{1/2}~f(E)~Z_j^{1/2}.
\label{F-ZZ}
\end{eqnarray}

The inclusive energy distribution in Eq.~(\ref{lsh:gen})
simplifies in the  $D^{*0} \bar D^0$ threshold region:
\begin{equation}
\frac{d\Gamma}{dE}[ B^+ \to K^+ + (1^{++})] \approx \Gamma_{B}^{K} 
\left| C_{B^+}^{K^+,0} 
- \frac{\gamma_1}{\gamma_1 - \kappa_{11}} C_{B^+}^{K^+,1}
- \frac{g^2 \gamma_1 \kappa_{11}}{2 (\gamma_1 - \kappa_{11}) \nu} C_{B^+}^{K^+,2} 
\right|^2 \, {\rm Im} f(E) ,
\label{lsh:gen-thresh}
\end{equation}
where ${\rm Im} f(E)$ is given in Eq.~(\ref{ImA-optical2}).
The coefficients $Z_j^{1/2}$ of 
$C_{B^+}^{K^+,j}$ have been simplified by taking the limit $\gamma \to 0$ 
and by neglecting the imaginary parts of the interaction parameters.
The result in Eq.~(\ref{lsh:gen-thresh}) agrees with the universal 
factorization formula in Eq.~(\ref{lsh1:nc}) 
after a redefinition of the short-distance coefficient $C_{B^+}^{K^+,0}$.
By using the expression for ${\rm Im} \gamma$ in Eq.~(\ref{Imgamma-ZR+R}),
the energy-dependent factor in Eq.~(\ref{lsh:gen-thresh})
can be expressed as a linear combination of the imaginary parts of
$\gamma_0$, $\gamma_1$, $\nu$, and $\kappa(E)$.
There are additional contributions to the coefficients of the imaginary parts 
of $\gamma_0$, $\gamma_1$, and $\nu$ that come from cuts through the 
factors of $Z_i^{1/2}$ and $Z_j^{1/2}$ in Eq.~(\ref{F-ZZ}),
but they are suppressed compared 
to those in Eq.~(\ref{lsh:gen-thresh}) by factors that are at least as small as
$\gamma/\kappa_{11}$ or $\kappa(E)/\kappa_{11}$.

\subsubsection{Resonance far above $D^* \bar D$ threshold}

In the region of parameter space in which $\nu$ is much larger 
than the energy scale $\nu_{11} = 8.1$~MeV of isospin splitting,
there is a second resonance $\chi$ well above the 
$D^{*+} D^-$ threshold in addition to the $X(3872)$ resonance
just below the $D^{*0} \bar D^0$ threshold.
In this case, the amplitudes $F_{ij}(E)$ can be simplified.

For energies $E$ in the $\chi$ resonance region,
the line shapes are dominated by a Breit-Wigner resonance 
whose energy and width are given by Eqs.~(\ref{EGam:F}).
The scattering amplitudes $f_{ij}(E)$ reduce near the resonance 
to the expressions in Eq.~(\ref{fij-hinu}).
The remaining amplitudes reduce to
\begin{subequations}
\begin{eqnarray}
F_{02}(E) \approx -F_{12}(E) &\approx& 
-\frac{i \gamma_0}{\sqrt{2 \mu E_\chi}}~f_\textrm{BW}(E) ,
\\
F_{22}(E) &\approx& 
f_\textrm{BW}(E) ,
\end{eqnarray}
\end{subequations}
where $f_\textrm{BW}(E)$ is the Breit-Wigner scattering amplitude 
in Eq.~(\ref{f-BW}).
Note that $F_{ij}(E)$ differs from $f_\textrm{BW}(E)$ by a factor of 
$-i \gamma_0/\sqrt{2 \mu E_\chi}$
for every subscript 0 and by a 
factor of $+i \gamma_0/\sqrt{2 \mu E_\chi}$
for every subscript 1.
Thus the expression in Eq.~(\ref{lsh:gen}) for the inclusive 
energy distribution reduces near the resonance to
\begin{eqnarray}
\frac{d\Gamma}{dE}[ B^+ \to K^+ + (1^{++})] \approx \Gamma_{B}^{K} 
\left| -\frac{i \gamma_0}{\sqrt{2 \mu E_\chi}} 
(C_{B^+}^{K^+,0} - C_{B^+}^{K^+,1})
+ C_{B^+}^{K^+,2} \right|^2 
\frac{g^2 \Gamma_\chi/4}{(E - E_\chi)^2 + \Gamma_\chi^2/4} .
\nonumber \\
\label{lsh:hinu}
\end{eqnarray}
Upon integrating over the energy, the partial width from the Breit-Wigner
resonance is
\begin{eqnarray}
\Gamma[ B^+ \to K^+ + \chi_{c1}'] \approx \frac{\pi g^2}{2}\Gamma_{B}^{K} 
\left| -\frac{i \gamma_0}{\sqrt{2 \mu E_\chi}} 
(C_{B^+}^{K^+,0} - C_{B^+}^{K^+,1})
+ C_{B^+}^{K^+,2} \right|^2 .
\end{eqnarray}

\subsubsection{Zero-range limit}

If the parameters satisfy $|\nu| \gg  g^2 |\gamma_0|$,
the Zero-Range+Resonance model reduces to the Zero-Range model 
in the energy region $|E| \ll |\nu|$.
The scattering amplitudes reduce to those in Eqs.~(\ref{fij-E}).
The amplitudes $F_{02}(E)$, $F_{12}(E)$, and $F_{22}(E)$
reduce to
\begin{subequations}
\begin{eqnarray}
F_{02}(E) &\approx&  
- \frac{g^2 \gamma_0 [- \gamma_1 + \kappa_1(E)]}{\nu D_0(E)}, 
\\
F_{12}(E) &\approx& 
\frac{g^2 \gamma_0 [- \gamma_1 + \kappa(E)]}{\nu D_0(E)},
\\
F_{22}(E) &\approx& 
\frac{g^2}{2 \nu}.
\end{eqnarray}
\label{Fij-zerorange}
\end{subequations}
They are suppressed by a factor of $g^2 \gamma_0/\nu$.
Thus the line shapes reduce to those of the Zero-Range model 
in Eq.~(\ref{lsh2:cc}).

\subsubsection{Flatt\'e limit}

If the parameters satisfy
$|\gamma_1| \gg \kappa_{11}$ and $\nu \ll g^2 |\gamma_0|,g^2 |\gamma_1|$,
the Zero-Range+Resonance model reduces to the Flatt\'e model 
in the energy region $|E| \ll |\gamma_1|^2/\mu,g^2 |\gamma_0|,g^2 |\gamma_1|$.
The scattering amplitudes $f_{ij}(E)$ reduce to those 
of the Flatt\'e model in Eqs.~(\ref{fij-Flatte}).
The remaining amplitudes $F_{02}(E)$, $F_{12}(E)$, and $F_{22}(E)$
reduce  to
\begin{equation}
F_{02}(E) \approx -F_{12}(E) \approx 
F_{22}(E) \approx f_\textrm{Flatt\'e}(E) .
\end{equation}
where $f_\textrm{Flatt\'e}(E)$ is the Flatt\'e scattering amplitude 
in Eq.~(\ref{f-Flatte}). Note that $F_{ij}(E)$ differs from $f_\textrm{Flatt\'e}(E)$ 
by a factor of $-1$ for every subscript 1.
Thus the expression in Eq.~(\ref{lsh:gen}) for the 
inclusive energy distribution reduces to
\begin{equation}
\frac{d\Gamma}{dE}[ B^+ \to K^+ + (1^{++})] \approx \Gamma_{B}^{K} 
\left| C_{B^+}^{K^+,0} - C_{B^+}^{K^+,1}
+ C_{B^+}^{K^+,2} \right|^2 ~{\rm Im} f_\textrm{Flatt\'e}(E). 
\label{lsh:noscat}
\end{equation}
The imaginary part of the Flatt\'e scattering amplitude
is given in Eq.~(\ref{Imf-Flatte}).  The inclusive 
energy distribution in Eq.~(\ref{lsh:noscat}) agrees with that
for the Flatt\'e model in Eq.~(\ref{lsh2:cc})
after a redefinition of the short-distance coefficient $C_{B^+}^{K^+,2}$.

If there is a resonance well above the $D^{*+} D^-$ threshold
that can be identified with $\chi_{c1}'$,
the Flatt\'e scattering amplitude $f_\textrm{Flatt\'e}(E)$  
in Eq.~(\ref{f-Flatte}) reduces near the resonance to the 
Breit-Wigner amplitude in Eq.~(\ref{f-BW}).
Upon integrating over the energy, the partial width 
from the Breit-Wigner resonance is
\begin{equation}
\Gamma[ B^+ \to K^+ + \chi_{c1}'] \approx \frac{\pi g^2}{2} \Gamma_{B}^{K} 
\left| C_{B^+}^{K^+,0} - C_{B^+}^{K^+,1}
+ C_{B^+}^{K^+,2} \right|^2 .
\label{pw:noscat}
\end{equation}

\section{Constraints on the $\bm{B \to K}$ transition coefficients}
\label{sec:BtoKcoeffs}

In this section, we analyze the constraints on the short-distance 
coefficients for the $B \to K$ transitions
from information about $B$ meson decays.

\subsection{$\bm{B}$ meson decays into $\bm{K + X(3872)}$}
\label{sec:BtoKX}

Products of the branching fractions for the decay 
$B \to K + X(3872)$ followed by the decay of $X(3872)$ 
into specific final states have been measured 
by the Belle and Babar Collaborations.  
The most precise measurements are for the final state 
$J/\psi \pi^+ \pi^-$ \cite{Aubert:2008gu,Belle:2008te}
and for the final states $D^0 \bar{D}^0 \pi^0$ \cite{Gokhroo:2006bt} and 
$D^0 \bar{D}^0 \gamma$ \cite{Aubert:2007rva,Belle:2008su}.
We denote the final states $D^0 \bar{D}^0 \pi^0$ and 
$D^0 \bar{D}^0 \gamma$ collectively by $(D^0 \bar D^0 \pi^0,D^0 \bar D^0 \gamma)$.
The measurements for the final state $D^0 \bar{D}^0 \pi^0$ 
in Ref.~\cite{Gokhroo:2006bt} can be interpreted 
as measurements for $(D^0 \bar D^0 \pi^0,D^0 \bar D^0 \gamma)$ multiplied by 
a 62\% branching fraction.  
Since $D^{*0}$ ultimately decays into $D^0 \pi^0$ or $D^0 \gamma$,
the measurements for the final states $D^{*0}\bar{D}^0$ and 
$D^0 \bar D^{*0}$ in Refs.~\cite{Aubert:2007rva,Belle:2008su}
can be interpreted as measurements for $(D^0 \bar D^0 \pi^0,D^0 \bar D^0 \gamma)$.
The measurements for $B^+$ decays are more precise than those 
for $B^0$ decays. We introduce a simple notation for the product
of branching fractions in $B^+$ decays:
\begin{equation}
\textrm{Br}_+[{\rm final}] \equiv
{\rm Br}[B^+ \to K^+ + X(3872) \to K^+ + {\rm final}] .
\label{BrBr:def}
\end{equation}
Another convenient observable is the ratio $R_{0+}$ of the products 
of branching fractions for $B^0$ decays and $B^+$ decays:
\begin{equation}
R_{0+}[{\rm final}] \equiv
\frac{{\rm Br}[B^0 \to K^0 + X(3872) \to K^0 + {\rm final}]}
    {{\rm Br}[B^+ \to K^+ + X(3872) \to K^+ + {\rm final}]}.
\label{R0+:def}
\end{equation}
The measurements of Br$_+$ and $R_{0+}$ 
are summarized in Table~\ref{tab:Xdata}.
To obtain the average values
of the measurements for $J/\psi \pi^+ \pi^-$ and
$(D^0 \bar D^0 \pi^0,D^0 \bar D^0 \gamma)$ in Table~\ref{tab:Xdata},
statistical and systematic errors were added in quadrature.
The ratio of $R_{0+}$ for $(D^0\bar{D}^0 \pi,D^0\bar{D}^0 \gamma)$
and $R_{0+}$ for $J/\psi \pi^+ \pi^-$ is $2.1 \pm 0.9$, 
which differs from 1 by more than 1 standard deviation.
The ratio of $\textrm{Br}_+$ for $(D^0\bar{D}^0 \pi,D^0\bar{D}^0 \gamma)$
and $\textrm{Br}_+$ for $J/\psi \pi^+ \pi^-$ is $11.0 \pm 2.5$.
For reasons that will be discussed below,
this result does not necessarily imply that the 
branching fraction for $X(3872)$ into $(D^0\bar{D}^0 \pi,D^0\bar{D}^0 \gamma)$
is an order of magnitude larger than for $J/\psi \pi^+ \pi^-$.

\begin{table}[t]
\begin{tabular}{l|cll}
Reference & Decay mode & $\textrm{Br}_+ \times 10^6$& $R_{0+}$  \\
\hline 
Babar \cite{Aubert:2008gu} & $J/\psi \pi^+ \pi^-$ & 
$8.4 \pm 1.5 \pm 0.7$ & $0.41 \pm 0.24 \pm 0.05$ \\
Belle \cite{Belle:2008te} & $J/\psi \pi^+ \pi^-$ & 
$8.10 \pm 0.92 \pm 0.66$ & $0.82 \pm 0.22 \pm 0.05$ \\
Average                  & $J/\psi \pi^+ \pi^-$ &
$8.2 \pm 0.9$ & $0.63 \pm 0.17$ \\
\hline 
Belle \cite{Gokhroo:2006bt} & $D^0\bar{D}^0 \pi^0$   & 
$102 \pm 31^{+21}_{-29}$ & $1.63 \pm 1.03$ \\
Babar \cite{Aubert:2007rva} & $D^{*0}\bar{D}^0,D^0 \bar D^{*0}$ & 
$167 \pm 36 \pm 47$ & $1.33 \pm 0.69 \pm 0.43$ \\
Belle \cite{Belle:2008su}   & $D^{*0}\bar{D}^0,D^0 \bar D^{*0}$ &  
$77 \pm 16 \pm 10$ & $1.26 \pm 0.65 \pm 0.06$ \\
Average                   &  $(D^0 \bar D^0 \pi^0,D^0 \bar D^0 \gamma)$ &
$90 \pm 17$ &$1.35 \pm 0.46$ \\
\end{tabular}
\caption{
Data on products of branching fractions for 
$B \to K + X(3872)$ followed by the decay of $X$ into specified decay modes.
The first column gives the reference for the experimental result 
by the Belle or Babar Collaboration.
The second column specifies the decay mode for $X(3872)$.
The third column gives the product $\textrm{Br}_+$ 
of the branching fractions for $B^+$ decays.
The fourth column gives the ratio $R_{0+}$ of the products 
of the branching fractions for $B^0$ decays and $B^+$ decays.
Results with both a statistical and a systematic error are 
experimental results.
Results with a single error were obtained by combining
statistical and systematic errors in quadrature.}
\label{tab:Xdata}
\end{table}

In the Zero-Range+Resonance model,
the products of branching fractions 
can be calculated in terms of the scattering parameters
$\gamma$, $\gamma_1$, $g$, and $\nu$
and the $B \to K$ transition coefficients
$C_{B^+}^{K^+,i}$, $i=1,2,3$.
The inclusive energy distribution for $1^{++}$ states 
is given in Eq.~(\ref{lsh:gen}).  The imaginary parts of the
amplitudes $F_{ij}(E)$ are given in Eqs.~(\ref{Imfijchi}),
(\ref{Im_FiRRi}), and (\ref{Im_FRR}).
The contribution from $(D^0 \bar{D}^0 \pi^0,D^0 \bar{D}^0 \gamma)$ 
is the sum of all the terms with the factor Im$\kappa(E)$.  
Since the decay of $X(3872)$ into $J/\psi \, \pi^+ \pi^-$
is dominated by the decay into $J/\psi$ and a virtual $\rho^0$,
this decay mode has isospin 1.
The contributions from all isospin-1 final states other than 
$D \bar{D} \pi$ and $D \bar{D} \gamma$ is the sum of all terms 
with the factor Im$\gamma_1$.
The contribution from $J/\psi \, \pi^+ \pi^-$ is obtained by replacing
Im$\gamma_1$ by a term 
$(\textrm{Im} \gamma_1)^{J/\psi \, \pi^+ \pi^-}$.

We will use several results from Table~\ref{tab:Xdata} to 
constrain the $B \to K$ transition coefficients.
We will use the average values of the ratios $R_{0+}$:
\begin{subequations}
\begin{eqnarray}
R_{0+}[J/\psi \, \pi^+ \pi^-] &=& 0.63 \pm 0.17,
\label{R0+Jpsi:data}
\\
R_{0+}[D^0 \bar D^0 \pi^0,D^0 \bar D^0 \gamma] &=& 1.35 \pm 0.46.
\label{R0+DD:data}
\end{eqnarray}
\label{R0+JpsiDD:data}
\end{subequations}
We will also use the product of the branching fractions 
for $B^+$ to decay into $K^+$ and $D^0 \bar D^0 \pi^0$ 
or $D^0 \bar D^0 \gamma$:
\begin{equation}
\textrm{Br}_+[D^0 \bar D^0 \pi^0,D^0 \bar D^0 \gamma] = 
(9.0 \pm 1.7) \times 10^{-5}.
\label{Br+DD:data}
\end{equation}
We will not use the product of the branching fractions for
$B^+$ to decay into $K^+$ and $J/\psi \, \pi^+ \pi^-$, 
because it would introduce the additional 
unknown parameter $(\textrm{Im} \gamma_1)^{J/\psi \, \pi^+ \pi^-}$.

We first consider the constraints on the parameters from  
$R_{0+}[J/\psi \, \pi^+ \pi^-]$.
Since $J/\psi \, \pi^+ \pi^-$ has isospin 1, 
the only resonant enhancement comes from the $X(3872)$ resonance
just below the $D^{*0} \bar D^0$ threshold.  The dominant 
contribution to the product of branching fractions 
will come from the  $D^{*0} \bar D^0$ threshold region,
where the line shape from the decay of $B^+$
can be approximated by the simple
expression in Eq.~(\ref{lsh:gen-thresh}).
The line shape from the decay of $B^0$
can be approximated by the same expression with the 
coefficients $C_{B^+}^{K^+,i}$ replaced by $C_{B^0}^{K^0,i}$.
According to the isospin symmetry relations in 
Eqs.~(\ref{CBK:isospin}) and (\ref{C3:isospin}), 
this is equivalent to the substitutions
$C_{B^+}^{K^+,0} \to C_{B^+}^{K^+,1}$ and
$C_{B^+}^{K^+,1} \to C_{B^+}^{K^+,0}$. 
Thus the ratio $R_{0+}$ for the final state
$J/\psi \, \pi^+ \pi^-$ is simply
\begin{equation}
R_{0+}[J/\psi \, \pi^+ \pi^-] =
\frac{\left| 2 (\gamma_1 - \kappa_{11}) \nu C_{B^+}^{K^+,1} 
     - 2 \gamma_1 \nu C_{B^+}^{K^+,0} 
     - g^2 \gamma_1 \kappa_{11} C_{B^+}^{K^+,2} \right|^2}
     {\left| 2 (\gamma_1 - \kappa_{11}) \nu C_{B^+}^{K^+,0} 
     - 2 \gamma_1 \nu C_{B^+}^{K^+,1} 
     - g^2 \gamma_1 \kappa_{11} C_{B^+}^{K^+,2} \right|^2} .
\label{R0+:Jpsi}
\end{equation}
This ratio is different from 1 only if 
$C_{B^+}^{K^+,1} \neq  C_{B^+}^{K^+,0}$, which means that the 
$I=1$ component of the amplitude for producing the charm mesons 
at short distance is nonzero. 
We will constrain this ratio to have the value in 
Eq.~(\ref{R0+Jpsi:data}).

If the line shape for $(D^0 \bar D^0 \pi^0,D^0 \bar D^0 \gamma)$
was completely dominated by the $X(3872)$ resonance
below the $D^{*0} \bar D^0$ threshold, the ratio
$R_{0+}[D^0 \bar D^0 \pi^0,D^0 \bar D^0 \gamma]$ would also be given by the
expression on the right side of Eq.~(\ref{R0+:Jpsi}).  
However the measured value of this ratio in Eq.~(\ref{R0+DD:data})
is significantly larger than that for $J/\psi \, \pi^+ \pi^-$
in Eq.~(\ref{R0+Jpsi:data}).  This is easy to understand.
Both channels have a peak just below the $D^{*0} \bar D^0$ threshold
from the $X(3872)$ resonance.  However the $(D^0 \bar D^0 \pi^0,D^0 \bar D^0 \gamma)$ 
channel also has a threshold enhancement above the 
$D^{*0} \bar D^0$ threshold.  At the peak of the threshold enhancement, 
which is very close to the threshold, the ratio in 
Eq.~(\ref{R0+:Jpsi}) may be a good approximation.  However,
above the peak in the threshold enhancement, 
the energy distribution decreases relatively slowly as $E$ increases.  
Because of this high-energy shoulder, there 
may be a substantial contribution from energies where the ratio 
$R_{0+}$ differs significantly from that in Eq.~(\ref{R0+:Jpsi}).
Some idea of the energy range over which the threshold 
enhancement extends can be obtained from measurements of the 
$X(3872)$ resonance in the $D^{*0} \bar D^0$ and $D^{0} \bar D^{*0}$ 
decay channels by the Belle and Babar Collaborations 
\cite{Aubert:2007rva,Belle:2008su}.
The average of their measurements of the position is $2.1 \pm 1.2$~MeV
above the $D^{*0} \bar D^0$ threshold.  
The average of their 
measurements of the width is $3.5 ^{+1.6}_{-1.0}$~MeV.
These measurements imply that most of the contribution to the threshold 
enhancement comes from the energy region below the 
$D^{*+} D^-$  threshold at $\nu_{11}= 8.1$~MeV.
These measurements of the position and the width in the $D^{*0} \bar D^0$ 
channel should not be interpreted as the actual position 
and the width of the $X(3872)$ resonance for two reasons.
First, the $X(3872)$ resonance and the $D^{*0} \bar D^0$ threshold
enhancement are smeared into a single peak by the experimental resolution.
This explains why the measured width in the $D^{*0} \bar D^0$ 
channel is larger than the upper bound 
in Eq.~(\ref{GamX-lim}) from measurements in the $J/\psi \, \pi^+ \pi^-$
channel.  This effect also biases the measurement 
of the position towards larger values.
Second, in the analysis procedures in Refs.~\cite{Aubert:2007rva,Belle:2008su},
a $D^0 \pi^0$ whose invariant mass is close to the mass of the $D^{*0}$ 
is constrained to have an invariant mass exactly equal to $M_{*0}$.  
This shifts $D^0 \bar D^0 \pi^0$ events just below the 
$D^{*0} \bar D^0$ threshold to above the threshold, which
further biases the measurement of the position 
towards larger values.  
These two biases explain why the measured resonance position in the 
$D^{*0} \bar D^0$ channel is larger than that in Eq.~(\ref{MX-ave})
by more than two standard deviations.

Our prescription for the branching fraction Br$_+$ for the final state
$(D^0 \bar D^0 \pi^0,D^0 \bar D^0 \gamma)$ is an integral over $E$ 
of the appropriate energy distribution:
\begin{equation}
\textrm{Br}_+[D^0 \bar D^0 \pi^0,D^0 \bar D^0 \gamma] = 
\frac{1}{\Gamma[B^+]} \int_{-\delta_{00}}^{+ \nu_{11}} dE~
\frac{d\Gamma}{dE}[B^+ \to K^+ + (D^0 \bar D^0 \pi^0,D^0 \bar D^0 \gamma)],
\label{B+DD}
\end{equation}
where $\Gamma[B^+]= 4.02 \times 10^{-10}$~MeV is the total width of the $B^+$.
The integrand is the contribution to the inclusive energy 
distribution in Eq.~(\ref{lsh:gen}) from terms with the factor
Im$\kappa(E)$.  We have chosen the integration region 
somewhat arbitrarily to extend from the $D^0 \bar D^0 \pi^0$ 
threshold at $-\delta_{00} = -7.1$~MeV to the 
$D^{*+} D^-$ threshold at $\nu_{11} = +8.1$~MeV.
We will constrain the branching fraction in Eq.~(\ref{B+DD})
to have the value in Eq.~(\ref{Br+DD:data}).

Our prescription for the ratio $R_{0+}$ for 
$(D^0 \bar D^0 \pi^0,D^0 \bar D^0 \gamma)$ is the ratio of two integrals
like the one in Eq.~(\ref{B+DD}):
\begin{equation}
R_{0+}[D^0 \bar D^0 \pi^0,D^0 \bar D^0 \gamma] = 
\frac{\int_{-\delta_{00}}^{+ \nu_{11}} dE~
     d\Gamma[B^0 \to K^0 + (D^0 \bar D^0 \pi^0,D^0 \bar D^0 \gamma)]/dE}
     {\int_{-\delta_{00}}^{+ \nu_{11}} dE~
     d\Gamma[B^+ \to K^+ + (D^0 \bar D^0 \pi^0,D^0 \bar D^0 \gamma)]/dE} .
\label{R0+DD}
\end{equation}
The integrand in the numerator differs from that in the denominator
by the substitutions $C_{B^+}^{K^+,0} \to C_{B^+}^{K^+,1}$ and 
$C_{B^+}^{K^+,1} \to C_{B^+}^{K^+,0}$.
We will constrain the ratio in 
Eq.~(\ref{R0+DD}) to have the value in Eq.~(\ref{R0+DD:data}).

\subsection{$\bm{B}$ meson decays into $\bm{K + \chi_{c1}'}$}
\label{sec:Kchi}

The $X(3872)$ could arise from a fine tuning of the 
P-wave charmonium state $\chi_{c1}' \equiv \chi_{c1}(2P)$
to near the $D^{*0} \bar D^0$ threshold.  However the $\chi_{c1}'$
could also be a separate resonance above the $D^{*+} D^-$ threshold. 
The Zero-Range+Resonance model of Section~\ref{sec:zero-range+resonance} 
allows for both possibilities.  If the mass of the $\chi_{c1}'$
is sufficiently far above the $D^{*+} D^-$ threshold, 
the primary effect of the coupling of the $\chi_{c1}'$ to the charm mesons 
is through their contribution to its width.  Thus information about the 
decay $B \to K + \chi_{c1}'$ in the absence of resonant interactions 
with charm mesons can be used to constrain the coefficients for 
the $B \to K$ transition.

In the decay $B \to K + \chi_{c1}'$, the momentum transferred to the 
mesons in the final state is about 1100~MeV.  This might be large enough 
that the decay can be treated using factorization methods that 
separate the hard momentum scales comparable to or larger 
than the momentum transfer from the soft hadronic momentum scales. 
There have been several studies of 
factorization for the decays $B \to K + \chi_{cJ}$, where
$\chi_{cJ} \equiv \chi_{cJ}(1P)$, $J=0,1,2$, is the lowest
multiplet of P-wave charmonium states.  One might expect 
the amplitude for this process to satisfy a simple factorization formula 
analogous to that for the decay of a $B$ meson into two light hadrons
\cite{Beneke:2000ry}.
In that factorization formula, the factor associated with $\chi_{cJ}$ 
is the light-front distribution amplitude for the charmonium state.
At leading order in the relative velocity of the charm quarks, 
this factor is proportional to $R_{\chi_c}'(0)$, the derivative 
of the radial wavefunction at the origin.
However, Chao, Song, and collaborators discovered infrared
divergences at next-to-leading order in $\alpha_s$ that are proportional
to $(m_c/m_b)^2$ \cite{Song:2002mh,Song:2003yc}. These infrared
divergences signal that the factorization formula holds, at best, only
up to corrections of order $(m_c/m_b)^2$.
Beneke and Vernazza studied the factorization of the decay amplitude 
in the limit $m_b,m_c \to \infty$ with $m_c/m_b$ fixed \cite{Beneke:2008pi}.
They demonstrated that all infrared divergences at next-to-leading order
in $\alpha_s$ can be absorbed into the matrix element of a color-octet
operator.  In the asymptotic limit in which $\chi_{cJ}$ is a
Coulombic charmonium state, the color-octet matrix element is 
proportional to $R_{\chi_c}'(0)$, but in general it is an independent
nonperturbative factor.
Bodwin, Tormo and Lee studied the factorization of the decay amplitude 
in the limit  $m_b \to \infty$ with $m_c$ fixed \cite{Bodwin:2010fi}.
They proved the factorization to all orders in $\alpha_s$ 
up to corrections suppressed by $m_c/m_b$.  
Their factorization formula has the same structure as that 
for decays into two light mesons.  The factors associated with the $\chi_{cJ}$ 
are matrix elements of local operators in nonrelativistic QCD.
The matrix element at leading order in the relative velocity 
of the charm quark is proportional to $R_{\chi_c}'(0)$.

There have been attempts to calculate the branching fraction for
$B^+ \to K^+ + \chi_{c1}'$.  Meng, Gao, and Chao used a naive factorization 
formula with an infrared cutoff on the divergences at order $\alpha_s$
and obtained the branching fraction $1.8 \times 10^{-4}$ \cite{Meng:2005er}.
They also applied their method to the decay into $K^+ + \chi_{c1}$
and obtained essentially the same branching fraction,
which is smaller by about a factor of 3 than the measured value.
Liu and Wang calculated the branching fraction for 
$B^+ \to K^+ + \chi_{c1}'$  \cite{Liu:2007uj} using a perturbative 
QCD method that incorporates Sudakov effects \cite{Li:1994cka}.
Their result was $(7.9^{+4.9}_{-3.8}) \times 10^{-4}$.
They did not calculate the corresponding result for $K^+ + \chi_{c1}$,
so the accuracy of their method can not be judged by comparing with data.

In the absence of any reliable calculations of the branching fraction 
for $B^+ \to K^+ + \chi_{c1}'$, we will use a simple phenomenological 
estimate.  We scale the measured branching fraction for 
$B^+ \to K^+ + \chi_{c1}$ by factors that take into account its 
dependence on the mass and the wavefunction of the charmonium state.
The dependence on the mass $M_{\chi_{c1}}$ is primarily through 
a multiplicative factor $\lambda^{3/2}(M_B,m_K,M_{\chi_{c1}})$ 
that comes from the phase-space integral 
and from the Lorentz structure of the decay amplitude. 
We assume that the dependence on the wave
function comes primarily from a multiplicative factor $|R'_{\chi_c}(0)|^2$,
where $R'_{\chi_c}(0)$ is the derivative 
of the radial wavefunction at the origin.  The factor $|R'_{\chi_c}(0)|^2$
for both the $2P$ and $1P$ multiplets has been calculated for 
four potential models in Ref.~\cite{Eichten:1995ch}.
The ratio of this factor for $2P$ and $1P$ ranges from 0.97 to 1.42.
We interpret this as a ratio 1.20 with a theoretical error
$\pm 0.23$ that can be added in quadrature with the experimental error
in the branching fraction for $B^+ \to K^+ + \chi_{c1}$.
The measured branching fraction for $B^+ \to K^+ + \chi_{c1}$
is $(4.9 \pm 0.5) \times 10^{-4}$.  Multiplying by our two scaling factors, 
our estimate for the branching fraction is
\begin{equation}
\textrm{Br}[B^+ \to K^+ + \chi_{c1}'] \approx 
\left[ (5.9 \pm 1.3) \times 10^{-4} \right] 
\frac{\lambda^{3/2}(M_B,m_K,M_{\chi_{c1}'})}
    {\lambda^{3/2}(M_B,m_K,M_{\chi_{c1}})} .
\label{BrBKchi}
\end{equation}
The last factor ranges from 0.53 if the $\chi_{c1}'$ is near
the $D^{*0} D^0$ threshold to 0.39 for $M_{\chi_{c1}'} = 4000$~MeV.

Our estimate for the branching fraction in Eq.~(\ref{BrBKchi})
is based only on information about charmonium states.  
Thus the appropriate limit of the Zero-Range+Resonance model in 
Sec.~\ref{sec:zero-range+resonance} is the Flatt\'e limit in which the
charm mesons scatter only through their coupling to the 
resonance.  In this limit, the partial width for 
$B^+ \to K^+ + \chi_{c1}'$ is
given by Eq.~(\ref{pw:noscat}), with the short-distance factor 
$\Gamma_B^K$ given by Eq.~(\ref{Gammaij}). 
The factor $\lambda^{3/2}(M_B,m_K,M_{*0}+M_0)$ in $\Gamma_B^K$
came from taking the limit $E \to 0$ in 
$\lambda^{3/2}(M_B,m_K,M_{*0}+M_0+E)$.  This factor matches the 
factor $\lambda^{3/2}(M_B,m_K,M_{\chi_{c1}'})$ in Eq.~(\ref{BrBKchi}).
Matching the other factors in Eqs.~(\ref{pw:noscat}) and (\ref{BrBKchi}),
we obtain the constraint
\begin{equation}
\left| C_{B^+}^{K^+,0} - C_{B^+}^{K^+,1}
+ C_{B^+}^{K^+,2} \right|^2  = (8.1 \pm 1.8) \times 10^{-13}.
\label{Ccons}
\end{equation}

\subsection{$\bm{B}$ meson decays into $\bm{K + D^{*0} \bar{D}^0}$}
\label{sec:BtoKDD}

The invariant mass distributions of charm meson pairs in the decays
$B \to K + (D^{*0} \bar D^0, D^0 \bar D^{*0})$ have been measured 
by the Babar and Belle Collaborations
from the $D^{*0} \bar D^0$ threshold near 3872~MeV up to 4000~MeV 
\cite{Aubert:2007rva,Belle:2008su}.
Measurements of the enhancement near the $D^{*0} \bar D^0$ threshold,
which is associated with the $X(3872)$,
provide the constraints on
$\textrm{Br}_+[D^0 \bar D^0 \pi^0,D^0 \bar D^0 \gamma]$ and 
$R_{0+}[D^0 \bar D^0 \pi^0,D^0 \bar D^0 \gamma]$ described in Section~\ref{sec:BtoKX}.  
The distributions from 3880~MeV to 4000~MeV provide additional
constraints.  In particular, they constrain the possibility of an 
additional resonance that could be associated with the $\chi_{c1}'$.

The Babar Collaboration used a data sample that consists of about 
$N_{B \bar B} = 3.83 \times 10^8$ $B \bar B$ pairs \cite{Aubert:2007rva}.
They combined both $B^+$ and $B^0$ decays
and the $D^0 \bar D^0 \pi^0$ and $D^0 \bar D^0 \gamma$ 
decay channels into single measurements in each energy bin.
Their efficiencies for $B^+$ and $B^0$ decays were
$\epsilon_+ = 4.3 \times 10^{-4}$ and $\epsilon_0 = 0.7 \times 10^{-4}$,
respectively.  The numbers of events were given in energy bins of 2.5~MeV.
The Belle Collaboration used a data sample that consists 
of about $N_{B \bar B} = 6.57 \times 10^8$ $B \bar B$ pairs \cite{Belle:2008su}.
They combined $B^+$ and $B^0$ decays into separate measurements 
for the $D^0 \bar D^0 \pi^0$ and $D^0 \bar D^0 \gamma$ 
decay channels in each energy bin.  From the efficiencies that were given, 
we infer that the efficiencies for $B^+$ and $B^0$ decays were
$\epsilon_+ = 3.93 \times 10^{-4}$ and $\epsilon_0 = 0.63 \times 10^{-4}$
in the $D^0 \bar D^0 \gamma$ channel and
$\epsilon_+ = 4.24 \times 10^{-4}$ and $\epsilon_0 = 0.69 \times 10^{-4}$
in the $D^0 \bar D^0 \pi^0$ channel.
The numbers of events were given in energy bins of 2~MeV.
In the Belle analysis, the energy resolution increased from about
0.5~MeV at 3880 MeV to about 1~MeV at 3900 MeV and then to about 
2~MeV at 4000~MeV.

\begin{figure}[t]
\centerline{\includegraphics*[width=15cm,angle=0,clip=true]{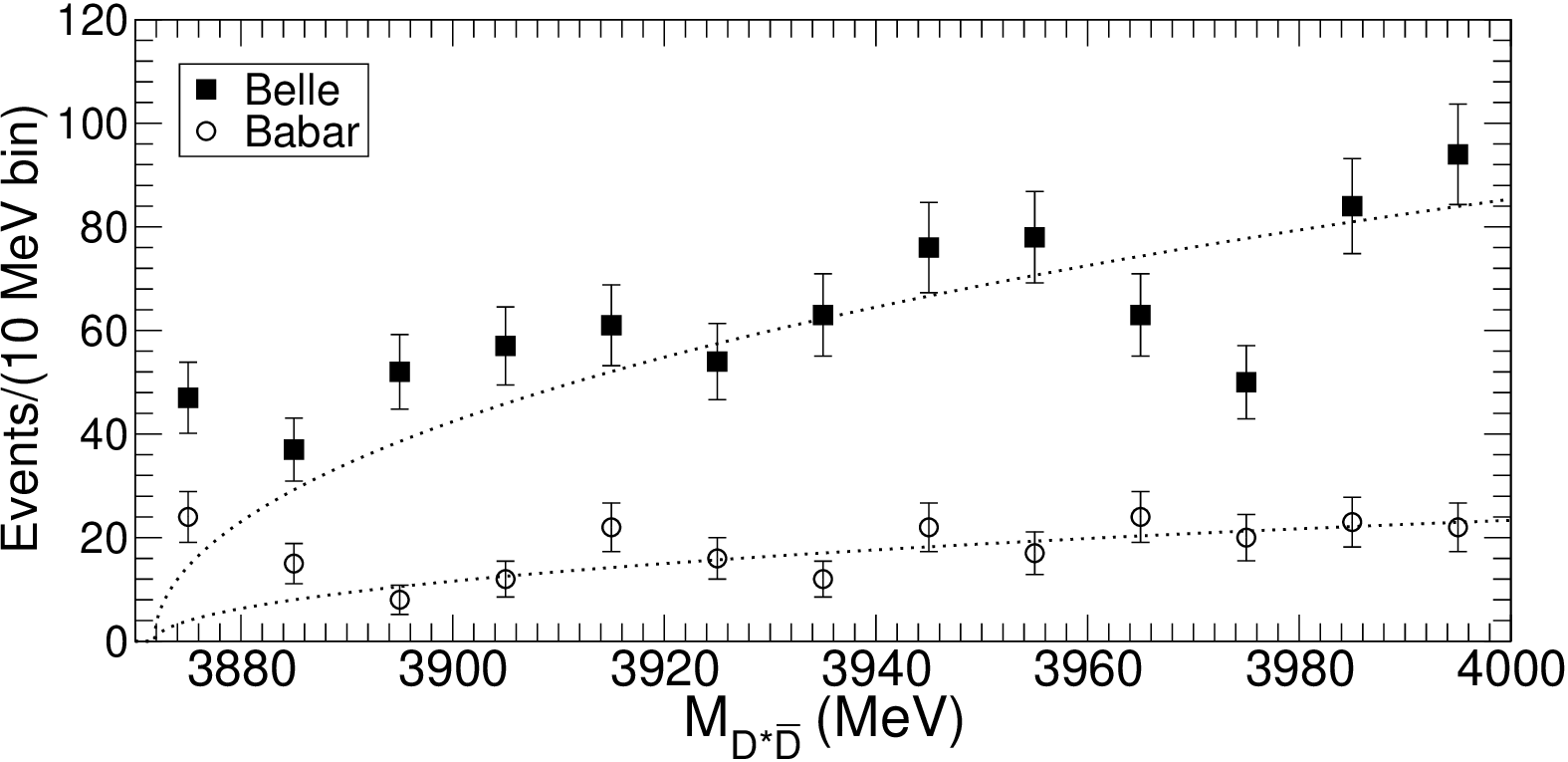}}
\vspace*{0.0cm}
\caption{$D^{*0} \bar D^0$ invariant mass distributions
from the decays $B \to K + (D^{*0} \bar D^0, D^{0} \bar D^{*0})$.
The data are the numbers of $(D^{*0} \bar D^0, D^{0} \bar D^{*0})$ events 
per 10~MeV bin observed by the Babar Collaboration \cite{Aubert:2007rva}
(open dots)
and by the Belle Collaboration \cite{Belle:2008su} (solid squares).
The dotted curves are the best fits to the data in the bins from
3900 MeV to 4000~MeV with no signal and a background
proportional to the phase-space volume.}
\label{fig:DDdata}
\end{figure}

The Babar and Belle measurements provide no evidence for a 
second resonance above 3880~MeV, but they also do not exclude it.  
To simplify our analysis, 
we combine adjacent bins into bins with 10~MeV width.
The minimum number of events in any bin is then 8.  This is large enough
that the experimental uncertainty in a bin with $N$ events 
can be approximated by $\sqrt{N}$.  The larger bins also decrease 
the sensitivity to the experimental energy resolution.
We add the number of Belle events in the $D^0 \bar D^0 \pi^0$ 
and $D^0 \bar D^0 \gamma$ channels to obtain the total number 
of $(D^0 \bar D^0 \pi^0,D^0 \bar D^0 \gamma)$ events in each energy bin.
The resulting Babar and Belle data sets are shown 
in Figure~\ref{fig:DDdata}.  The curves are the best fits to the data from
3900 MeV to 4000~MeV with no signal and with a background proportional to 
the phase-space volume.  The large excesses in the first bin 
are associated with the $X(3872)$.

The predicted number of $D^{*0} \bar D^0$ and $D^{0} \bar D^{*0}$
events in an energy bin
is the sum of the predicted numbers of signal events
$N_\textrm{sig}$ and background events $N_\textrm{bg}$.
We consider only the energy bins above 3880~MeV, 
so there is a negligible overlap with the $X(3872)$ signal region, 
which is taken into account through the constraints  on
$\textrm{Br}_+[D^0 \bar D^0 \pi^0,D^0 \bar D^0 \gamma]$ and 
$R_{0+}[D^0 \bar D^0 \pi^0,D^0 \bar D^0 \gamma]$ described in Section~\ref{sec:BtoKX}.
The predicted number of $(D^0 \bar D^0 \pi^0,D^0 \bar D^0 \gamma)$ signal events
from both $B^+$ decays and $B^0$ decays 
in an energy bin extending from $E_1$ to $E_2$ is
\begin{eqnarray}
N_\textrm{sig} &=& \frac{N_{B \bar B}}{\Gamma[B^+]}
\int_{E_1}^{E_2} dE~
\bigg( \epsilon_+~\frac{d \Gamma}{dE}[B^+ \to K^+ + (D^0 \bar D^0 \pi^0,D^0 \bar D^0 \gamma)]
\nonumber
\\ 
&& \hspace{3cm}
+ \epsilon_0~\frac{d \Gamma}{dE}[B^0 \to K^0 + (D^0 \bar D^0 \pi^0,D^0 \bar D^0 \gamma)]
\bigg).
\label{Nsig-DstarD}
\end{eqnarray}
We take the background to be incoherent and proportional to the 
phase-space volumes for the two successive 2-body decays 
$B \to K + (D^* \bar D)$ and $(D^* \bar D) \to D^{*0} + \bar D^0$.  
The predicted number of background events in an energy bin is
\begin{eqnarray}
N_\textrm{bg} = 
C_\textrm{bg} \int_{E_1}^{E_2} dE~\lambda^{1/2}(M_B,M_K,M_{*0} + M_0 + E) ~
\lambda^{1/2}(M_{*0} + M_0 + E,M_{*0},M_0),
\label{Nbg-DstarD}
\end{eqnarray}
where $C_\textrm{bg}$ is an adjustable constant that is different 
for the Babar data and for the Belle data.

The inclusive energy distribution $d \Gamma/dE$ for $B^+$ decay
is given by the factorization formula in Eq.~(\ref{lsh:gen}).
In the Zero-Range+Resonance model, the sums are over $i,j \in \{ 0,1,2 \}$.
The corresponding energy distribution for $B^0$ decay can be obtained by the 
interchange $C_{B^+}^{K^+,0} \leftrightarrow C_{B^+}^{K^+,1}$.
The energy-dependent functions Im$F_{ij}(E)$ are expressed as linear 
combinations of the imaginary parts of $\kappa(E)$, $\kappa_1(E)$,
$\gamma_0$, $\gamma_1$, and $\nu$ in Eqs.~(\ref{Imfijchi}), 
(\ref{Im_FiRRi}), and (\ref{Im_FRR}).
The energy distribution for $(D^0 \bar D^0 \pi^0,D^0 \bar D^0 \gamma)$ is the sum of 
the terms with the factor Im$\kappa(E)$.
The Zero-Range+Resonance model should give an accurate description 
of this energy distribution in the $D^* \bar D$ threshold region.
We will use this model all the way up to 4000~MeV, 
which is 120~MeV above the $D^{*+} D^-$ threshold.
At energies well above the $D^{*+} D^-$ threshold, 
the resonance term in the amplitude may still be accurate, but the
zero-range approximation to the non-resonance term becomes inadequate.
Thus in the high energy region, our model for the non-resonance 
amplitude can at best be regarded as illustrative.

\section{Numerical Analysis}
\label{sec:analysis}

In this section, we analyze the available data to determine whether 
they are able to discriminate between binding mechanisms for the
$X(3872)$.  Our strategy is to fix the interaction parameters
of the Zero-Range+Resonance model
and then to vary the $B \to K$ transition coefficients 
to obtain the best possible fit to all the constraints.
We repeat this procedure for various values of the interaction parameters 
to see if there are any parameters that are preferred by the constraints.

Given a set of interaction parameters, we determine 
the $B \to K$ transition coefficients by minimizing the
$\chi^2$ associated with the following constraints:
\begin{itemize}
\item
the experimental value for $R_{0+}[J/\psi \, \pi^+ \pi^-]$  
in Eq.~(\ref{R0+Jpsi:data}), which is set equal to
the theoretical result in Eq.~(\ref{R0+:Jpsi}).
\item
the experimental value for $R_{0+}[D^0 \bar D^0 \pi^0,D^0 \bar D^0 \gamma]$
in Eq.~(\ref{R0+DD:data}), which is set equal to
the theoretical result in Eq.~(\ref{R0+DD}).
\item
the experimental value for $\textrm{Br}_+[D^0 \bar D^0 \pi^0,D^0 \bar D^0 \gamma]$
in Eq.~(\ref{Br+DD:data}), which is set equal to
the theoretical result in Eq.~(\ref{B+DD}).
\item
the constraint in Eq.~(\ref{Ccons}), which follows from 
our phenomenological estimate for the branching fraction for
$B^+ \to K^+ + \chi_{c1}'$. 
\item
the numbers of $B \to K + (D^{*0} \bar D^0, D^0 \bar D^{*0})$ 
events observed by the Babar and Belle collaborations
in each of the 12 energy bins covering the range 3880 to 4000~MeV.
The two data sets are shown in Figure~\ref{fig:DDdata}. 
The predicted number of events in an energy bin is the sum of 
the signal given by Eq.~(\ref{Nsig-DstarD}) 
and the background given by Eq.~(\ref{Nbg-DstarD}).
\end{itemize}
The total number of constraints is 28.
The complex constants $C_{B^+}^{K^+,i}$ are determined up to an
overall phase by 5 real parameters.
The numbers of background events in the Babar and Belle data
are determined by 2 additional parameters.
These 7 parameters are then varied to minimize the $\chi^2$ 
associated with the 28 constraints.

In the Zero-Range+Resonance model,
the independent interaction parameters are $\gamma$, $\gamma_1$,
$g$, and $\nu$, with $\gamma_0$ given by Eq.~(\ref{eq:gamma0}).
The allowed region for the real and imaginary parts of $\gamma$ are given
in Eqs.~(\ref{gamma-re}) and (\ref{gamma-re*im}).
Our constraints on the energy distributions are insensitive 
to the value of $\gamma$, because they either involve integrals 
over the $D^* \bar D$ threshold region or they involve only higher energies.
We therefore set the real part of $\gamma$ to its central value
in Eq.~(\ref{gamma-re}) and its imaginary part to zero: $\gamma = 28$~MeV.
The resonance coupling constant was determined in 
Eq.~(\ref{g-ave}) to be $g = 0.40$.
The undetermined interaction parameters are the isovector inverse scattering 
length $\gamma_1$ and the resonance parameter $\nu$.
Our constraints are insensitive to the imaginary parts of 
$\nu$ and $\gamma_1$, so we set them to 0.
We expect the isovector inverse scattering length $\gamma_1$ 
to be large compared to the momentum scale $\kappa_{11} = 125$~MeV
associated with isospin splittings.  
We will therefore consider values in the region 
$|\gamma_1| > 2\kappa_{11} = 250$~MeV.
We consider values of $\nu$ in the range
$-10~\textrm{MeV} < \textrm{Re} \nu < 140~\textrm{MeV}$.
For $\nu < -10$~MeV, there would be a second resonance below 
the $D^{*0} \bar D^0$ threshold in addition to the $X(3872)$,
which is not observed.  For $\nu > 140$~MeV, the second resonance 
would be above 4000~MeV, beyond the region covered by the Belle and Babar data.

\begin{figure}[t]
\centerline{\includegraphics*[height=8cm,angle=0,clip=true]{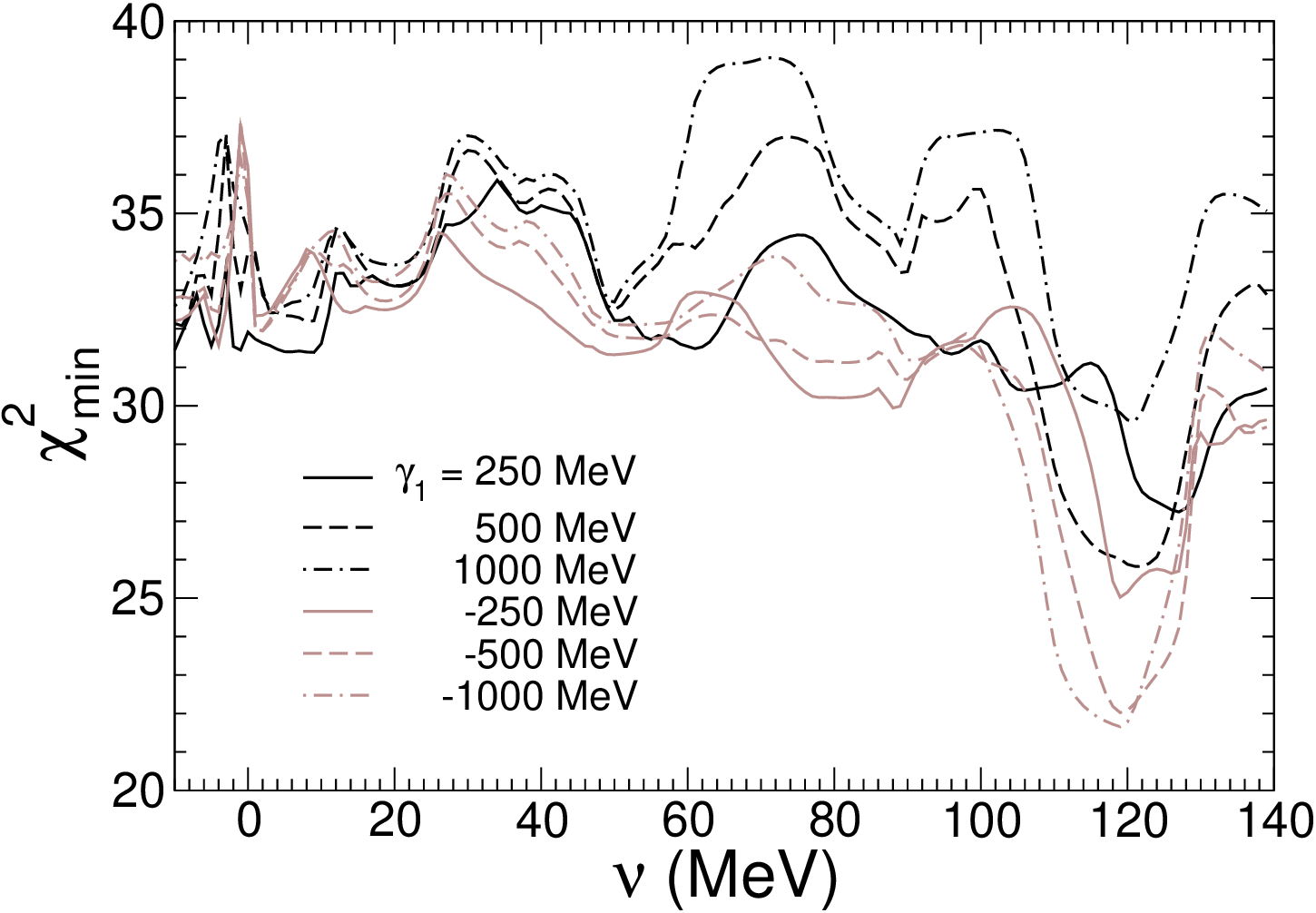}}
\vspace*{0.0cm}
\caption{The minimum of $\chi^2$ as a function of $\nu$
for $\gamma = 28$~MeV, $g = 0.4$, and various values of $\gamma_1$.}
\label{fig:chisq-gamma1}
\end{figure}

We have performed fits of the 7 parameters to the 28 constraints
for $\gamma = 28$~MeV, $g = 0.4$, and different values of $\nu$ and $\gamma_1$.
The parameter $\nu$ has been varied from $-10$ MeV up to $140$ MeV in
steps of 1~MeV. Since our results are less sensitive to $\gamma_1$, 
we have repeated the fit only for a few discrete values
of $\gamma_1$: $\pm 250$, $\pm 500$ and $\pm 1000$ MeV. 
The minimum values of $\chi^2$ for the various values of $\gamma_1$
are shown in Figure~\ref{fig:chisq-gamma1} as a function of $\nu$. 
For $\nu < 100$~MeV, the minimum $\chi^2$ 
lies in the range $30  < \chi_\textrm{min}^2 < 39$ for all values of $\gamma_1$, 
indicating that the quality of the fit is rather insensitive to 
$\gamma_1$ and $\nu$.  The minimum value of $\chi^2_\textrm{min}=21.7$ 
is reached at $\nu=119$ MeV and $\gamma_1=-1000$ MeV.
For these values of $\nu$ and $\gamma_1$, we verified that the fit 
and the value of $\chi^2_\textrm{min}$ are insensitive to variations in the 
real and imaginary parts of $\gamma$.
For $\nu > 130$~MeV, the minimum $\chi^2$ lies in the range 
$29  < \chi_\textrm{min}^2 < 36$, 
indicating again that the quality of the fit is rather insensitive 
to $\gamma_1$.

\begin{figure}[t]
\centerline{\includegraphics*[width=15cm,angle=0,clip=true]{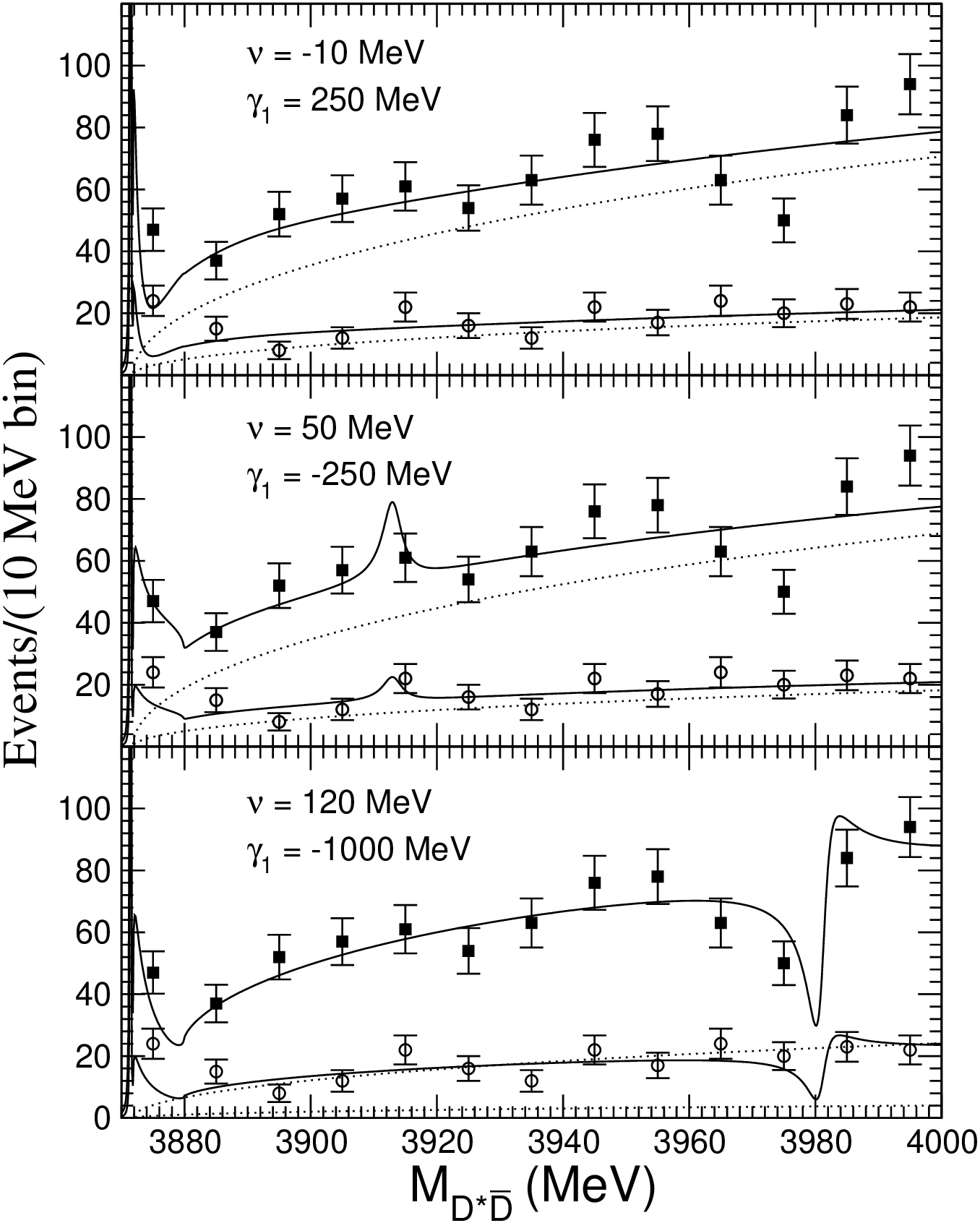}}
\vspace*{0.0cm}
\caption{
Numbers of $B \rightarrow K + (D^{*0} \bar D^0, D^{0} \bar D^{*0})$ events 
as functions of the invariant mass of the charm meson pair.
The data are the numbers of events observed 
by the Babar Collaboration \cite{Aubert:2007rva} (open circles)
and by the Belle Collaboration \cite{Belle:2008su} (solid squares) 
in 10~MeV bins. The predicted $(D^0 \bar D^0 \pi^0,D^0 \bar D^0 \gamma)$ 
distributions for Belle (upper solid lines) and for Babar (lower solid lines) 
are shown for $\gamma=28$~MeV, $g = 0.4$, 
and three sets of values of $(\nu,\gamma_1)$.
The background contributions are shown as dotted lines.
}
\label{fig:LineShape-glob}
\end{figure}

In Figure~\ref{fig:LineShape-glob}, we show the $D^{*0} \bar D^0$
invariant mass distributions for $\gamma=28$~MeV, $g = 0.4$, 
and three pairs of parameters $(\nu,\gamma_1)$
that are approximate local minima of $\chi^2$:
$(-10,250)$, $(50,-250)$, and $(120,-1000)$~MeV.
The third pair of parameters is close to the global minimum of $\chi^2$.
The predicted numbers of $D^{*0} \bar D^0$ events 
in a 10~MeV bin observed by the Belle (Babar) experiment
can be obtained by averaging the upper (lower) solid line over the bin.
The background contributions are shown as dotted lines.
For the first set of parameters, the $X(3872)$ arises from a fine-tuning 
of the energy of the $\chi_{c1}'$.
For the second and third sets of parameters, the $X(3872)$ arises from a fine-tuning 
of the strength of the interactions between $D^{*0}$ and $\bar D^0$.
For all three pairs of parameters, there is an $X(3872)$ resonance
below the $D^{*0} \bar D^0$ threshold and a $D^{*0} \bar D^0$ threshold
enhancement above the threshold.
For the first set of parameters, there are no additional peaks in 
the energy distribution.
For the second set of parameters, there is
an additional peak near 3913~MeV corresponding to the $\chi_{c1}'$ resonance.
For the third set of parameters, there is a peak near 3983~MeV
corresponding to the $\chi_{c1}'$ resonance, but it is preceded by a sharp decrease 
in the number of events down to near the background level.
This surprising behavior arises from destructive interference between 
the resonant and nonresonant amplitudes for producing charm meson pairs.

\begin{figure}[t]
\centerline{\includegraphics*[width=15cm,angle=0,clip=true]{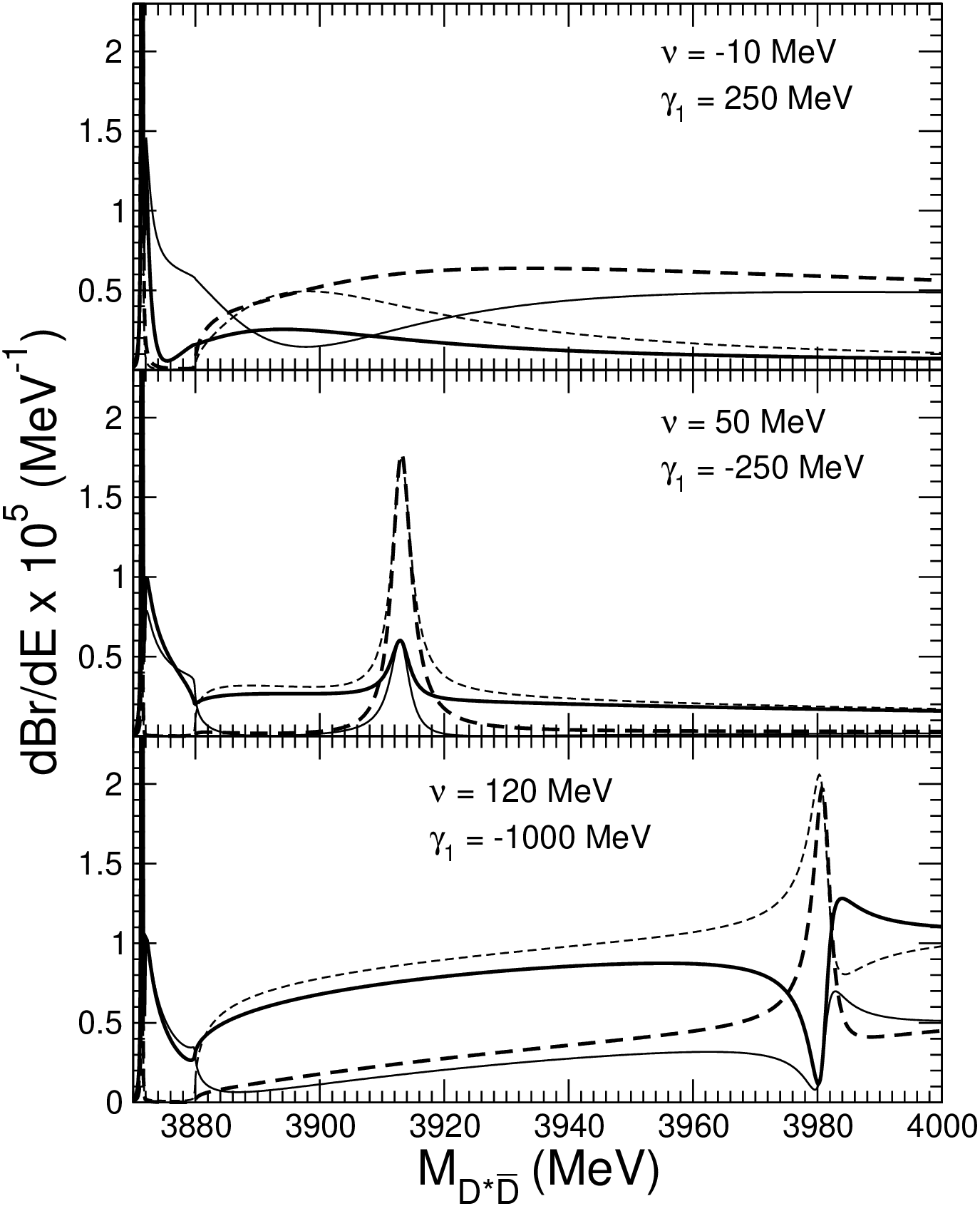}}
\vspace*{0.0cm}
\caption{
Energy distributions for the $(D^0 \bar D^0 \pi^0,D^0 \bar D^0 \gamma)$ signal 
(solid lines) and for the 
$(D^+ \bar D^0 \pi^-,D^0 D^- \pi^+,D^+ D^- \pi^0,D^+ D^- \gamma)$ signal 
(dashed lines) 
from both $B^+$ decay (thicker lines) and $B^0$ decays (thinner lines)
for $\gamma=28$~MeV, $g = 0.4$, 
and three sets of values of $(\nu,\gamma_1)$.
}
\label{fig:Signal-glob}
\end{figure}
 
In Figure~\ref{fig:Signal-glob}, 
we show the predicted energy distributions 
for the $(D^0 \bar D^0 \pi^0,D^0 \bar D^0 \gamma)$ signal and for the 
$(D^+ \bar D^0 \pi^-,D^0 D^- \pi^+,D^+ D^- \pi^0,D^+ D^- \gamma)$ signal
from both $B^+$ decay and $B^0$ decay for the same three sets of parameters 
as in Figure~\ref{fig:LineShape-glob}.
The $(D^+ \bar D^0 \pi^-,D^0 D^- \pi^+,D^+ D^- \pi^0,D^+ D^- \gamma)$ signal
is the sum of 
the terms in the factorization formula with a factor of Im$\kappa_1(E)$.
The two signals correspond to $(D^{*0} \bar D^0,D^0 \bar D^{*0})$ 
and to $(D^{*+} D^-,D^+ D^{*-})$ above the appropriate thresholds.
For all three sets of parameters, the energy distribution 
for $(D^+ \bar D^0 \pi^-,D^0 D^- \pi^+,D^+ D^- \pi^0,D^+ D^- \gamma)$
has an $X(3872)$ resonance below the $D^{*0} \bar D^0$ threshold, but there is 
no $D^{*0} \bar D^0$ threshold enhancement just above the threshold. 
The resonance should be interpreted as 
a contribution from $D^+ D^- \gamma$, because it is below the thresholds
for $D^+ \bar D^0 \pi^-$ and $D^+ D^- \pi^0$.
A quantitative treatment of this contribution would require using the 
energy-dependent width for the $D^{*+}$ defined in Ref.~\cite{Braaten:2007dw}.
For the second set of parameters, there is a peak at the $\chi_{c1}'$ 
resonance near 3913 MeV in both $D^{*0} \bar D^0$ and $D^{*+} D^-$,
but the peak is higher in $D^{*+} D^-$.
For the third set of parameters, there is constructive interference 
in $D^{*+} D^-$ at the $\chi_{c1}'$ resonance near 3980 MeV in contrast to 
the destructive  interference in $D^{*0} \bar D^0$.

\begin{figure}[t]
\centerline{\includegraphics*[width=15cm,angle=0,clip=true]{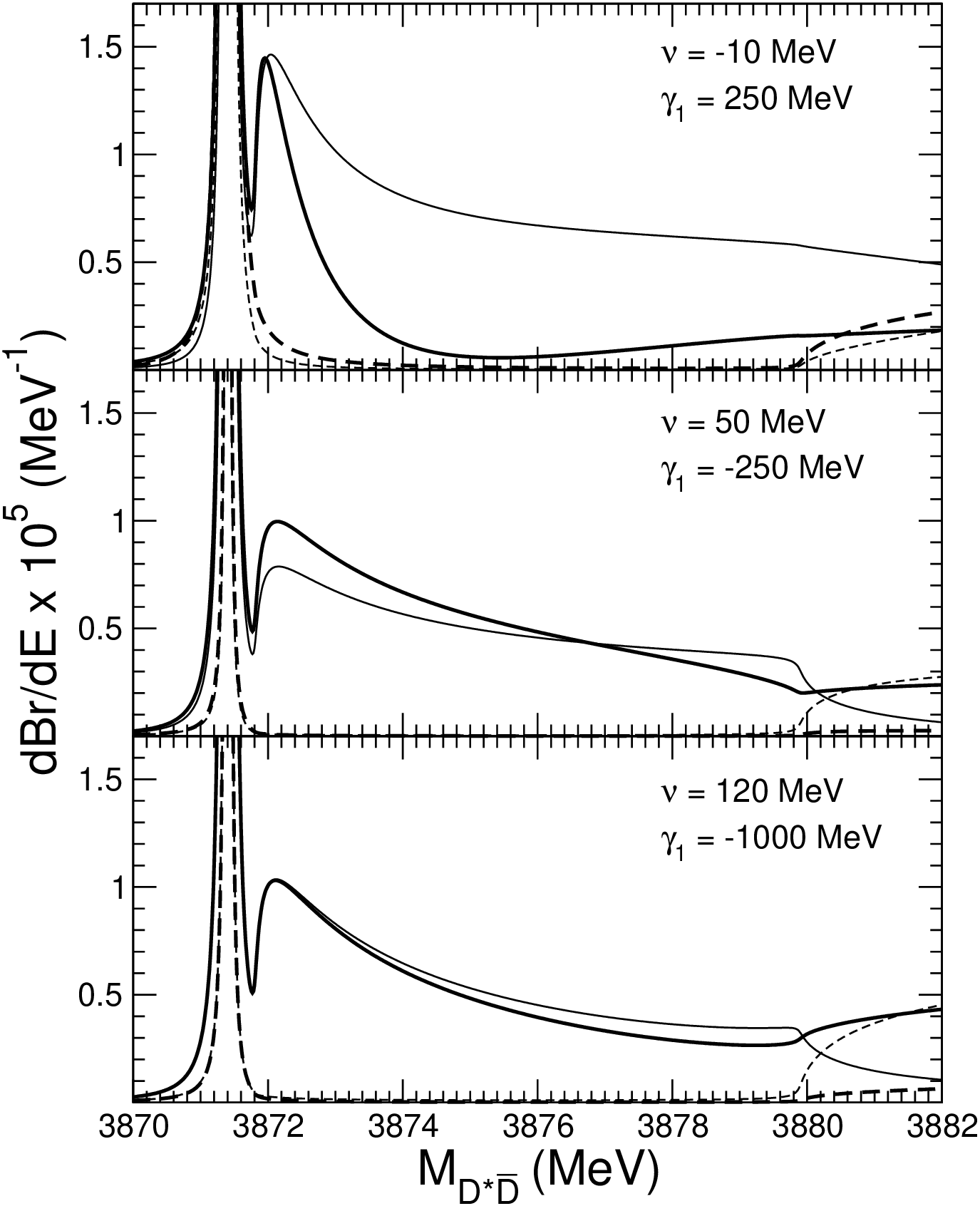}}
\vspace*{0.0cm}
\caption{
Line shapes in the $D^* \bar D$ threshold region
for the $(D^0 \bar D^0 \pi^0,D^0 \bar D^0 \gamma)$ signal 
(solid lines) and for the 
$(D^+ \bar D^0 \pi^-,D^0 D^- \pi^+,D^+ D^- \pi^0,D^+ D^- \gamma)$ signal 
(dashed lines) 
from both $B^+$ decay (thicker lines) and $B^0$ decays (thinner lines)
for $\gamma=28$~MeV, $g = 0.4$, 
and three sets of values of $(\nu,\gamma_1)$.
}
\label{fig:Signal-thresh}
\end{figure}

In Figure~\ref{fig:Signal-thresh}, we show the behavior of the line shapes 
in Figure~\ref{fig:Signal-glob} in the $D^* \bar D$ threshold region.
At the $X(3872)$ resonance, the line shapes in
$(D^0 \bar D^0 \pi^0,D^0 \bar D^0 \gamma)$ and in $D^+ D^- \gamma$
coming from $B^+$ decay and $B^0$ decay differ only in their normalizations.
The line shapes in $(D^0 \bar D^0 \pi^0,D^0 \bar D^0 \gamma)$
have a $D^{*0} \bar D^0$  threshold enhancement whose peak is about as far 
above the $D^{*0} \bar D^0$  threshold as the $X(3872)$ resonance is below 
the threshold.  
The two maxima are well-separated, because we have set Im$\gamma$ = 0.
If Im$\gamma$ is large enough, the resonance and the threshold enhancement
will be smeared into a line shape with a single maximum.
The lines shapes from $B^+$ decay and $B^0$ decay
can be quite different in the region up to $D^{*+} D^-$ threshold,
and they can change dramatically at that threshold.
The energy distributions in
$(D^+ \bar D^0 \pi^-,D^0 D^- \pi^+,D^+ D^- \pi^0,D^+ D^- \gamma)$ 
are close to zero between the $D^{*0} \bar D^0$ and $D^{*+} D^-$ thresholds,
but they can increase dramatically above the $D^{*+} D^-$ threshold.

\section{Summary}
\label{sec:summary}

We have derived general line shapes for the $X(3872)$ that can be used 
to discriminate between alternative binding mechanisms.
In the $D^{*0} \bar D^0$ threshold region,
which extends only to a few MeV from the threshold,
the line shapes are universal in the sense that they are determined 
only by the binding energy and width of the $X(3872)$ \cite{Braaten:2007ft}.
In the $D^{*} \bar D$ threshold region, which extends to tens of MeV from the
threshold, the line shapes depend on the binding mechanism.
The mechanism could be the fine-tuning of the 
interaction between the charm mesons to near the critical strength for  
a bound state, in which case the appropriate 
line shapes are the zero-range line shapes of 
Ref.~\cite{Braaten:2007ft}.
The mechanism could also be the fine-tuning of the energy of a resonance 
to near the $D^{*0} \bar D^0$ threshold, in which case the appropriate  
line shapes are the Flatt\'e line shapes.
The Flatt\'e line shapes are essentially those introduced in
Ref.~\cite{Hanhart:2007yq}, except that one cannot ignore the 
essential contribution to the line shape in the 
$(D^0 \bar D^0 \pi^0,D^0 \bar D^0 \gamma)$ channel 
from the $X(3872)$ resonance peak below the threshold.
Our general line shapes include the zero-range line shapes and the 
Flatt\'e line shapes as special cases.

Our general line shapes take into account scattering between 
two coupled channels with the same reduced mass $\mu$.
This is a good approximation if the masses of the pairs of 
particles in the two channels differ only by isospin splittings.
These line shapes could be applied to any weakly-bound 
hadronic molecule that has S-wave couplings to a pair of thresholds
separated by isospin splittings.
However the complications associated with the coupled channels 
are only relevant if the binding energy 
of the molecule and the widths of the constituents are all much smaller 
than the isospin splittings, which are typically less than 10~MeV. 
The $X(3872)$ may be the unique hadron that satisfies this requirements.
If these conditions are not satisfied, one might as well ignore 
the complications associated with the isospin splittings.
The zero-range approximations for the direct interaction between 
the pair of mesons may also be inadequate in this case.

A crucial ingredient in our general line shapes are the coupled-channel 
scattering amplitudes for the neutral and charged charm meson channels
$(D^{*0} \bar D^0)_+$ and $(D^{*+} D^-)_+$,
which are given in Eqs.~(\ref{fijchi-E}).
The resonance propagator is given in Eq.~(\ref{prop:resI}).
They depend on 4 interaction parameters: $\gamma_0$, $\gamma_1$, $\nu$,
and $g$.  The scattering amplitudes satisfy the constraints 
of unitarity exactly if these parameters are real.
The analytic continuation of the parameters to complex values can be used 
to take into account some of the effects of states that are not treated 
explicitly.  The line shapes in the universal region depend 
only on the inverse scattering length $\gamma$, which satisfies 
Eq.~(\ref{eq:gamZRR}).  It is therefore convenient to use
Eq.~(\ref{eq:gamma0}) to eliminate $\gamma_0$ in favor of $\gamma$.
The real and imaginary parts of $\gamma$ are constrained by measurements 
of the binding energy and width of the $X(3872)$.
The constraints are given in Eqs.~(\ref{gamma-re}) and (\ref{gamma-re*im}).
We expect the isovector inverse scattering length $\gamma_1$ to
be much larger than $\kappa_1 \approx 125$~MeV, but we do not have 
any useful quantitative constraints on this parameter.
If the resonance is identified with the P-wave charmonium state 
$\chi_{c1}' \equiv \chi_{c1}(2P)$, charmonium phenomenology can be used to constrain
the resonance parameters $\nu$ and $g$.  The coupling constant $g$
is determined from phenomenological models of the decays 
$\chi_{c1}' \to D^* \bar D$ to have the value $g= 0.4$
given in Eq.~(\ref{g-ave}).  The variations among potential models 
are sufficiently large that they do not provide any useful constraints 
on the energy parameter $\nu$.

The general line shapes for a weakly-bound hadronic molecule that is 
produced by a short-distance process can be expressed in terms of 
a factorization formula for the inclusive energy distribution.
In the case of the $X(3872)$, an example of a short-distance process 
is a $B \to K$ transition with momentum transfer of about 1100~MeV,
which can produce pairs of charm mesons in the $D^* \bar D$ threshold region.
The factorization formula for the inclusive energy distribution 
for states with quantum numbers $1^{++}$
that are produced by the $B^+ \to K^+$ transition is given in 
Eq.~(\ref{lsh:gen}).  All the dependence on the energy $E$ is in 
the long-distance factors Im$F_{ij}(E)$.
The short-distance coefficients $C_{B^+}^{K^+,i}$ are complex constants.
The factorization formula for another short-distance process 
differs only in the values of the short-distance coefficients.
The short-distance coefficients for the $B^0 \to K^0$ transition
are determined in terms of those for the $B^+ \to K^+$ transition
by the isospin-symmetry relations in Eqs.~(\ref{CBK:isospin}) and
(\ref{C3:isospin}).  These short-distance coefficients
are constrained by measurements of $B \to K + X(3872)$,
such as those in Eqs.~(\ref{R0+JpsiDD:data}) and (\ref{Br+DD:data}).
The additional constraint in Eq.~(\ref{Ccons}) was obtained
from a phenomenological estimate of the decay rate for 
$B \to K + \chi_{c1}'$.

In the factorization formula for the inclusive energy distribution
in Eq.~(\ref{lsh:gen}), the long-distance functions Im$F_{ij}(E)$
can be resolved
into contributions from different decay channels of the $X(3872)$
by expressing them in forms consistent with the Cutkosky cutting rules.
The functions Im$F_{ij}(E)$ are expressed as linear combinations of 
the imaginary parts of the functions $\kappa(E)$ and $\kappa_1(E)$
and the imaginary parts of the parameters $\gamma_0$, $\gamma_1$, 
and $\nu$ in Eqs.~(\ref{Imfijchi}), (\ref{Im_FiRRi}), and (\ref{Im_FRR}).
The terms proportional to Im$\kappa(E)$ can be interpreted as 
the contributions from $(D^0 \bar D^0 \pi^0,D^0 \bar D^0 \gamma)$.
The terms proportional to Im$\kappa_1(E)$ can be interpreted as 
the contributions from
$(D^+ \bar D^0 \pi^-,D^0 D^- \pi^+,D^+ D^- \pi^0,D^+ D^- \gamma)$.
The terms proportional to Im$\nu$, Im$\gamma_0$, and Im$\gamma_1$
can be interpreted as the contributions from other
decay modes of $\chi_{c1}'$, other channels with isospin 0, 
and other channels with isospin 1.

In Section~\ref{sec:analysis}, we used our line shapes to carry out a
phenomenological analysis of the data on $B \to K + X(3872)$ and data 
from the Belle and Babar Collaborations on $B \to K + D^{*0} \bar D^0$,
with the $D^{*0} \bar D^0$ invariant mass ranging up to 4000~MeV.
We assumed that the binding mechanism for the $X(3872)$ is either a
fine-tuning of the strength of the interaction between the charm mesons
or a fine-tuning of the energy of the P-wave charmonium state 
$\chi_{c1}'$.  We extrapolated our line shapes for the $D^* \bar D$ 
threshold region all the way up to 128~MeV above the  
$D^{*0} \bar D^0$ threshold. In the high energy region,
the nonresonant contributions to the production amplitudes,
which were derived using a zero-range approximation,
can at best be regarded as an illustrative model.
We found that the two mechanisms for the binding of the $X(3872)$ are 
both compatible with the data for $B \to K + D^{*0} \bar D^0$ 
and our other constraints.
One way to exclude the tuning of the $\chi_{c1}'$ energy as a 
binding mechanism for the $X(3872)$ is to observe the
$\chi_{c1}'$ resonance as a separate peak in the invariant mass distribution 
for $D^{*0} \bar D^0$ and $D^{*+} D^-$. In the Belle and Babar data,
there is no obvious peak in the $D^{*0} \bar D^0$ 
invariant mass distribution between 3880~MeV and 4000~MeV.
However, the width of this peak is determined by the resonance 
parameters $\nu$ and $g$ and its height is determined by the
short-distance coefficients $C_{B^+}^{K^+,i}$.
Our calculations show that a $\chi_{c1}'$ resonance in this region 
is compatible with the data for $B \to K + D^{*0} \bar D^0$ 
and with our constraints on the short-distance coefficients.
An interesting possibility for this $\chi_{c1}'$ resonance is 
to have destructive interference between the resonant and nonresonant
amplitudes in the $D^{*0} \bar D^0$ decay channel and
constructive interference in the $D^{*+} D^-$ decay channel.
This possibility is actually realized for the parameters that 
minimize the $\chi^2$ for our constraints, 
which give a $\chi_{c1}'$ resonance near 3980~MeV.

An alternative way to exclude the $\chi_{c1}'$ resonance
mechanism for the binding of the $X(3872)$ is to calculate
the $\chi_{c1}'$ mass using lattice gauge
theory.  The masses of excited charmonium states and states in the 
$c \bar c$ meson spectrum with exotic quantum numbers 
have been calculated by Dudek et al.\ using lattice gauge theory 
without dynamical light quarks \cite{Dudek:2007wv}.
The masses of the $2P$ charmonium multiplet are significantly higher 
than those of the $1D$ multiplet.  Their analysis suggests that the 
$2^{++}$ $c \bar c$ meson discovered near 3930~MeV is more likely 
to be the $^3D_2$ ground state than the first radial excitation of
the $^3P_2$.  This suggests that the $2P$ multiplet, including 
$\chi_{c1}'$, has higher mass.  If these results are confirmed by
lattice calculations with dynamical light quarks, it would exclude 
the  $\chi_{c1}'$ resonance mechanism.
Other resonance mechanisms, such as the tuning of the energy of 
a $1^{++}$ tetraquark $c \bar c$ meson, are not easily constrained by
phenomenology, but they can also be ultimately ruled out 
using lattice QCD calculations.

We used charmonium phenomenology to constrain the resonance 
parameters $g$ and $\nu$.  Phenomenological estimates 
for the scattering parameters $\gamma_0$ and $\gamma_1$ 
could be obtained from meson potential models, such as those in 
Refs.~\cite{Tornqvist:1993ng,Tornqvist:2004qy,%
Thomas:2008ja,Liu:2008fh,Liu:2008tn,Swanson:2004pp}, or
from meson scattering models, such as the one in Ref.~\cite{Gamermann:2009fv}.
Thus far, these models have been used primarily to calculate binding energies.
They could also be used to calculate scattering variables, 
such as $\gamma_0$ and $\gamma_1$.
A particularly convenient pair of scattering variables are
the residues $Z_0$ and $Z_1$ of the poles in the elastic scattering
amplitudes $f_{00}(E)$ for $D^{*0} \bar D^0$ 
and $f_{11}(E)$ for $D^{*+} D^-$ at the $X(3872)$ resonance.
For the Zero-Range+Resonance model,
the ratio $Z_{1}^{1/2}/Z_{0}^{1/2}$ is given in Eq.~(\ref{Z1/Z0:ZRR}).
It does not depend on the resonance parameters $g$ and $\nu$, so  
it is determined primarily by $\gamma_1$.

Our general line shapes could be used by experimentalists to carry out 
a global analysis of the energy distributions for the decay modes 
of the $X(3872)$ that is not biased towards a specific binding mechanism.
One complication is the number of independent parameters in the 
line shapes.  The line shapes in 
$(D^0 \bar D^0 \pi^0,D^0 \bar D^0 \gamma)$ depend essentially on 5 
interaction parameters:  the real parameters $\gamma_1$, $g$, and $\nu$
and the complex parameter $\gamma$.  For other decay channels,
there is also a normalization parameter, such as
$(\textrm{Im} \gamma_1)^{J/\psi \, \pi^+ \pi^-}$ for $J/\psi \, \pi^+ \pi^-$
and $(\textrm{Im} \gamma_0)^{J/\psi \, \pi^+ \pi^- \pi^0}$ 
for $J/\psi \, \pi^+ \pi^- \pi^0$.
The imaginary parts of $\gamma_0$, $\gamma_1$, and $\nu$ give contributions
to Im$\gamma$, as indicated by Eq.~(\ref{Imgamma-ZR+R}),
but their effects should otherwise be negligible in the
$D^* \bar D$ threshold region. 
In addition to the interaction parameters, there are 5 real parameters
associated with the short-distance 
coefficients for the production of the charm meson pairs 
and the resonance.  In an analysis of the line shapes of the $X(3872)$ 
produced by $B$ decays, the determination of the $5+$ interaction parameters 
and the 5 short-distance parameters would require analyzing several
decay channels in both $B^+$ and $B^0$ decays.

An alternative strategy would be to carry out two separate 
analyses assuming either the dynamical mechanism or the resonance 
mechanism for the binding of the $X(3872)$.
With the dynamical mechanism, one could use the zero-range line shapes,
which have $3+$ interaction parameters and 3 short-distance parameters.
With the resonance mechanism, one could use the Flatt\'e line shapes,
which have $3+$ interaction parameters and 1 short-distance parameter.
If one set of line shapes gives a significantly better global fit
to the data, it would be evidence in favor of the 
corresponding binding mechanism for the $X(3872)$.

\begin{acknowledgments}
This research was supported in part by the Department of Energy
under grant DE-FG02-91-ER40690 and by a joint grant from the 
Army Research Office and the Air Force Office of Scientific Research.
One of us (E.B.) would like to thank  C.~Hanhart for valuable discussions.

\end{acknowledgments}

\section*{Appendix: Renormalization of the Zero-Range+Resonance Model}

The Zero-Range+Resonance model considered
in Sections \ref{sec:zero-range+resonance} and \ref{sec:LSzero-range+resonance}
can be derived from a renormalizable quantum field theory with local interactions.
In this Appendix, we consider the renormalization of the interactions 
of this field theory and the renormalization of the local operators 
that describe production by a short-distance process.

\subsection{Renormalization of the Interactions}

The Zero-Range+Resonance model can be represented by a nonrelativistic field theory.
The fields are scalar fields for the 
spin-0 charm mesons $D^0$, $\bar D^0$, $D^+$, and $D^-$, vector fields for the 
spin-1 charm mesons $D^{*0}$, $\bar D^{*0}$, $D^{*+}$, and $D^{*-}$, 
and a vector field for the resonance $\chi$.
The vector fields, such as $\chi^m$, have a Cartesian vector index $m$.
The interaction terms in the Hamiltonian density are
\begin{eqnarray}
{\cal H}_{\textrm{int}} = 
\frac{2 \pi}{\mu} (D^* \bar{D})^{m\dagger} \Lambda_0(D^* \bar{D})^m
+\sqrt{\frac{2 \pi}{\mu}}
\left[ (D^* \bar{D})^{m \dagger} G_0 \chi^{m}
+ \chi^{m\dagger} G_0^{T} (D^* \bar{D})^m \right]
+ \nu_0 \,\chi^{m\dagger}\chi^{m}.
\nonumber \\
\label{Hint}
\end{eqnarray}
where $(D^* \bar{D})^m$ is a 2-component column vector with a 
Cartesian vector index $m$.  Its upper and lower entries are the
combinations of fields that annihilate pairs of charm mesons in the 
neutral channel $(D^{*0} \bar D^0)_+$ and in the
charged channel $(D^{*+} D^-)_+$ given in Eqs.~(\ref{D*Dbar}).
The bare parameters are the three independent entries of the 
symmetric matrix $\Lambda_0$,
the two entries of the column vector $G_0$, and $\nu_0$.
The factors of $\sqrt{2\pi/\mu}$ in the Hamiltonian density 
have been inserted to simplify the renormalization equations.
The $\chi^{m\dagger}\chi^m$ term has been included 
in the interaction Hamiltonian,
because its coefficient $\nu_0$ requires renormalization.

The matrix of transition amplitudes ${\cal A}(E)$ for charm meson pairs 
in the channels $(D^{*0} \bar D^0)_+$ and $(D^{*+} D^-)_+$ can be calculated 
by summing diagrams constructed out of the two types of vertices, 
resonance propagators, and loop subdiagrams involving pairs of charm mesons.
The ultraviolet divergences in the loop diagrams can be regularized 
with an ultraviolet momentum cutoff $\Lambda_\textrm{UV}$.
The matrix ${\cal A}(E)$ can also be derived by solving the Lippmann-Schwinger 
integral equation, as described in Ref.~\cite{Braaten:2007nq}.
It is convenient to express the solution in terms of the matrix of 
scattering amplitudes $f(E)$ defined by Eq.~(\ref{Aij-fij}),
which differs from ${\cal A}(E)$ by a factor of $\mu/(2\pi)$.
The solution for the inverse of $f(E)$ is 
\begin{eqnarray}
f(E)^{-1} =
-\left( \Lambda_0 + G_0 \frac{1}{E - \nu_0} G_0^T \right)^{-1}
- \frac{2}{\pi} \Lambda_\textrm{UV} I + K(E),
\label{f-inverse:bare}
\end{eqnarray}
where $I$ is the identity matrix 
and $K(E)$ is the diagonal matrix given in Eq.~(\ref{K-kappa}). 
The energy dependence on the right side of Eq.~(\ref{f-inverse:bare})
allows $f(E)^{-1}$ to be expressed 
in the form in Eq.~(\ref{f-inverse:res}), in which the dependence on 
$\Lambda_\textrm{UV}$ has been absorbed into renormalized parameters.
The renormalized parameters are the three independent entries of 
the symmetric matrix $\Lambda$, the two entries of the column vector $G$, 
and $\nu$.  The relations between the renormalized parameters 
and the bare parameters are given by
\begin{subequations}
\begin{eqnarray}
\Lambda&=&Z^{-1} \Lambda_0,
\label{Lambda-renorm}
\\
G&=&Z^{-1}  G_0,
\label{G-renorm}
\\
\nu&=&\nu_0 - G_0^T \left( 1 - Z^{-1} \right) \Lambda_0^{-1} G_0,
\label{nu-renorm}
\end{eqnarray}
\label{param-renorm}
\end{subequations}
where the renormalization matrix $Z$ is
\begin{eqnarray}
Z = I + \frac{2}{\pi} \Lambda_\textrm{UV} \Lambda_0.
\label{ZC}
\end{eqnarray}
The verification of the equality between the expressions for $f(E)^{-1}$ in 
Eqs.~(\ref{f-inverse:bare}) and (\ref{f-inverse:res}) 
can be simplified by taking advantage of the existence of
renormalization-invariant combinations of parameters.
Rewriting Eq.~(\ref{Lambda-renorm}) as $Z^{-1} = \Lambda \Lambda_0^{-1}$ 
and inserting it into Eqs.~(\ref{G-renorm}) and (\ref{nu-renorm}),
we find that the following combinations of parameters are 
renormalization invariants:
\begin{subequations}
\begin{eqnarray}
\Lambda_0^{-1} G_0 &=& \Lambda^{-1} G,
\label{invariants-0}
\\
G_0^T \Lambda_0^{-1} &=&  G^T \Lambda^{-1},
\label{invariants-1}
\\
\nu_0 - G_0^T \Lambda_0^{-1} G_0 &=& \nu -G^T \Lambda^{-1} G.
\label{invariants-2}
\end{eqnarray}
\label{invariants}
\end{subequations}
The following function of $E$ is also a renormalization invariant:
\begin{eqnarray}
\left( \Lambda_0 + G_0 \frac{1}{E - \nu_0} G_0^T \right)^{-1} - \Lambda_0^{-1} 
= \left( \Lambda + G \frac{1}{E - \nu} G^T \right)^{-1} - \Lambda^{-1}.
\label{invariants-E}
\end{eqnarray}
This can be verified by multiplying both sides on the left by
$\Lambda_0 + G_0 G_0^T/(E - \nu_0)$ and on the right by
$\Lambda + G G^T/(E - \nu)$ and then using the renormalization 
invariants in Eqs.~(\ref{invariants}).
The equality between the expressions for $f(E)^{-1}$ in 
Eqs.~(\ref{f-inverse:bare}) and (\ref{f-inverse:res})
follows immediately from Eq.~(\ref{invariants-E}).

The complete propagator for the resonance $\chi$ can be obtained
by summing the geometric series of self-energy corrections.
The self-energy can be obtained by summing a geometric series of
one-loop diagrams:
\begin{equation}
\Sigma(E) = - G_0^T 
\left( \frac{2}{\pi} \Lambda_\textrm{UV} I - K(E) \right)
\left[ I + \Lambda_0 \left(\frac{2}{\pi} \Lambda_\textrm{UV} I - K(E) \right) \right]^{-1} 
G_0 .
\label{Sigma:bare}
\end{equation}
The complete resonance propagator is therefore
\begin{equation}
P(E) =
\left[ E - \nu_0 + G_0^T 
\left( \frac{2}{\pi} \Lambda_\textrm{UV} I - K(E) \right)
\left( \Lambda_0^{-1}+\frac{2}{\pi} \Lambda_\textrm{UV} I -K(E) \right)^{-1} 
 \Lambda_0^{-1} G_0 \right]^{-1} .
\label{prop:bare}
\end{equation}
This can be expressed in the form
\begin{equation}
P(E) =
\left[ E - \nu_0 + G_0^T \Lambda_0^{-1} G_0
- G_0^T \Lambda_0^{-1}
\left( \Lambda^{-1} - K(E) \right)^{-1} 
 \Lambda_0^{-1} G_0 \right]^{-1} .
\label{prop:semi}
\end{equation}
By using the renormalization invariants in Eqs.~(\ref{invariants}),
this propagator can be expressed in terms of renormalized parameters 
and then simplified to the renormalized expression in Eq.~(\ref{prop:res}).

\subsection{Renormalization of the Production Operators}

The production of particles at short distances
can be represented in an effective field theory by local operators that create
the particles when acting on the vacuum.  The matrix elements of the 
local operators are in general ultraviolet divergent.  
The corresponding renormalized operators are linear combinations 
that have finite matrix elements.  The renormalized operators
can be determined by calculating matrix elements of the local operators 
and then constructing linear combinations whose matrix elements are finite. 
Alternatively, the renormalized operators can be deduced by 
inspired guesswork.

In the Zero-Range+Resonance model, the particle that are produced at 
short distances can be pairs of charm mesons or the resonance.
There are three leading operators: the two components of the column vector
$(D^* \bar{D})^{m}$ that appears in the interaction Hamiltonian in 
Eq.~(\ref{Hint}) and the resonance field $\chi^{m}$.  
They all have a Cartesian vector index $m$.  We denote the
corresponding renormalized operators by ${\cal O}_0^m$, ${\cal O}_1^m$, 
and ${\cal O}_2^m$. The Green's function $F_{ij}(E)$ for pairs of the 
operators ${\cal O}_i^{m}$ are defined in Eq.~(\ref{Fij}). 
The Green's function for a pair of operators $\chi^{m}$ is just 
$-P(E)$, where $P(E)$ is the resonance
propagator in Eq.~(\ref{prop:bare}). Since this propagator can be expressed 
in the renormalized form in Eq.~(\ref{prop:res}),
the resonance field has finite matrix elements.  We can therefore 
choose its hermitian conjugate as one of the renormalized operators:
\begin{eqnarray}
{\cal O}_2^{m\dagger} = \chi^m.
\label{O-2}
\end{eqnarray}
The matrix elements of $(D^* \bar{D})^{m}$ are ultraviolet divergent.
The renormalized operators ${\cal O}_0^m$ and ${\cal O}_1^m$ are 
linear combinations of the hermitian conjugates of all three local operators.  
A particular convenient choice
consists of the operators obtained by differentiating 
${\cal H}_{\textrm{int}}$ with respect to the components of $(D^* \bar{D})^{m}$.
The Green's functions for these operators are proportional to the 
scattering amplitudes.  Our choices for the last two renormalized operators are
\begin{eqnarray}
\left(
\begin{array}{c}
{\cal O}_0^{m\dagger} \\
{\cal O}_1^{m\dagger}
\end{array}
\right)
= \sqrt{\frac{2\pi}{\mu}}\, \Lambda_0(D^*\bar{D})^m+G_0\chi^m.
\label{O-01}
\end{eqnarray}
The normalization factor has been chosen so that the Green's functions for 
these operators are exactly equal to the scattering amplitudes $f_{ij}(E)$ 
for $i,j \in \{ 0,1 \}$.
The remaining Green's function $F_{i2}(E)$ for $i=0,1$ are then given by
\begin{eqnarray}
F_{i2}(E) &=& 
 \left( f(E)~
        [(E- \nu_0)\Lambda_0 + G_0 G_0^T]^{-1}~G_0 \right)_i ,
\label{FiR:bare}
\end{eqnarray}
where $f(E)$ is the matrix of scattering amplitudes.
The corresponding renormalized expression is obtained by replacing
$\nu_0$, $\Lambda_0$, and $G_0$ by $\nu$, $\Lambda$, and $G$.
The equality of the two expressions for $F_{i2}(E)$ follows
from the fact that $\Lambda_0^{-1} G_0 = \Lambda^{-1} G$
is an eigenvector of both $\nu_0 I - \Lambda_0^{-1}G_0 G_0^T$
and $\nu I - \Lambda^{-1}G G^T$:
\begin{subequations}
\begin{eqnarray}
\left( \nu_0 I - \Lambda_0^{-1}G_0 G_0^T \right) \Lambda_0^{-1} G_0 &=& 
\left( \nu_0 - G_0^T \Lambda_0^{-1} G_0 \right) \Lambda_0^{-1} G_0 ,
\\
\left( \nu I - \Lambda^{-1}G G^T \right) \Lambda^{-1} G &=& 
\left( \nu - G^T \Lambda^{-1} G \right) \Lambda^{-1} G .
\end{eqnarray}
\end{subequations}

The Green's functions $F_{ij}(E)$ for $i,j \in \{ 0,1 \}$
are not exactly equal to the scattering amplitudes $f_{ij}(E)$.
The diagrams contributing to the $F_{ij}(E)$ 
are the same as the diagrams for $f_{ij}(E)$ except that the 
leading order diagram for scattering of charm meson pairs is omitted.
Thus $F_{ij}(E)$ actually differs from 
$f_{ij}(E)$ by the additive constant $(\Lambda_0)_{ij}$.
This constant gives an energy-independent contribution 
to the inclusive energy distribution in Eq.~(\ref{lsh:gen}).
This additive contribution can be interpreted as a constant background 
in decay channels of the $X(3872)$ that are not treated explicitly.

The renormalized operators ${\cal O}_0^{m}$ and ${\cal O}_1^{m}$
defined by Eq.~(\ref{O-01}) have different dimensions from
the renormalized operator ${\cal O}_2^{m} = \chi^m$.
For some purposes, it is more convenient for all the renormalized
operators to have the same dimensions.  One way to 
arrange for ${\cal O}_2^{m}$ to have the same dimension
as ${\cal O}_0^{m}$ and ${\cal O}_1^{m}$ is to multiply $\chi^m$ by
a constant with the appropriate dimension.
In the text, the multiplying factor is chosen to be
$(G^T G/2)^{1/2}$, where $G$ is the two-component vector 
of renormalized interaction parameters.  With this choice,
the renormalized expressions for $F_{22}(E)$ and $F_{i2}(E)$
are given by Eqs.~(\ref{FRR}) and (\ref{FiR}).


\end{document}